\documentclass[fleqn,12pt]{article}
\usepackage{amssymb,graphics,epsfig,rotating,cite,subfigure}
\def\beq   {\begin{equation}}
\def\eeq   {\end{equation}}
\def\beqd  {\begin{displaymath}}
\def\eeqd  {\end{displaymath}}
\def\beqaa {\begin{eqnarray}}
\def\eeqaa {\end{eqnarray}}

\def\ti  {\tilde}

\def\sq  {\ti q}
\def\st  {\ti t}

\def\sl  {\ti \ell}

\def\b   {\beta}
\def\t   {\theta}

\def\sz{\ifmmode{\tilde{\chi}^0} \else{$\tilde{\chi}^0$} \fi}
\def\sw{\ifmmode{\tilde{\chi}} \else{$\tilde{\chi}$} \fi}

\newcommand{\be}[1]{\begin{equation} \label{(#1)}}
\newcommand{\ee}{\end{equation}}
\newcommand{\baq}[1]{\begin{eqnarray} \label{(#1)}}
\newcommand{\eaq}{\end{eqnarray}}

\newcommand{\ba}{\begin{array}}
\newcommand{\ea}{\end{array}}

\newcommand{\smaf}[2] {{\textstyle \frac{#1}{#2} }}
\begin{document}
%------------------------------------------------------------------------
\pagestyle{empty}

\begin{flushright}
CERN-PH-TH/2008-180\\
  DCPT-08-130 \\
  IPPP-08-65 \\
\end{flushright}

\vfill

\begin{center}

{\Large {\bf
%A T-odd asymmetry from stop production and cascade decay at the LHC
Measurement of CP Violation in Stop Cascade Decays at the LHC
}}

\vspace{10mm}

{\large
J.~Ellis$^a$, F.~Moortgat$^b$, G.~Moortgat-Pick$^c$, J.~M.~Smillie$^c$, 
J.~Tattersall$^c$
}

\vspace{6mm}

\begin{flushleft}
$^a${\it Theory Division, Physics Department, CERN, CH-1211 Geneva 23, Switzerland}\\
$^b${\it Department of Physics, ETH Honggerberg, CH-8093 Zurich, Switzerland}\\
$^c${\it IPPP, University of Durham, Durham DH1 3LE, U.K.}\\
\end{flushleft}

\end{center}

\vfill

\begin{abstract}

  We study the potential observation at the LHC of CP-violating effects in stop production
  and subsequent cascade decays, $g g \to \tilde{t}_i \tilde{t}_i$, $\tilde{t}_i \to t
  \tilde{\chi}^0_j$, $\tilde{\chi}^0_j \to \tilde{\chi}^0_1 \ell^+ \ell^-$, within the
  Minimal Supersymmetric Standard Model.  We study T-odd asymmetries based on triple
  products between the different decay products.  There may be a large CP asymmetry
  at the parton level, but there is a significant dilution at the hadronic level after
  integrating over the parton distribution functions.
  Consequently, even for scenarios where
  large CP intrinsic asymmetries are expected, the measurable asymmetry is rather small.  
  High luminosity and precise measurements of masses,
  branching ratios and CP asymmetries may enable measurements of the
  CP-violating parameters in cascade
  decays at the LHC.
%  originally proposed in \cite{Bartl:2004jj} in the context of a linear collider, and
%  recently studied for the LHC in \cite{Langacker:2007ur}.

\end{abstract}

\vfill

\newpage
\pagestyle{plain}

% --- introduction ---

\section{Introduction}

The Minimal Supersymmetric Standard Model (MSSM) is a particularly compelling extension of
the Standard Model, that may soon be explored at the Large Hadron Collider (LHC).
%due to begin collisions in the next few months.  
Current data suggest that, if the
MSSM is realised in Nature, the supersymmetry scale should easily be within reach of the
LHC design centre-of-mass energy of 14 TeV~\cite{Buchmueller:2007zk,Heinemeyer:2008fb}.
If supersymmetry is discovered, many studies will be required to determine the exact details of its
realisation.

The MSSM contains a large number of (as yet) undetermined parameters that may
have non-zero phases~\cite{Hesselbach:2004sp,Hesselbach:2007dq}.  Many of these phases are unphysical in the sense that
they can be absorbed into the definitions of the fields; however, not all phases can be
consistently removed in this way.  In the neutralino/chargino sector of the complex MSSM,
the phase of the SU(2) gaugino mass $M_2$ is usually
absorbed, whereas the phases of the U(1) gaugino mass $M_1$ and the Higgsino
mixing parameter $\mu$ are generally left manifest; this is the
parameterisation we use. The trilinear couplings $A_f$ can also be complex.
Studies of these CP-violating parameters in sparticle decays and 
via other properties measurable at the
LHC will be challenging~\cite{Williams:2005gg,Langacker:2007ur}. However, they are extremely important, and provide a
valuable training ground for exploring the limits of the LHC's capabilities.

Certain combinations of the CP-violating MSSM
phases are constrained by the experimental upper bounds on
the electric dipole moments (EDMs) of the electron, neutron and atoms, notably $^{205}$Tl and $^{199}$Hg.
Ignoring possible cancellations, the most severely constrained individual phase in the MSSM is that of $\mu$,
which contributes at the one-loop level. For \it{O}\rm(100) GeV masses, one must require $|\phi_{\mu}|\lesssim 0.1$.
However, this restriction can be relaxed if the masses of the first- and second-generation squarks are large ($>$ TeV) while the
third-generation masses remain relatively small ($<$ TeV), or in the presence of cancellations between the
contributions of different CP-violating phases. We note that $M_1$ also contributes at the one-loop level, but
again, if accidental cancellations are allowed between terms, it  remains essentially unconstrained.
The phases of the third-generation trilinear couplings, $\phi_{A_{t,b,\tau}}$ have weaker constraints,
as they contribute to EDMs only at the two-loop level. Again, accidental cancellations can occur that
weaken further the constraints: see~\cite{Ibrahim:1997gj,Brhlik:1998zn,Bartl:1999bc,Pilaftsis:2002fe,Bartl:2003ju,Barger:2001nu,Pospelov:2005pr,Olive:2005ru,Abel:2005er,YaserAyazi:2006zw,Ellis:2008zy}.
A comprehensive summary of the EDM constraints and other CP-violating effects in SUSY is given in~\cite{Kraml:2007pr}.
Here we study the complete range of CP phases in order to see the general dependences exhibited by our observables, and what luminosity might be required to observe these within the LHC environment. Therefore, we do not calculate explicitly which values of the other phases might be required for the points in our displayed scenarios to satisfy the EDM constraints. 

The precise determination of these phases is expected to be possible only at an $e^+e^-$ linear
collider, for instance at the planned International Linear Collider (ILC) or at the Compact Linear
Collider (CLIC). However, it
will be crucial for such future search strategies to use LHC data to learn as much as
possible, as early as possible.  Furthermore,
the combination of independent measurements at the LHC and a
linear collider will be important to determine the underlying model.

In this paper we concentrate on the potential for observing unique CP-violating
effects in decay chains at the LHC, and investigate the circumstances under which a 
determination of the complex MSSM phases may be achievable, possibly with the
support of other LHC measurements.

Specifically, we consider the LHC process $gg \to \st \bar{\st}$, with subsequent
decay $\st \to t \tilde \chi_2^0$, $\tilde \chi_2^0 \to \ell^+ \ell^- \tilde \chi_1^0$.
We consider the situation where the $\tilde \chi_2^0$ decay is a three-body decay; this
leads to CP violation as there is a non-negligible contribution from interference
diagrams.  This process involves the three phases $\phi_{M_1}, \phi_\mu$ and $\phi_{A_t}$; we
discuss below the combinations to which this process is sensitive.  We extract information on the
phases using triple products formed from the decay products of the stop.  Such T-odd
variables have also been studied in the context of heavy squark and stau decays 
in~\cite{Bartl:2002hi,Bartl:2003ck,Bartl:2004jr,Bartl:2006hh,Langacker:2007ur,Kiers:2006aq}.  Other
related studies are \cite{Bartl:2002uy,Bartl:2002bh,Bartl:2003he,Bartl:2004ws,Bartl:2003pd,Kizukuri:1990iy,Choi:1999cc,Choi:2004rf,Choi:2005gt,Gajdosik:2004ed,Ibrahim:2004gb}.

The first CP-odd asymmetry we consider is formed from $\mathcal{T}_t=\bf p_t\cdot(\bf
p_{\ell^+} \times \bf p_{\ell^-})$.  This quantity has been studied at the parton level 
in~\cite{Langacker:2007ur}, assuming pure gaugino-like neutralinos.  In our current study we
provide analytic expressions for the squared amplitude of the cascade process
including full spin correlations and general neutralino mixing, and also provide an analytic
expression for the phase space in the laboratory system. We also incorporate parton
density functions (pdfs) and discuss the CP-odd observables at both the parton and the
hadronic levels. Transition to the latter level has a big dilution effect on the measurability of
a CP-odd asymmetry. We include the possible LHC uncertainties in masses and asymmetries
and discuss the extent to which CP-violating phases may be constrained in such cascade 
decays at the LHC.

In \cite{Bartl:2004jr}, further CP sensitive asymmetries formed from the momentum of
the $b$ quark in the top decay
were studied under the assumption of 2-body neutralino decays into on-shell sleptons,
namely $\mathcal{T}_b=\bf p_b\cdot(\bf p_{\ell^+} \times \bf p_{\ell^-})$ and
$\mathcal{T}_{tb}=\bf p_b\cdot(\bf p_t \times \bf p_{\ell^{\pm}})$.  These variables are
sensitive to $\phi_{M_1}$ and $\phi_{A_t}$, but have different dependences on the
CP-violating phases as described in Section~\ref{sect:spincor}.  Therefore, a combination of
all three observables would in principle allow one to disentangle the influences of all three phases.

Since T-odd observables can also be generated by final-state interactions at the one-loop
level, one should in principle combine the asymmetry for a process with that one of its
charge-conjugated process.  If a non-zero asymmetry is then observed in this combination,
it must correspond to a violation of CP symmetry.  For the triple product,
$\mathcal{T}_b$, this is experimentally possible as long as the associated $W$ decays into
a final-state lepton, which enables us to determine the change of the $\st$. Regarding the
other triple products, we require information from the opposite decay chain to identify
the charge. In all the scenarios we consider, the decay $\st\rightarrow\tilde{\chi}^+_ib$, is
dominant, enabling charge identification in principle. However,
a detailed simulation including all
combinatorial aspects and also other background processes would be required to validate
this possibility.

We begin by describing the process under consideration in Section~\ref{sec:formalism},
including the phases involved and their various effects.  In Section~\ref{sec:results} we
present numerical results for three specific benchmark scenarios and discuss the potential for a
measurement at the LHC.  The Appendices contain details of the Lagrangian, the expression
for the squared amplitude including full spin correlations, and the kinematics of the phase
space in the laboratory system.

%%%%%%%%%%%%%%%%%%%%%%%%%%%%%%%%%%%%%%%%%%%%%%%%%%%%%%%%%%%%%%%%%%%%%%%%%%%%%%%%%%%%%%%%%%%

\section{Formalism}
\label{sec:formalism}

\subsection{The process studied and its squared amplitude}
\label{sec:process}

\begin{figure}[ht!]
\hspace{-0.2cm}
\begin{minipage}[t]{6cm}
\begin{center}
{\setlength{\unitlength}{0.4cm}
\begin{picture}(5,9)
%\put(-1.2,-0.2){\psfig{file=ggu.ps,scale=0.5}}
\put(-1.2,-0.2){\epsfig{file=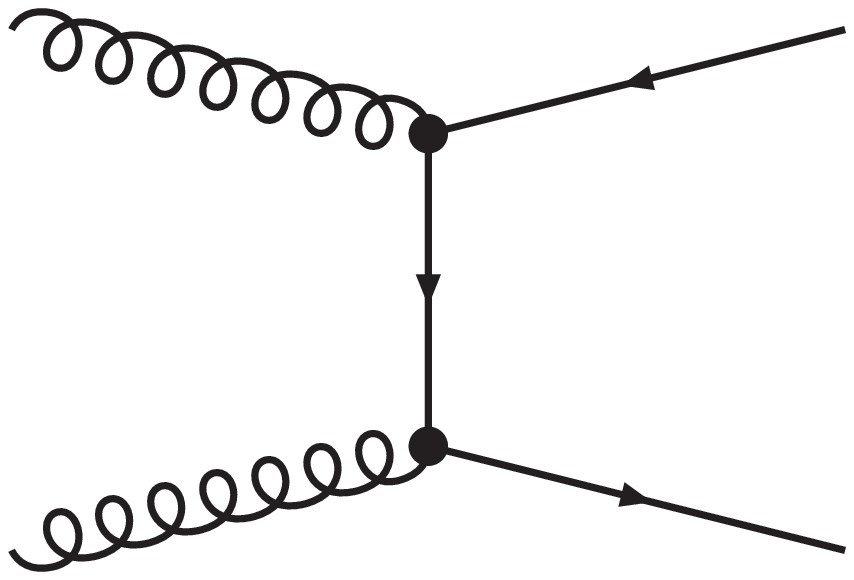,scale=0.5}}
 \put(0.2,0.1){{\small $g$}}
 \put(12,0.1){{\small $\overline{\st}_1$}}
 \put(0.2,6.7){{\small $g$}}
 \put(12,6.7){{\small $\st_1$}}
\end{picture}}
\end{center}
\end{minipage}
\hspace{1cm}
\begin{minipage}[t]{6cm}
\begin{center}
{\setlength{\unitlength}{0.4cm}
\begin{picture}(5,9)
\put(-1.3,-0.2){\epsfig{file=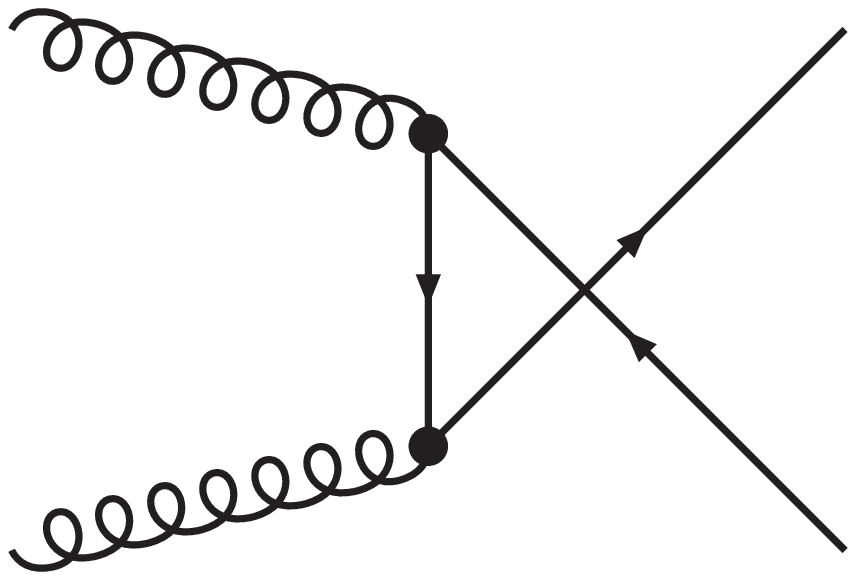,scale=0.5}}
 \put(0.2,0.1){{\small $g$}}
 \put(12,0.1){{\small $\overline{\st}_1$}}
 \put(0.2,6.7){{\small $g$}}
 \put(12,6.7){{\small $\st_1$}}
\end{picture}}
\end{center}
\end{minipage}

\vspace{0.5cm}
\hspace{-0.2cm}
\begin{minipage}[t]{6cm}
\begin{center}
{\setlength{\unitlength}{0.4cm}
\begin{picture}(5,9)
\put(0.6,0){\epsfig{file=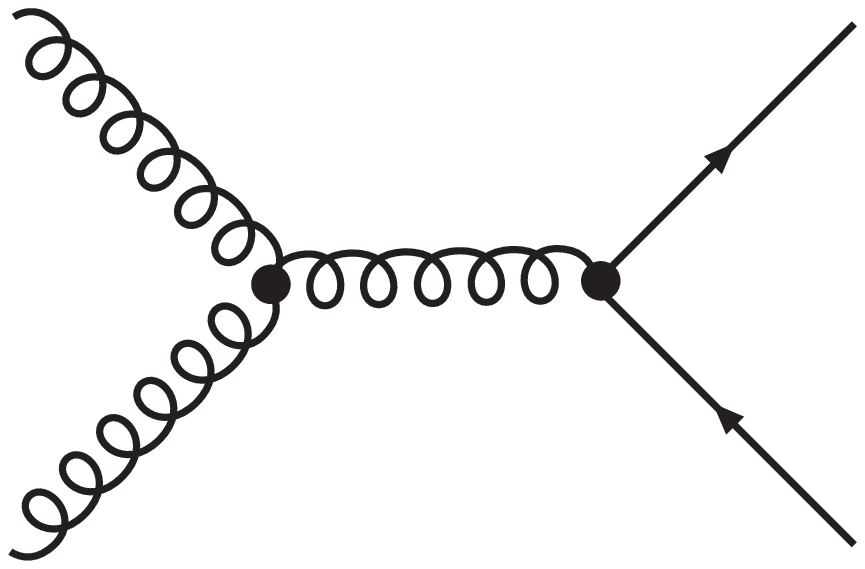,scale=0.5}}
 \put(0.2,0.1){{\small $g$}}
 \put(12,0.1){{\small $\overline{\st}_1$}}
 \put(0.2,6.7){{\small $g$}}
 \put(12,6.7){{\small $\st_1$}}
\end{picture}}
\end{center}
\end{minipage}
\hspace{1cm}
\begin{minipage}[t]{6cm}
\begin{center}
{\setlength{\unitlength}{0.4cm}
\begin{picture}(5,9)
\put(0.6,0){\epsfig{file=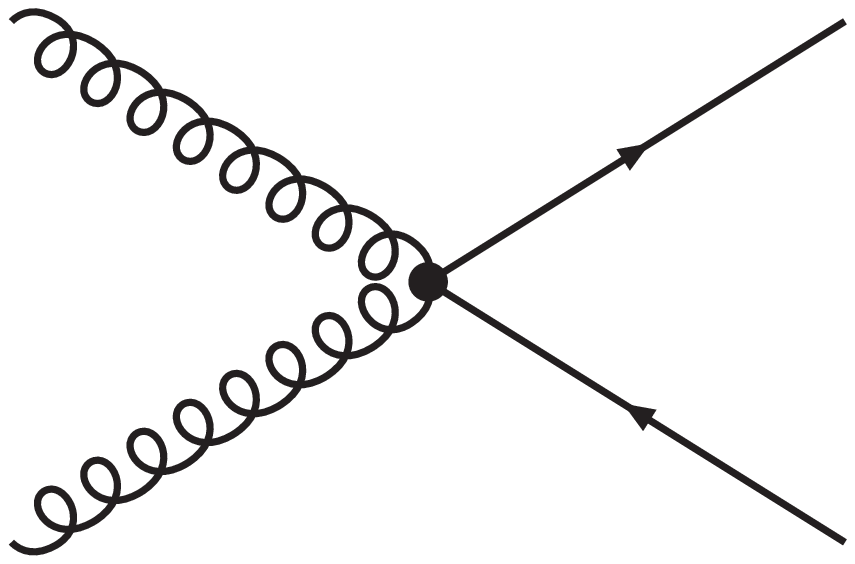,scale=0.5}}
 \put(0.2,0.1){{\small $g$}}
 \put(12,0.1){{\small $\overline{\st}_1$}}
 \put(0.2,6.7){{\small $g$}}
 \put(12,6.7){{\small $\st_1$}}
\end{picture}}
\end{center}
\end{minipage}
%\vspace{1cm}
\caption{\label{Fig:FeynProd}Feynman diagrams for the production process
  $gg\to\st_1\overline{\st}_1$.}
\vspace{1cm}
\end{figure}

We study the dominant stop production process at the LHC, namely
\begin{equation}
gg \to \tilde{t}_i \tilde{\bar{t}}_i,
\label{eq-prod}
\end{equation}
with the subsequent decay chain
\begin{eqnarray}
&&\tilde{t}_i \to \tilde{\chi}^0_j + t \to \tilde{\chi}^0_1 \ell^+\ell^- + W b.
\label{eq-decays} 
\end{eqnarray}
At tree level, the production process (\ref{eq-prod})  proceeds via $g$ exchange 
in the direct channel and $\st$
exchange in the crossed channel, and via a quartic coupling, as shown in
Fig.~\ref{Fig:FeynProd}.  
Another possible source of $\st_1$s is their production in gluino decays,
$\tilde{g} \to \st t$. However this leads to an experimentally more complex topology than the
direct production and consequently we do not investigate this channel.
The Lagrangian and the resulting neutralino and stop mixings and couplings are 
described in Appendix~\ref{sec:lagrangian-couplings}.  

Since gluons do not couple to off-diagonal combinations
of stop mass eigenstates of opposite chirality, and similarly for stop exchange and
the quadratic couplings, $\tilde{t}_1\bar{\tilde{t}}_2$ production occurs only at the
loop level, and we do not consider it here. We focus here on
$\tilde{t}_1\bar{\tilde{t}}_1$ production, since the reconstruction of full decay chains of
$\tilde{t}_1$ seems achievable, even in the complex experimental environment at the LHC.
With the exception of the stop mass eigenvalues, see
Appendix~\ref{sec:lagrangian-couplings}, no effects from supersymmetric CP-violating couplings 
occur in the tree-level production process.

\begin{figure}[ht!]
\hspace{-0.5cm}
\begin{minipage}[t]{5cm}
\begin{center}
{\setlength{\unitlength}{0.8cm}
\begin{picture}(5,7)
\put(-5.2,-5.2){\includegraphics{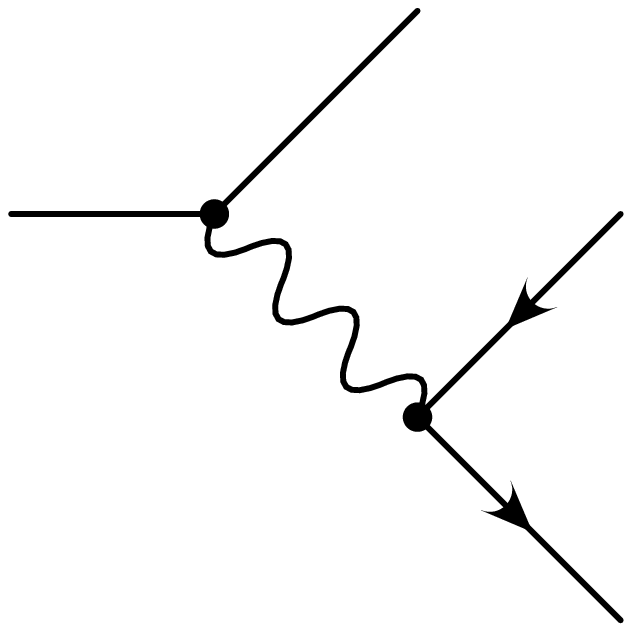}}
\put(4.7,1.7){{\small $\ell^{-}$}}
\put(0.3,5.5){{\small $\tilde{\chi}^0_{i}$}}
\put(4.7,5.5){{\small $\ell^{+}$}}
\put(3.7,7.1){{\small $\tilde{\chi}^0_{k}$}}
\put(2.2,3.8){{\small $Z^0$}}
\end{picture}}
\end{center}
\end{minipage}
\hspace{0.cm}
\begin{minipage}[t]{5cm}
\begin{center}
{\setlength{\unitlength}{0.8cm}
\begin{picture}(5,7)
\put(-5.2,-6){\includegraphics{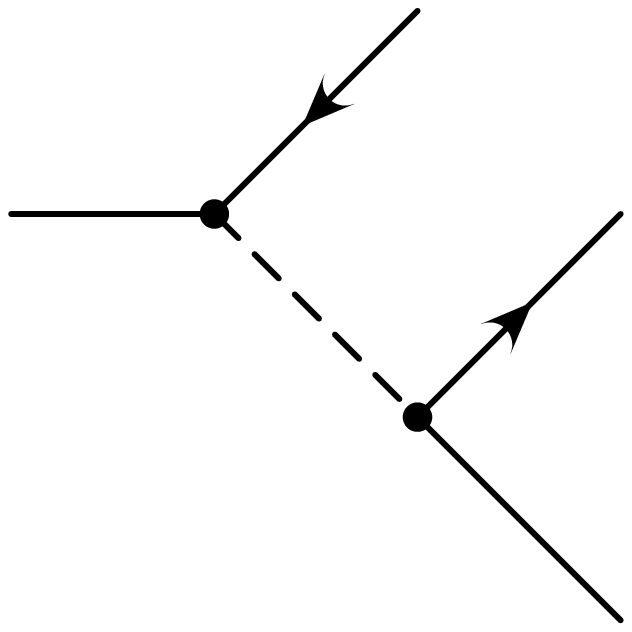}}
\put(4.7,5.5){{\small $\ell^{-}$}}
\put(3.7,7.1){{\small $\ell^{+}$}}
\put(4.7,1.5){{\small $\tilde{\chi}^0_{k}$}}
\put(0.3,5.5){{\small $\tilde{\chi}^0_{i}$}}
\put(2.1,3.9){{\small $\tilde{\ell}_{L,R}$}}
\end{picture}}
\end{center}
\end{minipage}
\begin{minipage}[t]{5cm}
\begin{center}
{\setlength{\unitlength}{0.8cm}
\begin{picture}(5,7)
\put(-5.2,-6){\includegraphics{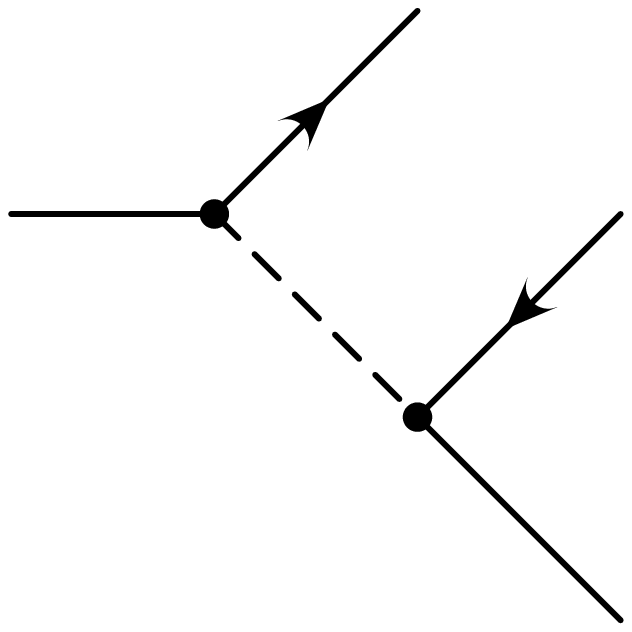}}
\put(3.7,7.1){{\small $\ell^{-}$}}
\put(4.7,5.5){{\small $\ell^{+}$}}
\put(0.3,5.5){{\small $\tilde{\chi}^0_{i}$}}
\put(4.7,1.5){{\small $\tilde{\chi}^0_{k}$}}
\put(2.1,3.9){{\small $\tilde{\ell}_{L,R}$}}
\end{picture}}
\end{center}
\end{minipage}
\vspace{-1.2cm}
\caption{\label{Fig:FeynDecayA}Feynman diagrams for the three-body decays
$\tilde{\chi}^0_i\to \tilde{\chi}^0_k\ell^{+}\ell^{-}$.}
\end{figure}

The first step in the cascade decay chain is the two-body process $\tilde{t}_i \to t
\tilde{\chi}^0_2$.  Here CP-violating couplings of the $\tilde{t}_1$ enter as well as those of the
$\tilde{\chi}^0_2$, and are dominated by the phases $\phi_{A_t}$ and $\phi_{M_1}$, see
Appendix~\ref{sect:stopdecay}~\footnote{Their structure has also been studied in detail
in~\cite{Bartl:2004jr}.}.  Since constraints from electric dipole measurements constrain
strongly $\phi_{\mu}$, we set $\phi_{\mu}=0$ in our 
study~\cite{Baker:2006ts,Romalis:2000mg,Regan:2002ta}.
We consider spectra where the
second steps in the cascade decay chains are the three-body decays of the neutralino,
$\tilde{\chi}^0_2 \to \tilde{\chi}^0_1 \ell^+ \ell^-$ (cf. Appendix~\ref{sect:neutdecay})
and the dominant top decay $t \to W b$ (cf. Appendix~\ref{sect:topdecay}).  The neutralino decay
occurs via $Z^0$ exchange in the direct channel and via $\tilde{\ell}_{L,R}$ exchanges in the
crossed channels, cf. Fig.~\ref{Fig:FeynDecayA}.  It is very sensitive to CP-violating
supersymmetric couplings, and its structure has been studied in detail 
in~\cite{Bartl:2004jj,MoortgatPick:1999di}.  The phase $\phi_{M_1}$ (and also $\phi_{\mu}$,
which has been set to zero here) affects the mass of the $\tilde \chi_2^0$, 
as well as its couplings and decay rates.

\vspace{1cm}

Using the formalism of~\cite{MoortgatPick:1999di,Haber:1994pe},
the squared amplitude $|T|^2$ of the full process can be factorized into the 
processes of production $gg\to \tilde{t}_1\bar{\tilde{t}}_1$ and the subsequent decays 
$\tilde{t}_1 \to t \tilde{\chi}^0_2$, 
$\tilde{\chi}^0_2\to\tilde{\chi}^0_1\ell^+\ell^-$ and $t\to W b$, with
the second $\tilde{t}_1$ being unobserved. We apply the narrow-width approximation for the
masses of the intermediate particles, $\tilde{t}_1$, $\tilde{\chi}^0_2$ and $t$, which is appropriate since
the widths of the respective particles are in all cases much smaller than their masses, cf. 
Table~\ref{tab:scentab}. the squared amplitude can then be expressed in the form
\begin{eqnarray}
|T|^2 &=& 4 |\Delta(\tilde{t}_1)|^2 |\Delta(\tilde{\chi}^0_2)|^2  |\Delta(t)|^2
 P(\tilde{t}_1\tilde{t}_1) \Large\{ P(\tilde{\chi}^0_2 t)
 D(\tilde{\chi}^0_2) D(t)
             +
            \sum^3_{a=1}\Sigma^a_P(\tilde{\chi}^0_2)
                    \Sigma^a_D(\tilde{\chi}^0_2)D(t)\nonumber\\
      &&       +
            \sum^3_{b=1}\Sigma^b_P(t)
                    \Sigma^b_D(t)D(\tilde{\chi}^0_2)
             +
            \sum^3_{a,b=1}\Sigma^{ab}_P(\tilde{\chi}^0_2t)
                    \Sigma^a_D(\tilde{\chi}^0_2) \Sigma^b_D(t)
\Large\},
\label{Tsquared}
\end{eqnarray}
where $a=1,2,3$ refers to the polarisation states of the neutralino $\tilde{\chi}^0_i$ and
top quark, which are described by the polarisation vectors $s^a(\tilde{\chi}^0_i)$,
$s^b(t)$ given in Appendix~\ref{sect:stopdecay}.  In addition,
\begin{itemize}
\item $\Delta(\tilde{t}_1)$, $\Delta(\tilde{\chi}^0_2)$ and $\Delta(t)$ are the
  `propagators' of the intermediate particles which lead to the factors
  $E_{\tilde{t}_1}/m_{\tilde{t}_1}\Gamma_{\tilde{t}_1}$,
  $E_{\tilde{\chi}^0_2}/m_{\tilde{\chi}^0_2}\Gamma_{\tilde{\chi}^0_2}$ and
  $E_t/m_t\Gamma_{t}$ in the narrow-width approximation.
\item $P(\tilde{t}_1\tilde{t}_1)$, $P(t\tilde{\chi}^0_2)$, $D(\tilde{\chi}^0_{i})$
and $D(t)$ are the terms in the
  production and decay that are independent of the polarisations of the decaying 
  neutralino and top,  whereas
\item $\Sigma^a_P(\tilde{\chi}^0_{i})$, $\Sigma^b_P(t)$,
  $\Sigma^{ab}_P(\tilde{\chi}^0_2t)$ and $\Sigma^a_D(\tilde{\chi}^0_{i})$, $\Sigma^b_D(t)$
  are the terms containing the correlations between production and decay spins of the $\tilde{\chi}^0_2$ and $t$.
\end{itemize}

According to our choice of the polarisation vectors $s^a(\tilde{\chi}^0_i)$ [$s^b(t)$], see
eqs.~(\ref{eq-s1chi})--(\ref{eq-s3t}) in Appendix \ref{sect:stopdecay}, $\Sigma^3_P/P$ is
the longitudinal polarisation, $\Sigma^1_P/P$ is the transverse polarisation in the
production plane, and $\Sigma^2_P/P$ is the polarisation perpendicular to the reference
plane of the neutralino $\tilde{\chi}^0_i$ [top quark $t$].

%%%%%%%%%%%%%%%%%%%%%%%%%%%%%
\subsection{Cross section for the whole process at parton level}
%%%%%%%%%%%%%%%%%%%%%%%%%%%%%
The differential cross section in the laboratory
system is
\begin{equation}
d\sigma =
 \frac{1}{8E^2_b}|T|^2 
 (2\pi)^4 \delta^4(p_1+p_2-\sum_n p_n)d \mbox{lips}(p_n), 
\end{equation}
where $E_b$ is the beam energy of the gluons, $p_1$ and $p_2$ are the momenta of the incoming gluons,
the $p_n$ are the momenta of the outgoing particles and $d\mbox{lips}(p_n)$ is the 
Lorentz-invariant phase space element. Integrating over all angles, all
spin-dependent contributions are cancelled and the cross section for the combined process
of production and decay is given by
\begin{eqnarray}
\mbox{\hspace{-1cm}}
\sigma & = & \sigma(gg \to \tilde{t}_1\bar{\tilde{t}}_1)
\times BR(\tilde{t}_1\to t \tilde{\chi}^0_2)\times
BR(\tilde{\chi}^0_2 \to \tilde{\chi}^0_1 \ell^+ \ell^-) \times BR(t \to W b) 
\nonumber\\
 & = & \frac{|\Delta(\tilde{t}_1)|^2|\Delta(\tilde{\chi}^0_2)|^2|\Delta(t)|^2}{2 E^2_b}
\nonumber\\
&&
\times \int  P(\tilde{t}_1\tilde{t}_1)
P(\tilde{\chi}^0_2 t) D(\tilde{\chi}^0_2) D(t)
 (2\pi)^4\delta^4(p_1+p_2-\sum_np_n)d \mbox{lips}(p_n).
\label{cross}
\end{eqnarray}
The explicit expression for the phase space in the laboratory system is given in Appendix~\ref{sec:kinematics}.

%%%%%%%%%%%%%%%%%%%%%%%%%%
\subsection{Structure of the T-odd asymmetries \label{sect:spincor}}
%%%%%%%%%%%%%%%%%%%%%%%%%%
Suitable tools to study CP-violating effects are T-odd observables
based on triple products of momenta or spin vectors of the involved
particles. In this paper we study the following T-odd observables:
\begin{eqnarray}
\mathcal{T}_t&=&\vec{p}_{t}
 \cdot (\vec{p}_{\ell^+} \times \vec{p}_{{\ell}^-})~,
\label{eq_tpt}\\
\mathcal{T}_b&=&\vec{p}_{b}\cdot (\vec{p}_{\ell^+}\times\vec{p}_{\ell^-})~,
\label{eq_tpb}\\
\mathcal{T}_{tb}&=&\vec{p}_{t}\cdot (\vec{p}_{b}\times\vec{p}_{\ell^\pm}).
\label{eq_tptb}
\end{eqnarray}
The T-odd asymmetries are defined as
\begin{equation}
\label{Asy}
\mathcal{A}_{T_f} = 
\frac{N_{\mathcal{T}_f+}-N_{\mathcal{T}_f-}}{N_{\mathcal{T}_f+}+N_{\mathcal{T}_f-}} =
\frac{\int\mathrm{sign}\{ \mathcal{T}_f\}
  |T|^2d\mbox{lips}}{{\int}|T|^2d\mbox{lips}},\quad\quad f=t,b \mbox{ and } tb ,
\end{equation}
where $N_{\mathcal{T}_f+}$, $N_{\mathcal{T}_f-}$ are the numbers of events for which
$\mathcal{T}_f$ is positive and negative respectively, and the second denominator in
(\ref{Asy}), ${\int}|T|^2d\mbox{lips}$, is proportional to the corresponding
cross section, namely $\sigma(g g \to
\tilde{t}_1\tilde{\bar{t}}_1 \to t \tilde{\chi}^0_1 \ell^+ \ell^-)$ in eq.(\ref{eq_tpt})
and $\sigma(g g \to \tilde{t}_1\tilde{\bar{t}}_1 \to W b \tilde{\chi}^0_1 \ell^+ \ell^-)$
in eqs.~(\ref{eq_tpb}) and (\ref{eq_tptb}).  In the second numerator
in (\ref{Asy}), only the triple-product correlations enter via the spin-dependent terms,
as explained in eq.~(\ref{CPinT}) and the following sections.
 
%It is essential to include the spin correlations between production
%and decay, because otherwise the numerator and hence the
%asymmetry $A_T$ would vanish.
The observable $\mathcal{A}_{T_b}$ has the advantage that it is not
necessary to reconstruct the momentum of the decaying $t$ quark. 
However, as explained below, in order to disentangle the effects of 
both phases of 
$A_t$ and $M_1$, it will be necessary to study all possible observables.

The asymmetry $\mathcal{A}_{T_f}$, eq.~(\ref{Asy}), is odd under the na\"ive time-reversal
operation.  It is the difference of the number of events with the final top quark or $b$-jet
above and below the plane spanned by $\vec{p}_{\ell^+}\times \vec{p}_{\ell^-}$ in
eqs.~(\ref{eq_tpt}) and (\ref{eq_tpb}), and by $\vec{p}_{b}\times \vec{p}_{\ell^{\pm}}$ in
eq.~(\ref{eq_tptb}), normalised by the sum of these events.

As can be seen from the numerator of $\mathcal{A}_{T_f}$, in order to identify the T-odd
contributions, we have to identify those terms in $|T|^2$, eq.~(\ref{Tsquared}), which
contain a triple product of the form shown in eqs.~(\ref{eq_tpt})--(\ref{eq_tptb}).
Triple products follow from expressions $i\epsilon_{\mu\nu\rho\sigma}a^\mu b^\nu c^\rho
d^\sigma$, where $a$, $b$, $c$, $d$ are 4-momenta and spins of the particles involved,
which are non-zero only when the momenta are linearly independent.  The expressions
$i\epsilon_{\mu\nu\rho\sigma}a^\mu b^\nu c^\rho d^\sigma$ are imaginary and when
multiplied by the imaginary parts of the respective couplings they yield terms that
contribute to the numerator of $A_{T_f}$, eq.~(\ref{Asy}).  In our process, T-odd terms
with $\epsilon$-tensors are only contained in the spin-dependent contributions to the
production, $\Sigma^{ab}_P(\tilde{\chi}^0_j t)$, and in the spin-dependent terms in neutralino
decay, $\Sigma^a_D(\tilde{\chi}^0_j)$.  It is therefore convenient to split
$\Sigma^{ab}_P(\tilde{\chi}^0_j t)$ and $\Sigma^a_D(\tilde{\chi}^0_j)$ into T-odd terms
$\Sigma^{ab,\mathrm{O}}_{P}(\tilde{\chi}^0_j t)$ and
$\Sigma^{a,\mathrm{O}}_{D}(\tilde{\chi}^0_j)$ containing the respective triple products,
and T-even terms $\Sigma^{ab,\mathrm{E}}_{P}(\tilde{\chi}^0_j t)$ and
$\Sigma^{a,\mathrm{E}}_{D}(\tilde{\chi}^0_j)$ without triple products:
\begin{equation}
 \Sigma^{ab}_P(\tilde{\chi}^0_j t) =
  \Sigma^{ab,\mathrm{O}}_P(\tilde{\chi}^0_j t) +
  \Sigma^{ab,\mathrm{E}}_P(\tilde{\chi}^0_j t), \qquad
 \Sigma^a_D(\tilde{\chi}^0_j) =
  \Sigma^{a,\mathrm{O}}_D(\tilde{\chi}^0_j) +
  \Sigma^{a,\mathrm{E}}_D(\tilde{\chi}^0_j).
\end{equation}
The other spin-dependent contributions $\Sigma^{a}_P(\tilde{\chi}^0_j)$ and
$\Sigma^{b}_P(t)$, as well as $\Sigma^{b}_D(t)$, are T-even.

When multiplying these terms together and composing a T-odd quantity, 
the only terms of $|T|^2$, eq.~(\ref{Tsquared}), which
contribute to the numerator of $\mathcal{A}_{T_f}$ are therefore
\begin{equation} \label{CPinT}
\mbox{\hspace{-1.1cm}}
 |T|^2 \supset 
  \sum^3_{a,b=1} \left[\Sigma^{ab,\mathrm{O}}_P(\tilde{\chi}^0_j t)
                    \Sigma^{a,\mathrm{E}}_D(\tilde{\chi}^0_j)\Sigma^{b}_D(t)
   + \Sigma^{a,\mathrm{E}}_P(\tilde{\chi}^0_j)
       \Sigma^{a,\mathrm{O}}_D(\tilde{\chi}^0_j)
   + \Sigma^{ab,\mathrm{E}}_P(\tilde{\chi}^0_j t)
                    \Sigma^{a,\mathrm{O}}_D(\tilde{\chi}^0_j)\Sigma^{b}_D(t) \right].
\end{equation}
The first term in eq.~(\ref{CPinT}) is sensitive to
the T-odd contributions from the production 
of the top and the neutralinos $\tilde{\chi}^0_j$. 
Comparing eq.(\ref{eq_prod-o}) with eqn.(\ref{eq_dssum-e}) and (\ref{eq_stdecay}) leads to the following
possible combination of contributing momenta 
\begin{equation}
\Sigma^{ab,\mathrm{O}}_P(\tilde{\chi}^0_j t)
                    \Sigma^{a,\mathrm{E}}_D(\tilde{\chi}^0_j)\Sigma^{b}_D(t)\sim 
 \epsilon_{\mu\nu\rho\sigma}s^{a,\mu}(\tilde{\chi}^0_j)p^{\nu}_{\tilde{\chi}^0_j}
s^{b,\rho}(t)p^{\sigma}_t \times
(p_{[\ell^{+},\ell^-]} s^{a})(p_{[b,W]}s^b).
\label{eq_term1}
\end{equation}

The second term and third terms in eq.~(\ref{CPinT}) are only sensitive to T-odd contributions from the
neutralino $\tilde{\chi}^0_j$ decay. The second term depends only on the polarization of 
$\tilde{\chi}^0_j$, comparing eq.(\ref{eq_dssum-o}) 
with eq.(\ref{eq_prod-ea}) therefore leads to the only possible
combination of momenta
\begin{equation}
  \Sigma^{a,\mathrm{E}}_P(\tilde{\chi}^0_j)
       \Sigma^{a,\mathrm{O}}_D(\tilde{\chi}^0_j) \sim 
 (p_t s^a) \times 
\epsilon_{\mu \nu \rho \sigma}
 s^{a \mu} p_{\tilde{\chi}^0_j}^{\nu} p_{\ell^-}^{\rho} p_{\ell^+}^{\sigma}. 
\label{eq_term2}
\end{equation}
Since the third term depends on the polarization of both fermions, $\tilde{\chi}^0_j$ and $t$,
the possible combinations, comparing eq.(\ref{eq_dssum-o}) with eqn.(\ref{eq_prod-e}) and (\ref{eq_stdecay}), 
are
\begin{eqnarray}
 \Sigma^{ab,\mathrm{E}}_P(\tilde{\chi}^0_j t)
\Sigma^{a,\mathrm{O}}_D(\tilde{\chi}^0_j)\Sigma^{b}_D(t)
&\sim& (p_t s^a)(p_{\tilde{\chi}^0_j} s^b) (s^b p_{[b,W]})
\times \epsilon_{\mu \nu \rho \sigma}
 s^{a \mu} p_{\tilde{\chi}^0_j}^{\nu} p_{\ell^-}^{\rho} p_{\ell^+}^{\sigma}
\label{eq_term3}
\\
&{\mbox{\rm and}}& (s^a s^b) (s^b p_{[b,W]})\times 
\epsilon_{\mu \nu \rho \sigma}
 s^{a \mu} p_{\tilde{\chi}^0_j}^{\nu} p_{\ell^-}^{\rho} p_{\ell^+}^{\sigma}
\label{eq_term3-ss}.
\end{eqnarray}

%In the following we derive explicitly the T-odd
%contributions to the spin density matrices of production and decay.

As can be seen by substituting eqs.~(\ref{eq-s1chi})--(\ref{eq-s3t}) into
eq.~(\ref{eq_prod-o}) in Appendix \ref{sect:stopdecay}, $\Sigma^{ab,O}_P(\tilde{\chi}^0_jt)$ 
vanishes for the combinations $(ab)=(11),(22),(33),(13),(31)$, because they contain
cross products of three linearly-dependent vectors.  Only for the remaining combinations,
$(ab)=(12),(21),(23),(32)\,$, do we get a T-odd contribution to the production density
matrix.

Similarly, the expression for the T-even contributions, $\Sigma^{ab,E}_P(\tilde{\chi}^0_j t)$, 
eq.~(\ref{eq_prod-e}) in Appendix
\ref{sect:stopdecay}, has non-zero components for $a=1,3$ but vanishes
when $a=2$.  These expressions are multiplied by $\Sigma^{a,\mathrm{O}}_D(\tilde{\chi}^0_j)$,
eq.~(\ref{eq_dssum-o}), and therefore only
$\Sigma^{1,\mathrm{O}}_D(\tilde{\chi}^0_j)$ and
$\Sigma^{3,\mathrm{O}}_D(\tilde{\chi}^0_j)$ contribute.

In the following section we derive the three triple products, 
%eqs.(\ref{eq_tpt})--(\ref{eq_tptb}),
study their different dependence on phases 
and provide explicitly a strategy for determining $\phi_{A_{t}}$ and 
$\phi_{M_{1}}$ and 
disentangling their effects.

%%%%%%%%%%%%%%%%%%%%%%%%%%%%%%%%%%%%%%%%%%%%%%%%%%%%%%%%%%%
\subsection{Strategy for determining $\phi_{A_t}$ and $\phi_{M_1}$}

\subsubsection{Derivation of the triple products \label{sect:241}}
%%%%%%%%%%%%%%%%%%%%%%%%%%%%%%%%%%%%%%%%%%%%%%%%%%
In order to describe the spin of a fermion $f$ in general, we 
introduce three four vectors, $s^a_{\mu}(f)$, $a=1,2,3$, such that  
the $s^a$ and the momentum and mass of the fermion $p/m$ form an orthonormal set of
four-vectors~\cite{Haber:1994pe};
\begin{eqnarray}
p \cdot s^a &=& 0,\\
s^a \cdot s^b &=& -\delta^{ab},\\
s^a_{\mu} s^a_{\nu} &=& -g_{\mu\nu} + \frac{p_{\mu} p_{\nu}}{m^2}, \label{eq_tensor} 
\end{eqnarray}
where repeated indices are implicitly summed over.

Applying eq.(\ref{eq_tensor}) on eqs.(\ref{eq_term1})--(\ref{eq_term3-ss}) lead to 
kinematic expressions that contain only explicit momenta. Expanding terms with
$\epsilon_{\mu\nu\rho\sigma}$ in  time- and space- components gives
scalar triple products between three momenta. 

In our process we can classify the terms of eq.(\ref{CPinT}) as follows:
\begin{itemize}
\item The terms of eq.(\ref{eq_term1}) lead to a combination between 
$\mathcal{T}_{tb}$ and $\mathcal{T}_{b}$.
\item The terms of eq.(\ref{eq_term2}) lead only to $\mathcal{T}_{t}$.
\item The terms of eq.(\ref{eq_term3}) lead again only 
to $\mathcal{T}_{t}$ but terms of eq.(\ref{eq_term3-ss}) produce
$\mathcal{T}_{t}$ as well as $\mathcal{T}_{b}$, due to interference 
effects between both spin vectors of $p_t$ and $p_{\tilde{\chi}^0_j}$.
\end{itemize}

%%%%%%%%%%%%%%%%%%%%%%%%%%%%%%%%%%%%%%%%%%%%%%%%%%%%%%%%%%
\subsubsection{T-odd terms sensitive to $\mathcal{T}_t$}
%%%%%%%%%%%%%%%%%%%%%%%%%%%%%%%%%%%%%%%%%%%%%%%%%%%%%%%%%%
\label{sec:t-odd-prod}
We consider first $\mathcal{T}_t$, eq.~(\ref{eq_tpt}).  As this includes the reconstructed
top quark momentum, there are no spin terms from the decay of the top quark 
and
the contributing terms are the second and third term in eq.~(\ref{CPinT}) as explained in 
the previous paragraphe. 
The CP-sensitive
terms of the decay density matrix are given by eqs.~(\ref{eq_dssum-o})--(\ref{eq_dselel_to})
and 
%but the only kinematical factor that can generate this triple product is $g_4^a$ for $a=1$,
the contributing kinematical factor is $g_4^a$, eq.(\ref{eq_dssub6}),
\begin{equation}
g_4^{a}=i m_k \epsilon_{\mu\nu\rho\sigma} s^{a\mu} p_{\tilde{\chi}^0_j}^{\nu}
p_{\ell^-}^{\rho} p_{\ell^+}^{\sigma}. \label{g4-decay}
\end{equation}
We note that $g^{a}_4$ is purely imaginary.
When inserted, for instance, in eq.~(\ref{eq_dszel_to})
it is multiplied by the factor
$i\cdot Im\{f^L_{\ell j} f^{L*}_{\ell k} O^{''L*}_{kj}\}$,
which depends on the phases $\phi_{M_1}$ (and $\phi_{\mu}$)
and contributes to $\Sigma_D^{a, \mathrm{O}}$.
Analogous contributions follow from 
eqs.~(\ref{eq_dszz_to}) and (\ref{eq_dselel_to}).
The corresponding T-even terms of the production
density matrix also entering in eq.~(\ref{CPinT}) are
obtained from eq.~(\ref{eq_prod-ea}).

%%%%%%%%%%%%%%%%%%%%%%%%%%%%%%%%%%%%%%%%%%%%%%%%%%%%%%%%%%
\subsubsection{T-odd terms sensitive to $\mathcal{T}_{tb}$} 
%%%%%%%%%%%%%%%%%%%%%%%%%%%%%%%%%%%%%%%%%%%%%%%%%%%%%%%%%%
\label{sec:t-odd-decay}
For the triple product $\mathcal{T}_{tb}$, eq.~(\ref{eq_tptb}), only the first term in 
eq.~(\ref{CPinT})
contributes, but the kinematics is complicated by the fact that we need to include
the decay of the $t$ in addition to that of the $\tilde{\chi}^0_2$. This comes from
the fact that the kinematical term that generates the triple product is $f_4^{ab}$, 
eq.(\ref{eq_f4}):
\begin{equation}
 f_4^{ab}= \epsilon_{\mu\nu\rho\sigma}s^{a,\mu}(\tilde{\chi}^0_j)p^{\nu}_{\tilde{\chi}^0_j}
s^{b,\rho}(t)p^{\sigma}_t.
\end{equation}
As both $s^{a,\mu}(\tilde{\chi}^0_j)$ and $s^{b,\rho}(t)p^{\sigma}_t$ are contained in this
term, we need to include their decays in order to to produce a non-zero contribution.

This term occurs only once in the $\st$ decay amplitude,
eq.~(\ref{eq_prod-o}), and is multiplied by the complex pre-factor
$g^2Im(a_{ij}b^*_{ij})$, eq.~(\ref{eq_Imab}). Both $a_{ij}$ and
$b_{ij}$ contain terms from the $\st$ and $\tilde{\chi}^0_j$ mixing matrices,
and so are sensitive to both the phases $\phi_{A_t}$ and $\phi_{M_1}$ (and $\phi_{\mu}$).

%%%%%%%%%%%%%%%%%%%%%%%%%%%%%%%%%%%%%%%%%%%%%%%%%%%%%%%%%%
\subsubsection{T-odd terms sensitive to $\mathcal{T}_{b}$} 
%%%%%%%%%%%%%%%%%%%%%%%%%%%%%%%%%%%%%%%%%%%%%%%%%%%%%%%%%%

The triple product $\mathcal{T}_b$, eq.~(\ref{eq_tpb}), is the most
complicated, as it contains contributions from both the $\st$ and
$\tilde{\chi}^0_2$ decays (the first and third terms in eq.~(\ref{CPinT})). The
kinematics is rendered more complex by the need to multiply each T-odd
contribution by the terms from the other two decays. Each T-odd
component is generated through $g_4^a$ and $f_4^{ab}$, as for the
other two triple products. As a consequence of having a dependence on
both the $\st$ and $\tilde{\chi}^0_2$ decays, $\mathcal{T}_b$ is also
sensitive to both phases $\phi_{A_t}$ and $\phi_{M_1}$ (and $\phi_{\mu}$).

%%%%%%%%%%%%%%%%%%%%%%%%%%%%%%%%%%%%%%%%%%%%%%%%%%%%%%%%%%%%%%%%%%%%%%%%%%%
\subsubsection{Disentangling of effects of $\phi_{A_t}$ and $\phi_{M_1}$ 
\label{subsec:disentangling}}
%%%%%%%%%%%%%%%%%%%%%%%%%%%%%%%%%%%%%%%%%%%%%%%%%%%%%%%%%%%%%%%%%%%%%%%%%%%
The T-odd asymmetries, eq.(\ref{Asy}), are determined by those CP-violating couplings that
are multiplied with the respective triple product.  
Under the assumption that $\phi_{\mu}$ is small, the neutralino sector depends
only on $\phi_{M_1}$ and the stop sector only on $\phi_{A_t}$. 
Since the involved triple product momenta show different dependence on the CP-violating 
phases, as discussed above, it is possible in principle to disentangle
the effects of $\phi_{A_t}$ and $\phi_{M_1}$ in our process and to
determine the phases separately.

The decoupling is possible as the triple
product $\mathcal{T}_{t}$, can only be produced by the term,
$\Sigma^{a,\mathrm{E}}_P(\tilde{\chi}^0_j)\Sigma^{a,\mathrm{O}}_D(\tilde{\chi}^0_j)$,
cf. section~\ref{sect:241}. 
The T-odd contribution in this term comes from the decay of the
$\tilde{\chi}^0_j$ and consequently is only sensitive to the phase
$\phi_{M_1}$. 
Once we have used the triple product $\mathcal{T}_{t}$
to determine the phase $\phi_{M_1}$ we can then use the value as an
input for the triple products, $\mathcal{T}_{tb}$ and $\mathcal{T}_b$,
in order to determine the phase $\phi_{A_t}$.

%If we are able to measure the T-odd asymmetries to a statistically
%significant level, we can then look at decoupling the phases to
%determine their values. 

%%%%%%%%%%%%%%%%%%%%%%%%%%%%%%%%%%%%%%%%%%%%%%%%%%%%%%%%%%%%%%%%%%%%%%%%%%%%%%%%%%%%%%%%%%
% test
%\clearpage
%%%%%%%%%%%%%%%%%%%%%%%%%%%%
\section{Results}
%%%%%%%%%%%%%%%%%%%%%%%%%%%%
\label{sec:results}

\subsection{Scenarios}
\label{sec:numerical-results}

In this Section we analyse numerically the various triple-product asymmetries introduced
in eqs.~(\ref{eq_tpt})--(\ref{eq_tptb}) at both the parton level and with the inclusion
of parton distribution functions (pdfs) to study the discovery potential at the LHC.  
In particular, we study the
dependences of these triple-product asymmetries on the MSSM parameters
$M_1=|M_1|e^{i\phi_{M_1}}$ and $A_t=|A_t|e^{i\phi_{A_t}}$. We also analyse the effects of
these parameters on the masses and branching ratios of the particles involved in our
process.

\begin{table}[!ht]
\renewcommand{\arraystretch}{1.3}
\begin{center}
\begin{tabular}{|c||c|c||c|c||c|c|} \hline
 Scenario & \multicolumn{2}{c||}{A: Reference} & \multicolumn{2}{c||}{B: NUHM -
$\gamma$} & \multicolumn{2}{c|}{C: Higgsino}\\ \hline\hline $M_1$ &
\multicolumn{2}{c||}{109} & \multicolumn{2}{c||}{97.6} &
\multicolumn{2}{c|}{105}\\ \hline $M_2$ & \multicolumn{2}{c||}{240} &
\multicolumn{2}{c||}{184} & \multicolumn{2}{c|}{400}\\ \hline $\mu$ &
\multicolumn{2}{c||}{220} & \multicolumn{2}{c||}{316} &
\multicolumn{2}{c|}{-190}\\ \hline $\tan{\beta}$ &
\multicolumn{2}{c||}{10} & \multicolumn{2}{c||}{20} &
\multicolumn{2}{c|}{20}\\ \hline $M_L$ & \multicolumn{2}{c||}{298} &
\multicolumn{2}{c||}{366} & \multicolumn{2}{c|}{298}\\ \hline $M_E$ &
\multicolumn{2}{c||}{224} & \multicolumn{2}{c||}{341.7} &
\multicolumn{2}{c|}{224}\\ \hline $M_{Q3}$ & \multicolumn{2}{c||}{511}
& \multicolumn{2}{c||}{534.5} & \multicolumn{2}{c|}{511}\\ \hline
$M_{U3}$ & \multicolumn{2}{c||}{460} & \multicolumn{2}{c||}{450} &
\multicolumn{2}{c|}{460}\\ \hline $A_t$ & \multicolumn{2}{c||}{-610} &
\multicolumn{2}{c||}{-451.4} & \multicolumn{2}{c|}{-610}\\ \hline\hline

 $M_{\st_1}$ & \multicolumn{2}{c||}{396.5} & \multicolumn{2}{c||}{447.8} & \multicolumn{2}{c|}{402.6}\\ \hline
 $M_{\st_2}$ & \multicolumn{2}{c||}{595} & \multicolumn{2}{c||}{609.6} & \multicolumn{2}{c|}{591.6}\\ \hline
 $M_{\tilde{\chi}^{\pm}_1}$ & \multicolumn{2}{c||}{177} & \multicolumn{2}{c||}{172.8} & \multicolumn{2}{c|}{186.3}\\ \hline
 $M_{\tilde{\chi}^{\pm}_2}$ & \multicolumn{2}{c||}{301.6} & \multicolumn{2}{c||}{346.05} & \multicolumn{2}{c|}{421.1}\\ \hline
 $m_{\tilde{\ell}_L}$ & \multicolumn{2}{c||}{302.4} & \multicolumn{2}{c||}{369.8} & \multicolumn{2}{c|}{303.1}\\ \hline
 $m_{\tilde{\ell}_R}$ & \multicolumn{2}{c||}{229.2} & \multicolumn{2}{c||}{345.2} & \multicolumn{2}{c|}{229.2}\\ \hline \hline

 $\phi_{M_1}$ & 
   \makebox[15mm]{$0$} & \makebox[15mm]{$\pi$} &
   \makebox[15mm]{$0$} & \makebox[15mm]{$\pi$} &
   \makebox[15mm]{$0$} & \makebox[15mm]{$\pi$} \\ \hline
 $m_{\tilde{\chi}^0_1}$ & 100.8 & 106.1 & 94.8 & 96.3 & 99.2 & 97.6 \\ \hline
 $m_{\tilde{\chi}^0_2}$ & 177.0 & 171.3 & 167.1 & 166.6 & 186.2 & 179.8 \\ \hline
 $m_{\tilde{\chi}^0_3}$ & 227.9 & 231.8 & 323.8 & 325.5 & 199.4 & 206.2 \\ \hline
 $m_{\tilde{\chi}^0_4}$ & 299.1 & 297.6 & 343.4 & 341.8 & 419 & 418.9 \\ \hline
$\Gamma_{\tilde{t}_1}$ & 3.88 & 3.88 & 3.48 & 3.48 & 5.29 & 5.29 \\\hline
$\Gamma_{\tilde{\chi}^0_2}$ & 1.4$\times 10^{-4}$ & 1.4$\times 10^{-4}$ & 2.3$\times
10^{-5}$ & 2.3$\times10^{-5}$ & 3.0$\times 10^{-3}$ & 3.0 $\times 10^{-3}$ \\ \hline
\end{tabular}\\[0.5ex]
\label{tab:scentab}
\caption{Parameters and spectra for the three scenarios A, B, C considered in this paper.
We display the input parameters $|M_1|$, $M_2$, $|\mu|$, $\tan\beta$, $m_{\tilde{\ell}_L}$ and
  $m_{\tilde{\ell}_R}$ and the resulting masses $m_{\tilde{\chi}^0_i}$, $i=1,\ldots,4$,
  for $(\phi_{M_1},\phi_{\mu}) = (0.5\pi,0)$ and $(0.5\pi,0.5\pi)$.  The parameters $M_2$,
  $|\mu|$ and $\tan\beta$ in scenario B are chosen as for the scenario SPS1a in
  \cite{Allanach:2002nj}. We used $m_t=171.2$~GeV \cite{Amsler:2008zz} and the SM value
  for the top width $\Gamma_t\sim 1.5$~GeV~\cite{Hoang:1999zc} for our study. 
%The current experimental upper bound is $\Gamma_t<12.4$~GeV \cite{topwidth}. 
 All masses and widths are 
  given in GeV.}
\end{center}
\end{table}

For our numerical analysis we study in detail at both the partonic and pdf level
a reference scenario, A, where the $\tilde{\chi}_1^0$ is a gaugino-higgsino mixture.  For comparison, we also study 
at the partonic level a
non-universal Higgs masses (NUHM) scenario, B, and a third scenario, C, in which the $\tilde{\chi}_2^0$ is higgsino-like. The particle
spectra for these scenarios have been computed with the program \texttt{SPheno}~\cite{Porod:2003um}.  These three scenarios have been chosen to have similar masses, as displayed in
Table~\ref{tab:scentab}, so
that the kinematic effects are similar in each case.  We perform our studies using our
own program based on the analytic formulae we have derived for the various cross sections and
spin correlations. The program uses the \texttt{VEGAS}~\cite{Lepage:1978,Lepage:1980} routine
to perform the multi-dimensional phase-space integral. We constrain
ourselves to cases where $m_{\tilde{\chi}^0_2}<m_{\tilde{\chi}^0_1}+m_{Z^0}$ and
$m_{\tilde{\chi}^0_2}<m_{\sl_{L,R}}$, so as to forbid the two-body decay of the 
$\tilde{\chi}^0_2$. The branching
ratios for both processes have been calculated with 
\texttt{Herwig++}~\cite{Bahr:2008pv,Bahr:2008tx}~\footnote{Beyond the Standard Model physics 
was produced using the
algorithm of \cite{Gigg:2007cr} and, in the running of $\alpha_{EM}$, the parametrization
of \cite{KLEISSCERN9808v3pp129} was used.}.

The feasibility of measuring these asymmetries at the LHC depends heavily on the
integrated luminosity at the LHC. For this reason we look closely at the cross section,
$\sigma=\sigma(gg \to \tilde{t}_1\bar{\tilde{t}}_1)
\times BR(\tilde{t}_1\to t \tilde{\chi}^0_2)\times
BR(\tilde{\chi}^0_2 \to \tilde{\chi}^0_1 \ell^+ \ell^-) \times BR(t \to W b)$ and determine the
nominal luminosity required to observe a statistically significant result.  

\subsection{CP asymmetry at the parton level}

\subsubsection{Dependence of $m_{\tilde \chi_1^0}$ and $\mathcal{A}_{T}$ on $\phi_{M_1}$ and $\phi_{A_t}$}
\label{sec:depend-m_tilde-chi_1}

\begin{figure}[ht!]
\subfigure{
\vspace{0.5cm}
\begin{picture}(16,8)
 \put(2.5,8){$\mathcal{A}_{\mathcal{T}_t}$, $\sqrt{\hat{s}}\sim2m_{\st}$}
  \put(10.7,8){$m_{\tilde{\chi}^0_i}$, Scenario A}
  \put(6,2.5){$\phi_{M_1}/\pi$}
 \put(14,2.5){$\phi_{M_1}/\pi$}
 \put(0,7.5){(a)}
 \put(6.2,7.1){\tiny{A}}
 \put(6.2,6.3){\tiny{C}}
 \put(6.2,3.7){\tiny{B}}
 \put(8,7.5){(b)}
  \put(12,4.45){\tiny{$m_{\tilde{\chi}^0_1}$}}
  \put(12,5.2){\tiny{$m_{\tilde{\chi}^0_2}$}}
  \put(12,6){\tiny{$m_{\tilde{\chi}^0_3}$}}
  \put(12,6.7){\tiny{$m_{\tilde{\chi}^0_4}$}}
\put(0,8){\epsfig{file=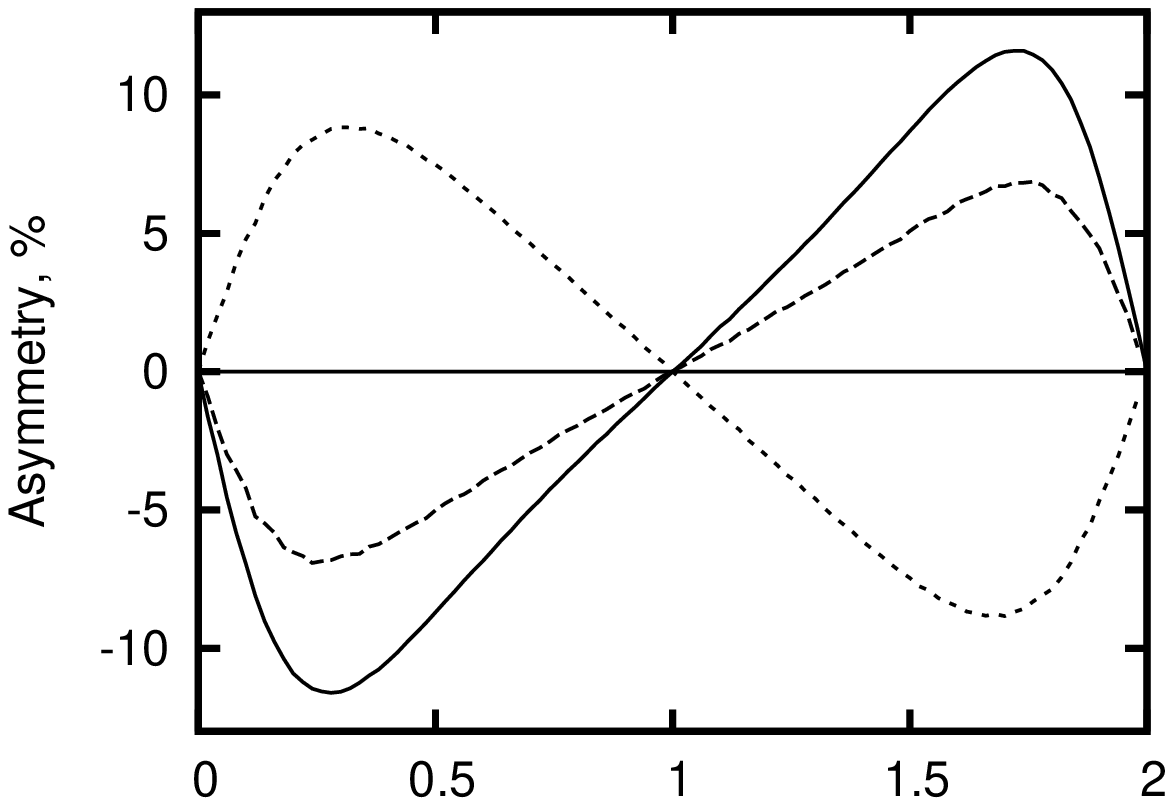,scale=0.6,angle=270}}
\put(8,8){\epsfig{file=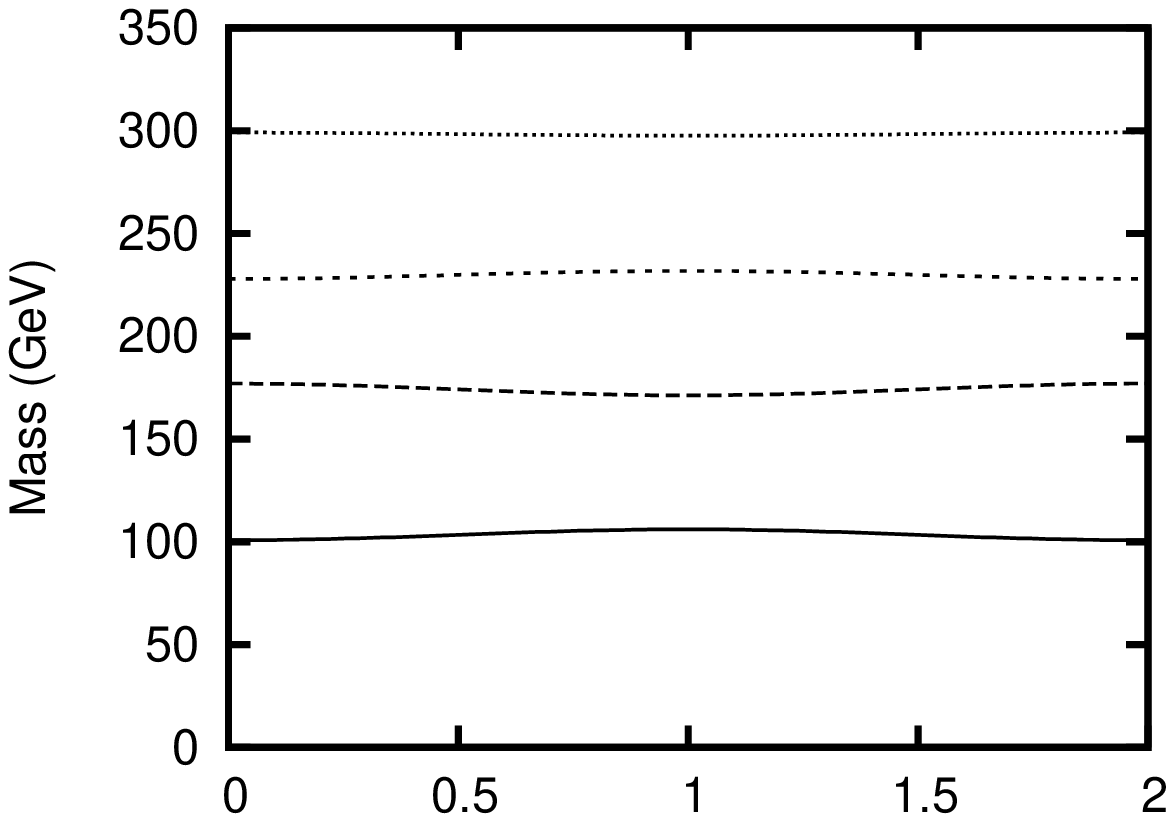,scale=0.6,angle=270}}
\end{picture}}
\vspace{-3cm}
\caption{\label{fig:Part_t_b} The asymmetry at threshold for the
  production process $gg\longrightarrow\st\overline{\st}$ for scenarios 
  A, B and C for (a) $\mathcal{A}_{\mathcal{T}_t}$ as a function of $\phi_{M_1}$,
  and (b) the masses of the neutralinos as functions of $\phi_{M_1}$.}
%\end{figure}

%\begin{figure}[ht!]
\vspace{0.5cm}
\subfigure{
\begin{picture}(16,8)
 \put(10.5,8){$\mathcal{A}_{\mathcal{T}_{tb}}$, $\sqrt{\hat{s}}\sim2m_{\st}$}
  \put(2.7,8){$\mathcal{A}_{\mathcal{T}_b}$, $\sqrt{\hat{s}}\sim2m_{\st}$}
  \put(6,2.5){$\phi_{A_t}/\pi$}
 \put(14,2.5){$\phi_{A_t}/\pi$}
 \put(5.5,7.3){\tiny{A}}
 \put(5.5,6.5){\tiny{C}}
 \put(5.5,4.6){\tiny{B}}
 \put(0,7.5){(a)}
 \put(10.7,7.1){\tiny{A}}
 \put(10.7,6.5){\tiny{C}}
 \put(10.7,4.7){\tiny{B}}
 \put(8,7.5){(b)}
  \put(0,8){\epsfig{file=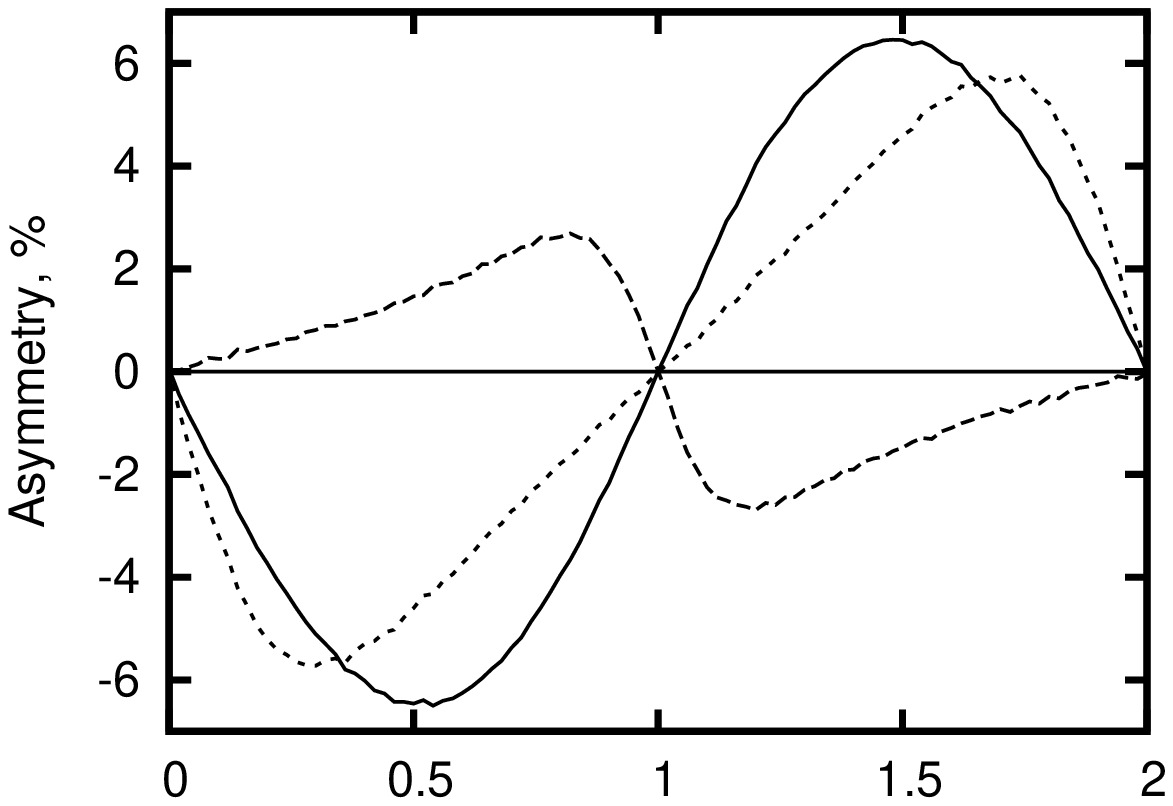,scale=0.6,angle=270}}
  \put(8,8){\epsfig{file=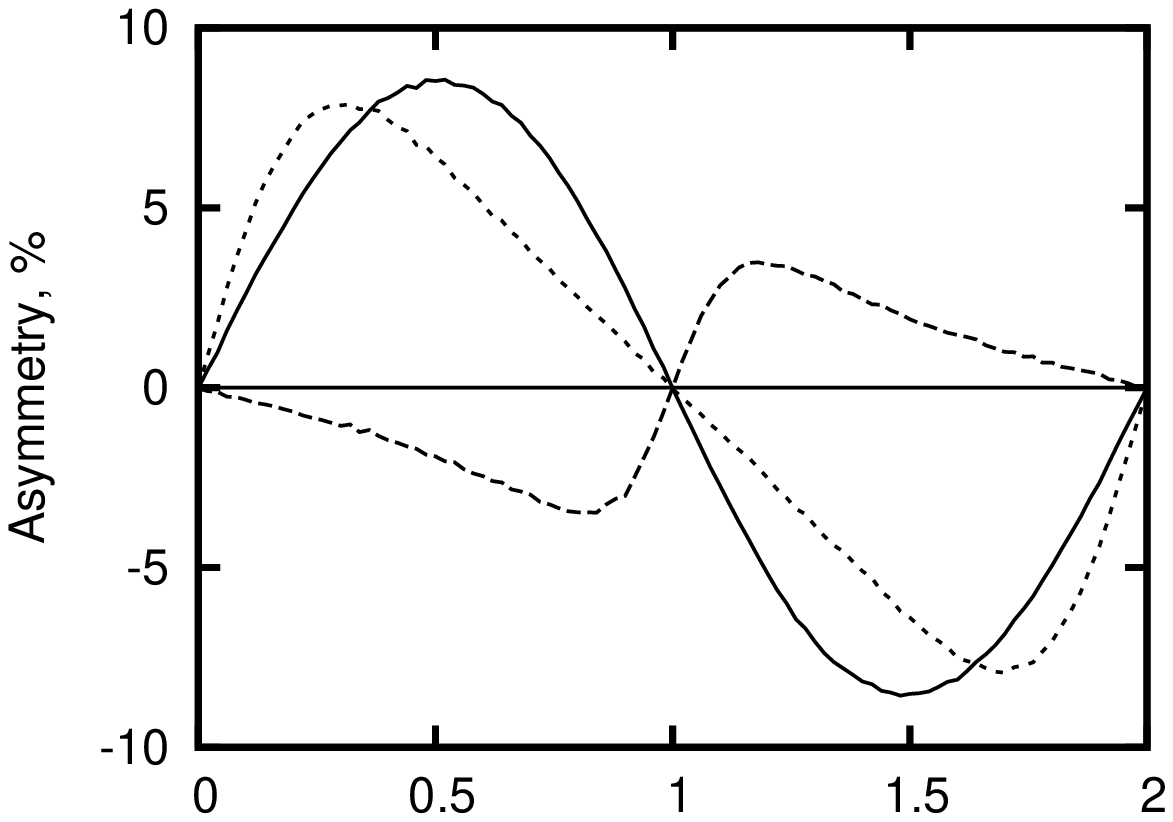,scale=0.6,angle=270}}
\end{picture}}
\vspace{-3cm}
% \caption{\label{fig:GaugeMass} (a) The asymmetry $\mathcal{A}_{\mathcal{T}_{tb}}$ at threshold for the
%   production process $gg\longrightarrow\st\overline{\st}$ for scenarios A, B and C, 
%   and (b) the asymmetry $\mathcal{A}_{\mathcal{T}_b}$ at threshold, both as functions
%   of the variable $\phi_{A_t}$.}
\caption{\label{fig:GaugeMass} (a) The asymmetry $\mathcal{A}_{\mathcal{T}_{b}}$ at threshold for the
  production process $gg\longrightarrow\st\overline{\st}$ for scenarios A, B and C, 
  and (b) the asymmetry $\mathcal{A}_{\mathcal{T}_b}$ at
  threshold, both as functions of $\phi_{A_t}$.}
\end{figure}

We start by discussing the dependence on $M_1=|M_1|e^{i\phi_{M_1}}$ of the parton-level
asymmetries for each of the three scenarios. In order to see the maximum dependence upon
$\phi_{M_1}$, we use the reconstructed $t$ quark momentum and the triple product
$\mathcal{T}_t=\vec{p}_{t}\cdot (\vec{p}_{\ell^+} \times \vec{p}_{\ell^-})~ $. It should
be noted from the following plots that the asymmetry is obviously a CP-odd quantity that
in addition to a measurement of the phase, also gives the sign,
as seen in Fig.~\ref{fig:Part_t_b}(a). In comparison, using CP-even
quantities, for example the mass, it is not possible to determine if the phase is positive
or negative, as seen in Fig.~\ref{fig:Part_t_b}(b).

We see in Fig.~\ref{fig:Part_t_b}(a) that the biggest asymmetry appears in scenario
A, which attains $|\mathcal{A}_{T_t}|_{\rm max}\approx12\%$ when
$\phi_{M_1}\approx0.3\pi$. One aspect of the plot that may be surprising is that the
asymmetry is not largest at the maximal value of the phase ($\phi_{M_1}=\frac{\pi}{2}$).
This is due to the coupling combinations and interferences and can be seen from the
equations in Sections \ref{sec:t-odd-prod}.  In
Fig.~\ref{fig:Part_t_b}(b), the dependence of the masses of the neutralinos is shown.  It
can be seen clearly that the variations are too small to be used to determine the phase.

In the cases of the two other scenarios shown in Fig.~\ref{fig:Part_t_b}(a), the dependence of the
asymmetry on the phase $\phi_{M_1}$ is similar but slightly smaller. In the case of
scenario B (NUHM), the peak asymmetry is $|\mathcal{A}_{T_t}|_{\rm max}\approx 9\%$ when
$\phi_{M_1}\approx 0.3\pi$ and in scenario C (Higgsino) it is $|\mathcal{A}_{T_t}|_{\rm
  max}\approx 7\%$ when $\phi_{M_1}\approx 0.25\pi$. Again, the asymmetry does not peak when
the phase is maximal.

To study the dependence upon $\phi_{A_t}$ we need to use the triple products sensitive to
this phase, $\mathcal{T}_b=\vec{p}_{t}\cdot (\vec{p}_{\ell^+} \times \vec{p}_{\ell^-})~ $
and $\mathcal{T}_{tb}=\vec{p}_{t}\cdot (\vec{p}_{b} \times \vec{p}_{\ell^\pm})~
$. Fig.~\ref{fig:GaugeMass}(a) shows $\mathcal{A}_{\mathcal{T}_b}$ and we see that the
biggest asymmetry again occurs in Scenario A, but the maximal asymmetry is only about half
of $|\mathcal{A}_{\mathcal{T}_t}|_{\rm max}$ with $|\mathcal{A}_{\mathcal{T}_b}|_{\rm
  max}\approx6\%$. Scenario C produces a very similar asymmetry to Scenario A, with
$|\mathcal{A}_{\mathcal{T}_b}|_{\rm max}\approx5.5\%$, whereas the asymmetry in Scenario B
is much smaller: $|\mathcal{A}_{\mathcal{T}_b}|_{\rm
  max}\approx2.5\%$. Fig.~\ref{fig:GaugeMass}(b) shows that the general shape of the
asymmetries for $\mathcal{A}_{\mathcal{T}_{tb}}$ is similar to that of
$\mathcal{A}_{\mathcal{T}_b}$ apart from a difference in sign and that all the asymmetries
are actually slightly larger. In fact, for Scenario C, the largest asymmetry is generated
using $\mathcal{T}_{tb}$ with $\mathcal{A}_{\mathcal{T}_{tb}}\approx 8\%$ when
$\phi_{A_t}\approx0.3\pi$.

In the subsequent analysis, we concentrate on the favourable Scenario A, 
with just a few remarks on the others.

\subsubsection{Contour Plots of $\mathcal{A}_{T_t}$ and $\mathcal{A}_{T_{tb}}$ for Variable $M_1$ and $A_t$}
\label{sec:cont-plots-mathc}

\begin{figure}[ht!]
\vspace{0.5cm}
\begin{picture}(16,8)
 \put(1,8){$\mathcal{A}_{\mathcal{T}_t}$, Scenario A ($\phi_{A_t}=0$), $\sqrt{\hat{s}}\sim2m_{\st}$}
  \put(9,8){$\mathcal{A}_{\mathcal{T}_{tb}}$, Scenario A ($\phi_{M_1}=0$), $\sqrt{\hat{s}}\sim2m_{\st}$}
\put(0,7.5){(a)}
 \put(-0.2,5.4){$\phi_{M_1}/\pi$}
 \put(6,2.5){$M_1/\mathrm{GeV}$}
\put(-1,9.7){\epsfig{file=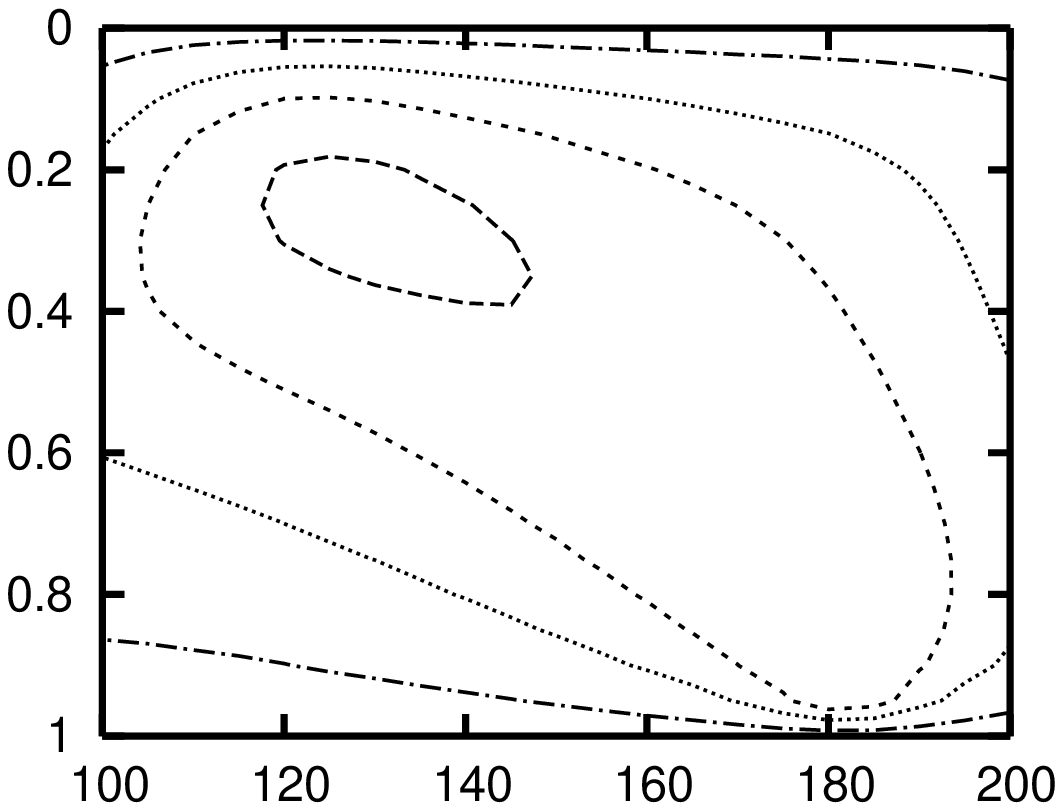,scale=0.6,angle=270}}
\put(3,6.2){\tiny{14}}
\put(3,5.4){\tiny{10}}
\put(3,4.6){\tiny{6}}
\put(3,3.8){\tiny{2}}
 \put(8,7.5){(b)}
\put(7.8,5.4){$\phi_{A_t}/\pi$}
 \put(14,2.5){$M_1/\mathrm{GeV}$}
\put(7,9.7){\epsfig{file=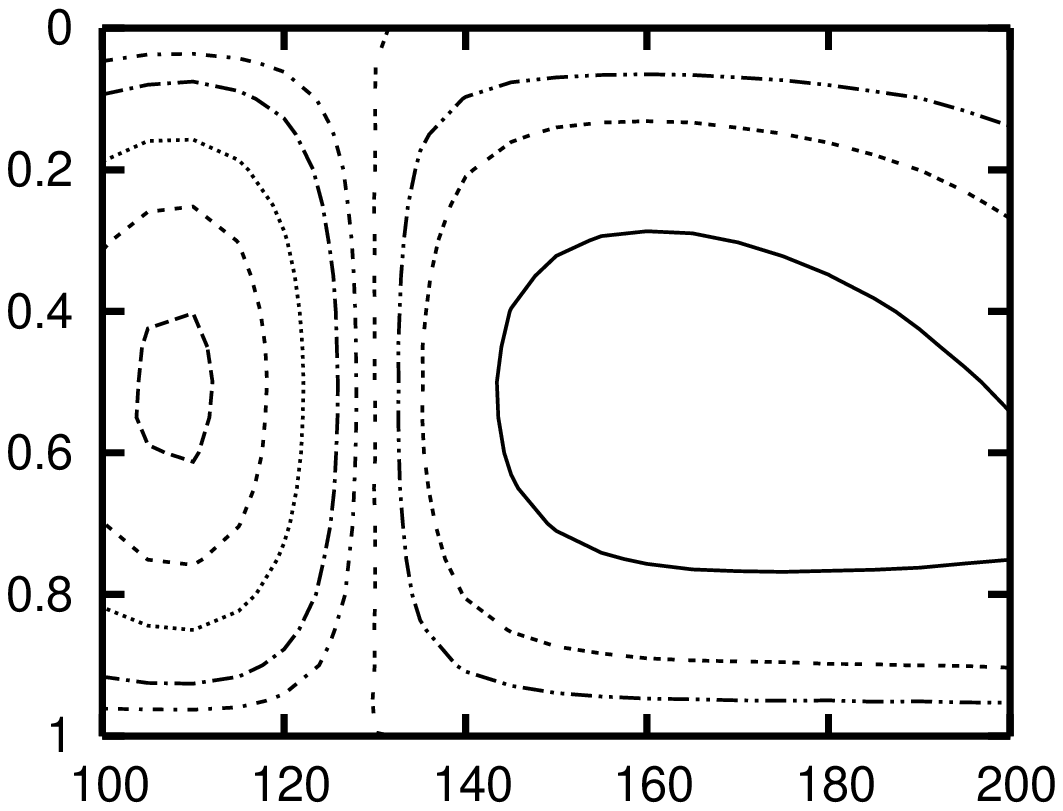,scale=0.6,angle=270}}
\put(12.1,5.2){\tiny{-4}}
\put(11.5,5.2){\tiny{-2}}
\put(11.1,5.2){\tiny{0}}
\put(10.9,5.2){\tiny{2}}
\put(10.65,5.2){\tiny{4}}
\put(10.4,5.2){\tiny{6}}
\put(10.1,5.2){\tiny{8}}
\end{picture}
\vspace{-3cm}
\caption{\label{fig:M1AtCon} Contours in scenario A (in \%) of the parton-level asymmetries
(a) $\mathcal{A}_{T_{t}}$ for the triple product $\mathcal{T}_t=\bf p_{t}\cdot(\bf p_{\ell^+} \times \bf
  p_{\ell^-})$, as functions of the variables $M_1$ and $\phi_{M_1}$, and (b)
  $\mathcal{A}_{T_{tb}}$ for the triple product $\mathcal{T}_{tb}=\bf
  p_{\ell^+}\cdot(\bf p_{t} \times \bf p_{b})$, as functions of the variables $M_1$ and $\phi_{A_t}$.}
\end{figure}

If we now lift the restriction of the GUT relation for $|M_1|$, we can see how the
asymmetry varies with $|M_1|$ while leaving all the other parameters the same, for Scenario
A. Fig. \ref{fig:M1AtCon}(a) shows that the asymmetry peaks at $|M_1|\approx130$ GeV and
$\phi_{M_1}\approx0.25\pi$ when $|\mathcal{A}_{T_t}|\approx15\%$. Importantly though, the
asymmetry can remain above 10\% between $|M_1|=110$ GeV and $|M_1|=190$ GeV, which is most
of the range allowed in this scenario.

\begin{figure}[ht!]
\vspace{0.5cm}
\begin{picture}(16,8)
 \put(1,8){$\mathcal{A}_{\mathcal{T}_b}$, Scenario A ($\phi_{M_1}=0$), $\sqrt{\hat{s}}\sim2m_{\st}$}
\put(0,7.5){(a)}
 \put(-0.2,5.4){$\phi_{A_t}/\pi$}
 \put(6,2.5){$M_1/\mathrm{GeV}$}
\put(-1,9.7){\epsfig{file=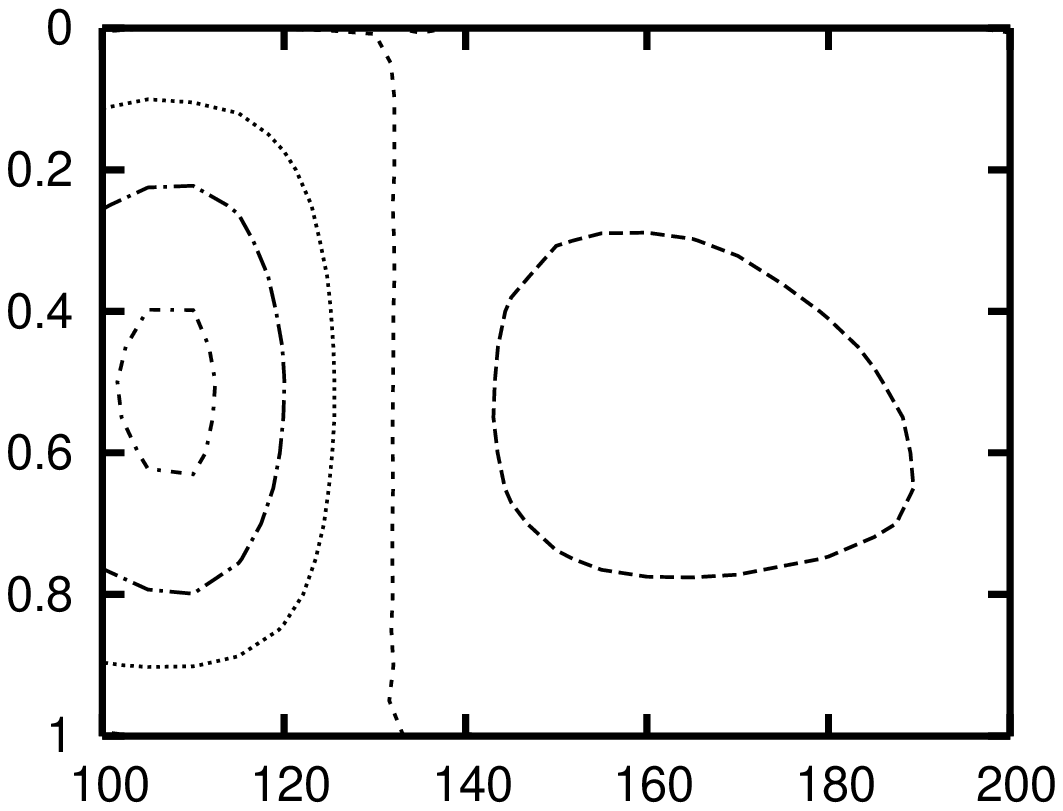,scale=0.6,angle=270}}
\put(3.65,5.2){\tiny{-2}}
\put(3.15,5.2){\tiny{0}}
\put(2.7,5.2){\tiny{-2}}
\put(2.4,5.2){\tiny{-4}}
\put(2.0,5.2){\tiny{-6}}
   \put(9,8){$\mathcal{A}_{\mathcal{T}_{tb}}$, Scenario A ($\phi_{M_1}=0$), $\sqrt{\hat{s}}\sim2m_{\st}$}
 \put(7.8,5.4){$\phi_{A_t}/\pi$}
\put(8,7.5){(b)}
 \put(14,2.5){$A_t/\mathrm{GeV}$}
\put(7,9.7){\epsfig{file=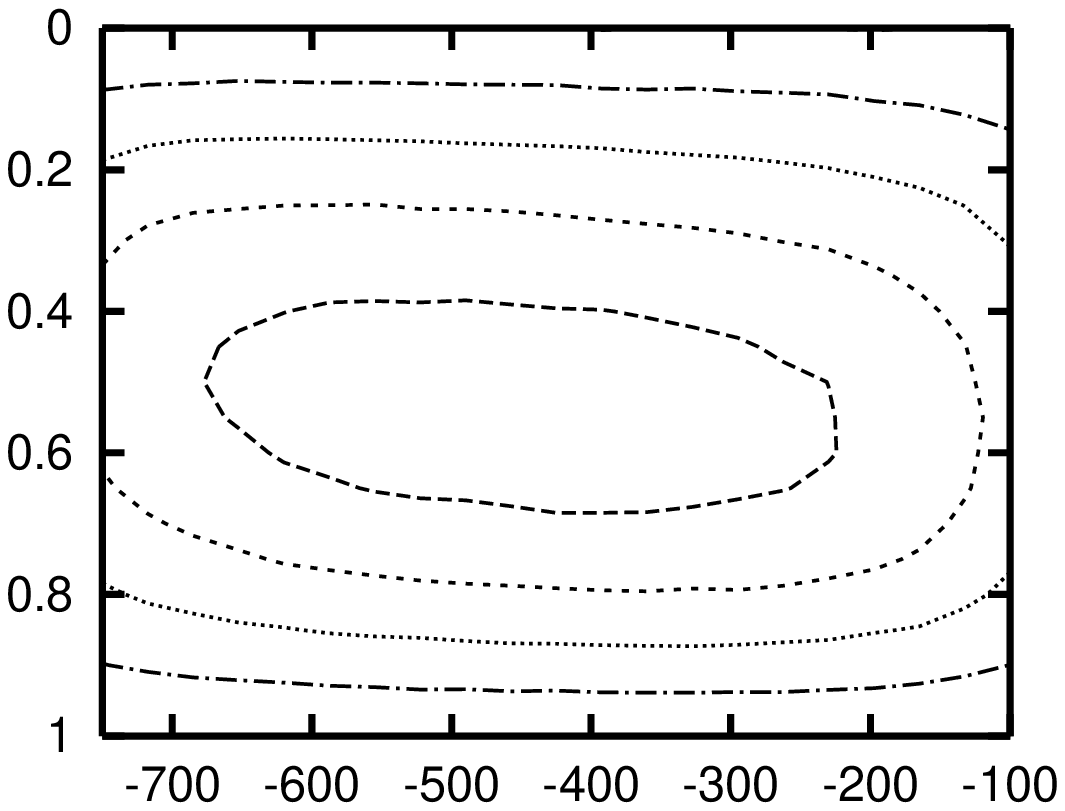,scale=0.6,angle=270}}
\put(11,4.9){\tiny{8}}
\put(11,4.4){\tiny{6}}
\put(11,4.0){\tiny{4}}
\put(11,3.65){\tiny{2}}
\end{picture}
\vspace{-3cm}
\caption{\label{fig:AtDoubleCon} Contours in scenario A (in \%) of the asymmetries (a) $\mathcal{A}_{b}$ for the triple product $\mathcal{T}=\bf p_{b}\cdot(\bf p_{\ell^-} \times \bf p_{\ell^+})$, as functions of the
  variables $M_1$ and $\phi_{A_t}$ and (b) $\mathcal{A}_{tb}$ for the triple product $\mathcal{T}=\bf p_{\ell^+}\cdot(\bf p_{t} \times \bf p_{b})$, as functions of the common
  variables $A=A_t=A_b=A_\tau$ and $\phi_{A_t}$.}
\end{figure}

By including the decay of the $t$ quark that was produced in the $\st$ decay, we can also study
the effect of $\phi_{A_t}$ on our asymmetries. As the spin-correlation information is now
carried by the $t$ quark, we have to change the triple product used to measure the asymmetry, eq.
(\ref{Asy}). It is found that the largest asymmetry can be measured using the triple
product, $\mathcal{T}_{tb}=\bf p_{\ell^+}\cdot(\bf p_{t} \times \bf p_{b})$ where
$|\mathcal{A}_{T_{tb}}|_{\rm max}\approx8.5\%$ when $\phi_{A_t}\approx0.5\pi$ in Scenario A,
as seen in Fig.~\ref{fig:M1AtCon}(a).
It may be noted that this asymmetry is slightly smaller than those of \cite{Bartl:2004jr} that can be reconstructed experimentally.  In that paper scenarios were chosen
where the $\tilde{\chi}^0_2$ decays via a two-body process, whereas here we concentrate on
scenarios where the $\tilde{\chi}^0_2$ decays via a three-body process, so to maximise the sensitivity
to $\phi_{M_1}$. This phase dependence can also be seen with the triple product
$\mathcal{T}_{b}=\bf p_{b}\cdot(\bf p_{\ell^+} \times \bf p_{\ell^+})$ although the asymmetry
is found to be smaller here with $|\mathcal{A}_{T_{b}}|_{\rm max}\approx6\%$, see Fig.~\ref{fig:AtDoubleCon}(a).

We have also considered the dependence of the asymmetry on a common trilinear
coupling, $A=A_t=A_b=A_\tau$, in scenario A, as shown in Fig.~\ref{fig:AtDoubleCon}(b). It can be seen that the
asymmetry is stable for the bulk of the region scanned, and only decreases near the edge of
the acceptable region for our scenario. The peak is now $|\mathcal{A}_{T_{tb}}|_{\rm
  max}\approx9\%$, when $A_t\approx-500$ GeV, and the region where
$|\mathcal{A}_{T_{tb}}|>8\%$ extends from $A_t\approx-650$ GeV to $A_t\approx-250$ GeV.

\begin{figure}[ht!]
\vspace{0.5cm}
\begin{picture}(16,8)
 \put(1,8){$\mathcal{A}_{\mathcal{T}_b}$, Scenario A, $\sqrt{\hat{s}}\sim2m_{\st}$}
  \put(9,8){$\mathcal{A}_{\mathcal{T}_{tb}}$, Scenario A, $\sqrt{\hat{s}}\sim2m_{\st}$}
 \put(-0.2,5.4){$\phi_{A_t}/\pi$}
 \put(6,2.5){$\phi_{M_1}/\pi$}
 \put(0,7.5){(a)}
\put(-1,9.7){\epsfig{file=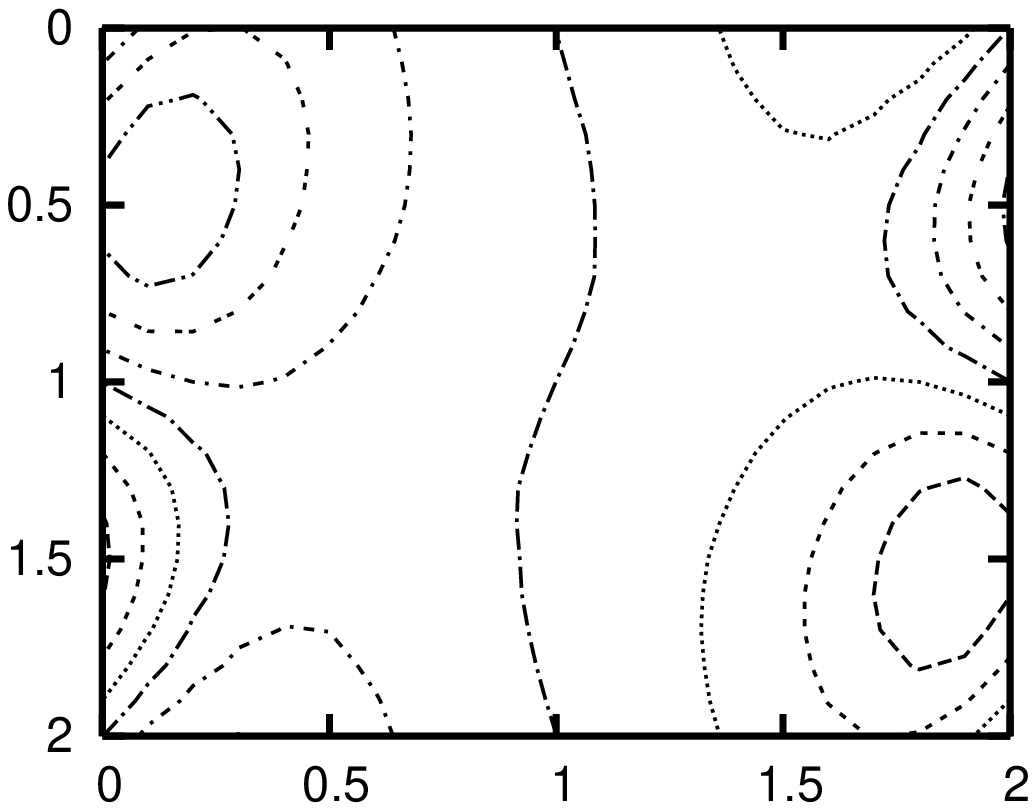,scale=0.6,angle=270}}
\put(4.1,5.4){\tiny{0}}
\put(2.2,4.6){\tiny{0}}
\put(6.4,6.2){\tiny{0}}
\put(3,6.1){\tiny{-2}}
\put(2.5,6.3){\tiny{-4}}
\put(2.1,6.4){\tiny{-6}}
\put(1.9,4.6){\tiny{2}}
\put(1.6,4.6){\tiny{4}}
\put(2.8,3.6){\tiny{-2}}
\put(5.4,4.6){\tiny{2}}
\put(6,4.4){\tiny{4}}
\put(6.4,4.3){\tiny{6}}
\put(6.6,6.2){\tiny{-2}}
\put(6.9,6.2){\tiny{-4}}
\put(5.6,7.2){\tiny{2}}
 
 \put(7.8,5.4){$\phi_{A_t}/\pi$}
 \put(14,2.5){$\phi_{M_1}/\pi$}
 \put(8,7.5){(b)}
\put(7,9.7){\epsfig{file=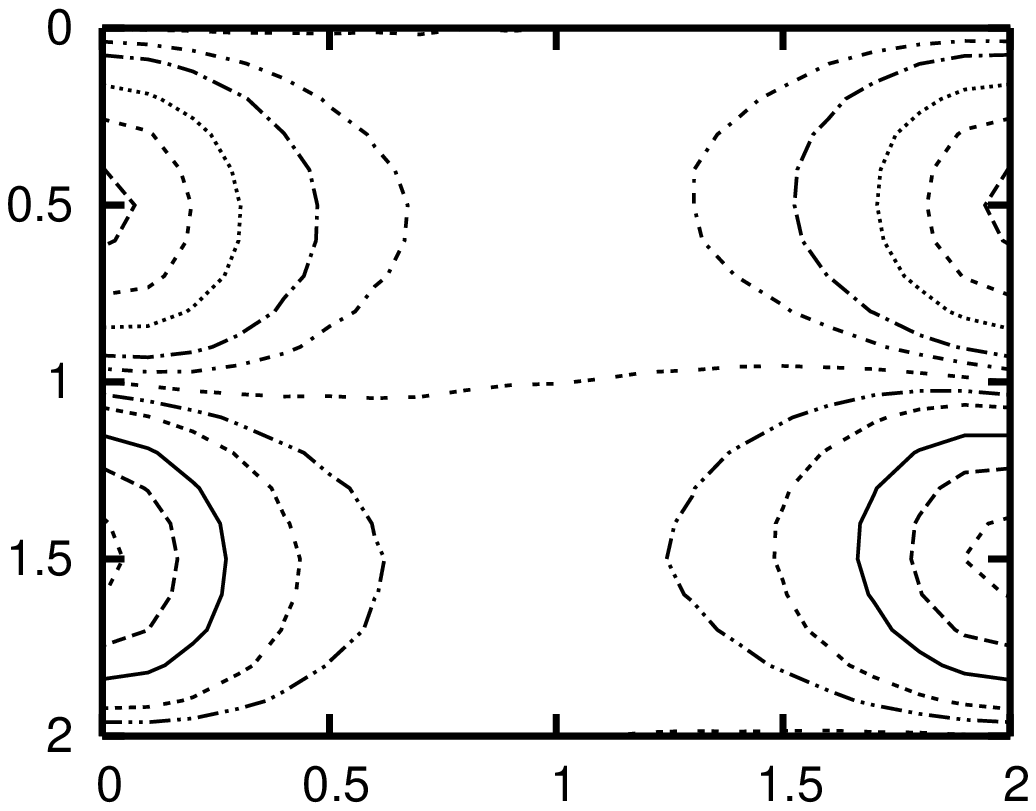,scale=0.6,angle=270}}
\put(9.6,4.3){\tiny{-8}}
\put(10.0,4.3){\tiny{-6}}
\put(10.3,4.3){\tiny{-4}}
\put(10.8,4.3){\tiny{-2}}
\put(11.4,4.3){\tiny{-1}}
\put(9.8,6.5){\tiny{8}}
\put(10.1,6.5){\tiny{6}}
\put(10.4,6.5){\tiny{4}}
\put(10.9,6.5){\tiny{2}}
\put(11.5,6.5){\tiny{1}}
\put(12.7,4.3){\tiny{-1}}
\put(13.3,4.3){\tiny{-2}}
\put(13.8,4.3){\tiny{-4}}
\put(14.2,4.3){\tiny{-6}}
\put(14.6,4.3){\tiny{-8}}
\put(12.9,6.5){\tiny{1}}
\put(13.5,6.5){\tiny{2}}
\put(14.0,6.5){\tiny{4}}
\put(14.4,6.5){\tiny{6}}
\put(14.8,6.5){\tiny{8}}
\end{picture}
\vspace{-3cm}
\caption{\label{fig:M1vAt109} Contours (in \%) of the asymmetry  at the parton level 
in scenario A with
  $M_1=109$~GeV for the triple products (a) $\mathcal{T}_{b}=\bf p_b\cdot(\bf p_{\ell^+}
  \times \bf p_{\ell^-})$ and (b) $\mathcal{T}_{tb}=\bf p_{\ell^+}\cdot(\bf p_{b} \times
  \bf p_{t})$ for varying phases $\phi_{M_1}$ and $\phi_{A_t}$.}
\end{figure}

We now consider the effect on the asymmetry of varying simultaneously both the phases
$\phi_{M_1}$ and $\phi_{A_t}$. The triple products $\mathcal{T}_b=\vec{p}_{b}\cdot
(\vec{p}_{\ell^+}\times\vec{p}_{\ell^-})~$ and $\mathcal{T}_{tb}=\vec{p}_{t}\cdot
(\vec{p}_{b}\times\vec{p}_{\ell^\pm})$ can have contributions from both phases, so we
concentrate on these. For $\mathcal{T}_b$, Fig.~\ref{fig:M1vAt109}(a) shows that the area
of parameter space where $\phi_{M_1}$ and $\phi_{A_t}$ constructively interfere is
actually quite small and peaked around $\phi_{M_1}\approx0.2\pi$ and
$\phi_{A_t}\approx0.5\pi$. Apart from this area, varying both phases generally results in
a reduction in the asymmetry observed, caused by the neutralino and squark mixing entering
the couplings, Section \ref{sec:t-odd-decay}. Importantly when $\phi_{M_1}\approx\pi$ or
$\phi_{A_t}\approx\pi$ the asymmetry vanishes, as it should. Fig.~\ref{fig:M1vAt109}(b)
demonstrates that, for this scenario, $\phi_{M_1}$ generates virtually no asymmetry for
$\mathcal{T}_b$. However, $\phi_{M_1}$ can still significantly reduce the asymmetry that
$\phi_{A_t}$ can produce and, again, when $\phi_{M_1}\approx\pi$ we see that
$|\mathcal{A}_{T_{tb}}|\approx0$ as expected.

\begin{figure}[ht!]
\vspace{0.5cm}
\begin{picture}(16,8)
 \put(1,8){$\mathcal{A}_{\mathcal{T}_b}$, Scenario A, $\sqrt{\hat{s}}\sim2m_{\st}$}
  \put(9,8){$\mathcal{A}_{\mathcal{T}_{tb}}$, Scenario A, $\sqrt{\hat{s}}\sim2m_{\st}$}
 \put(-0.2,5.4){$\phi_{A_t}/\pi$}
 \put(6,2.5){$\phi_{M_1}/\pi$}
\put(0,7.5){(a)}
\put(-1,9.7){\epsfig{file=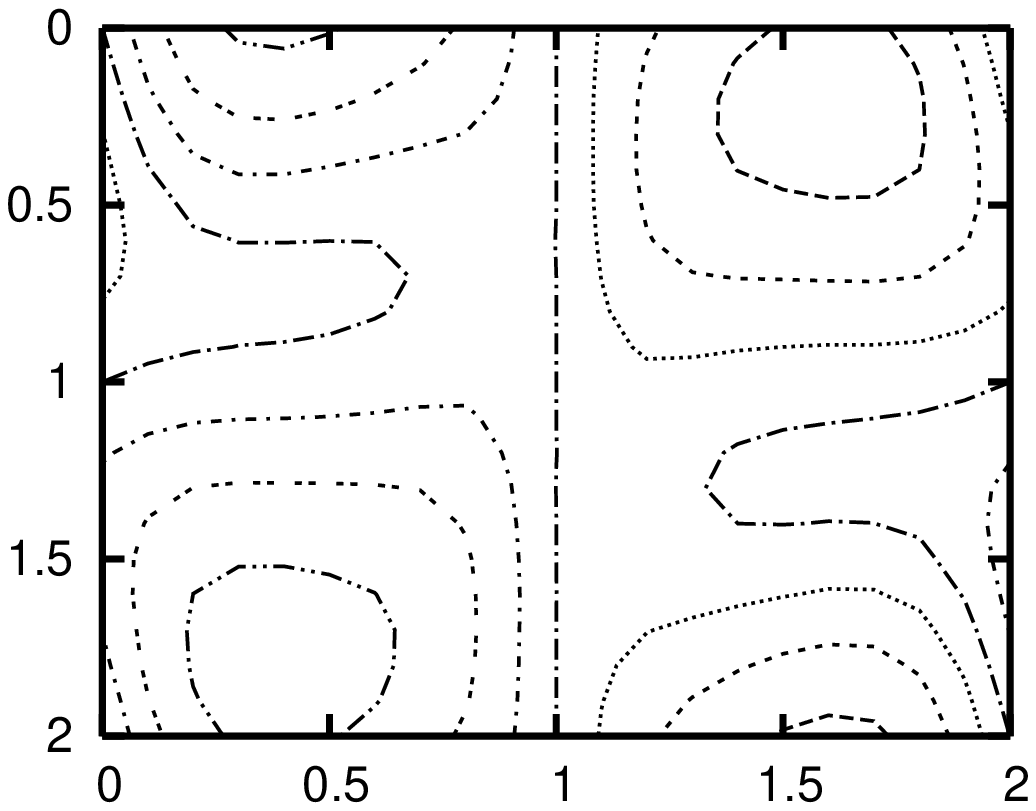,scale=0.6,angle=270}}
\put(2.6,7.3){\tiny{-6}}
\put(2.6,6.9){\tiny{-4}}
\put(2.6,6.5){\tiny{-2}}
\put(2.6,6.1){\tiny{0}}
\put(2.6,5.3){\tiny{-2}}
\put(2.6,4.9){\tiny{-4}}
\put(2.6,4.4){\tiny{-6}}
\put(1.7,6.1){\tiny{2}}
\put(5.9,6.4){\tiny{6}}
\put(5.9,5.9){\tiny{4}}
\put(5.9,5.5){\tiny{2}}
\put(5.9,5.0){\tiny{0}}
\put(5.9,4.25){\tiny{2}}
\put(5.9,3.9){\tiny{4}}
\put(5.9,3.45){\tiny{6}}
\put(6.7,4.6){\tiny{-2}}

 \put(7.8,5.4){$\phi_{A_t}/\pi$}
 \put(14,2.5){$\phi_{M_1}/\pi$}
 \put(8,7.5){(b)}
\put(7,9.7){\epsfig{file=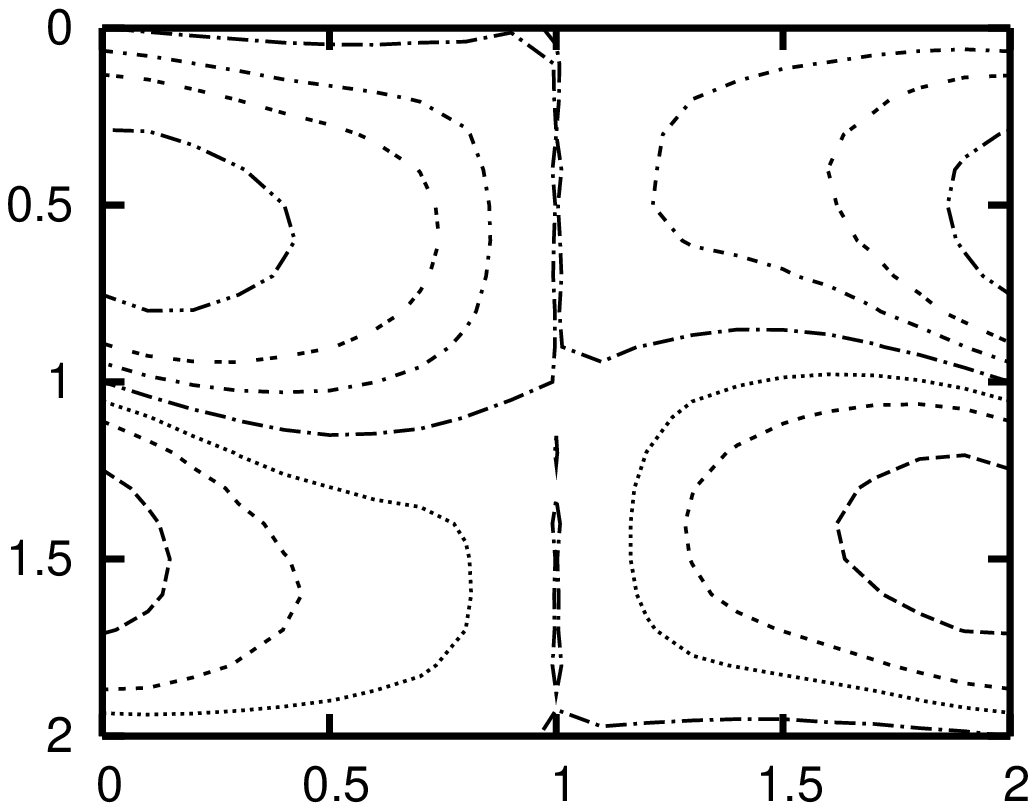,scale=0.6,angle=270}}
\put(10.1,4.3){\tiny{4}}
\put(10.8,4.3){\tiny{2}}
\put(11.85,4.3){\tiny{1}}
\put(10.7,6.5){\tiny{-4}}
\put(11.55,6.5){\tiny{-2}}
\put(11.9,6.5){\tiny{-1}}
\put(12.6,4.3){\tiny{1}}
\put(13.0,4.3){\tiny{2}}
\put(13.9,4.3){\tiny{4}}
\put(12.7,6.5){\tiny{-1}}
\put(13.65,6.5){\tiny{-2}}
\put(14.4,6.5){\tiny{-4}}
\end{picture}
\vspace{-3cm}
\caption{\label{fig:M1vAt160}
Contours (in \%) of the asymmetries in scenario A at the parton level with $M_1=160$~GeV for the
triple products (a) $\mathcal{T}_{b}=\bf p_b\cdot(\bf p_{\ell^+} \times \bf p_{\ell^-})$ and 
(b) $\mathcal{T}_{tb}=\bf p_{\ell^+}\cdot(\bf p_{b} \times \bf
p_{t})$, as fucntions of the varying phases $\phi_{M_1}$ and $\phi_{A_t}$.}
\end{figure}

If we now modify scenario A slightly by setting $|M_1|=160$~GeV, this results in a
more interesting scenario as the phases $\phi_{M_1}$ and $\phi_{A_t}$ can interfere
constructively to produce an asymmetry larger than that seen before. When
$\phi_{M_1}\approx0.4\pi$ and $\phi_{A_t}\approx1.8\pi$, we observe a peak asymmetry,
$|\mathcal{A}_{T_b}|\approx 7\%$ for the triple product $\mathcal{T}_b$, as seen in
Fig.~\ref{fig:M1vAt160}.

\subsection{Dependences of branching ratios on $\phi_{M_1}$ and $\phi_{A_t}$}
\label{sec:depend-branch-rati}

\begin{figure}[ht!]
\vspace{0.5cm}
\begin{picture}(16,8)
  \put(1,8){$BR(\tilde{t}_1 \to \tilde{\chi}^0_2 t)$, Scenario A ($\phi_{A_t}=0$)}  
  \put(9,8){$BR(\tilde{\chi}^0_2 \to \tilde{\chi}^0_1 \ell^+ \ell^-)$, Scenario A ($\phi_{A_t}=0$)} 
 \put(-0.2,5.4){$\phi_{M_1}/\pi$}
 \put(6,2.5){$M_1/\mathrm{GeV}$}
 \put(7.8,5.4){$\phi_{M_1}/\pi$}
 \put(14,2.5){$M_1/\mathrm{GeV}$} 
 \put(0,7.5){(a)}
 \put(8,7.5){(b)}
\put(-1,9.7){\epsfig{file=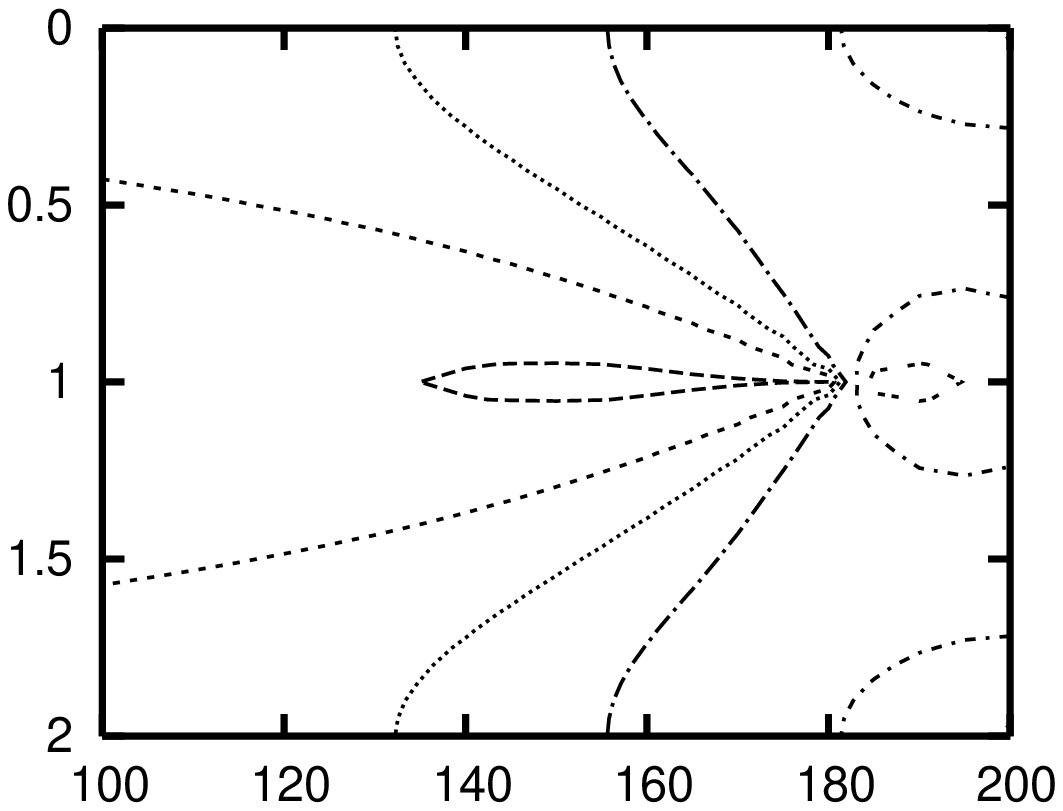,scale=0.6,angle=270}}
\put(7,9.7){\epsfig{file=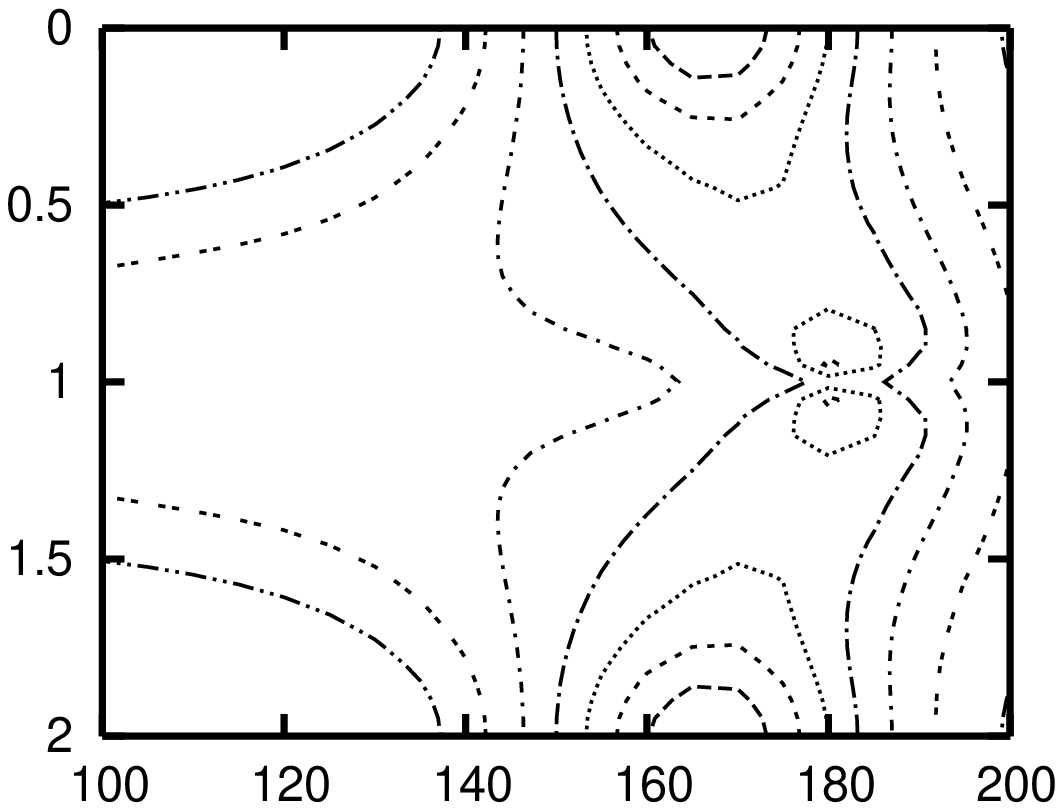,scale=0.6,angle=270}}
\put(3.4,5.5){\tiny{11}}
\put(3.9,4.8){\tiny{10}}
\put(4.2,4.3){\tiny{9}}
\put(4.8,4.0){\tiny{8}}
\put(6.1,3.7){\tiny{6}}
\put(6.2,4.8){\tiny{6}}
\put(6.8,5.5){\tiny{4}}
\put(10.1,4.3){\tiny{3}}
\put(10.8,4.6){\tiny{4}}
\put(11.9,4.9){\tiny{5}}
\put(12.7,4.6){\tiny{6}}
\put(12.9,4.1){\tiny{7}}
\put(13.1,3.9){\tiny{8}}
\put(13.2,3.4){\tiny{9}}
\end{picture}
\vspace{-3cm}
\caption{\label{fig:BRM1}
Contours (in \%) of branching ratios in scenario A as functions of $M_1$ and $\phi_{M_1}$: (a)
$BR(\tilde{t}_1 \to \tilde{\chi}^0_2 t)$  and (b) 
$BR(\tilde{\chi}^0_2 \to \tilde{\chi}^0_1 \ell^+ \ell^-)$, $\ell=e$ or $\mu$.}
\end{figure}

\begin{figure}[ht!]
\vspace{0.5cm}
\begin{picture}(16,8)
  \put(1,8){$BR(\tilde{t}_1 \to \tilde{\chi}^0_2 t)$, Scenario A ($\phi_{M_1}=0$)}  
  \put(9,8){$BR(\tilde{t}_1 \to \tilde{\chi}^0_2 t)$, Scenario A ($\phi_{M_1}=0$)}
 \put(-0.2,5.4){$\phi_{A_t}/\pi$}
 \put(6,2.5){$M_1/\mathrm{GeV}$}
 \put(7.8,5.4){$\phi_{A_t}/\pi$}
 \put(14,2.5){$A_t/\mathrm{GeV}$}
 \put(9.8,3.6){\tiny{4}}
 \put(10.4,3.9){\tiny{8}}
 \put(10.9,4.2){\tiny{12}}
 \put(11.2,4.7){\tiny{16}}
 \put(10.8,5.0){\tiny{20}}
 \put(10.1,5.5){\tiny{24}}   
 \put(6.5,3.6){\tiny{4}}
 \put(5.3,4.2){\tiny{8}}
 \put(4.5,4.5){\tiny{12}}
 \put(4.0,4.8){\tiny{16}}
 \put(3.3,5.1){\tiny{20}}
 \put(2.2,5.5){\tiny{24}}
  \put(0,7.5){(a)}
 \put(8,7.5){(b)}
\put(-1,9.7){\epsfig{file=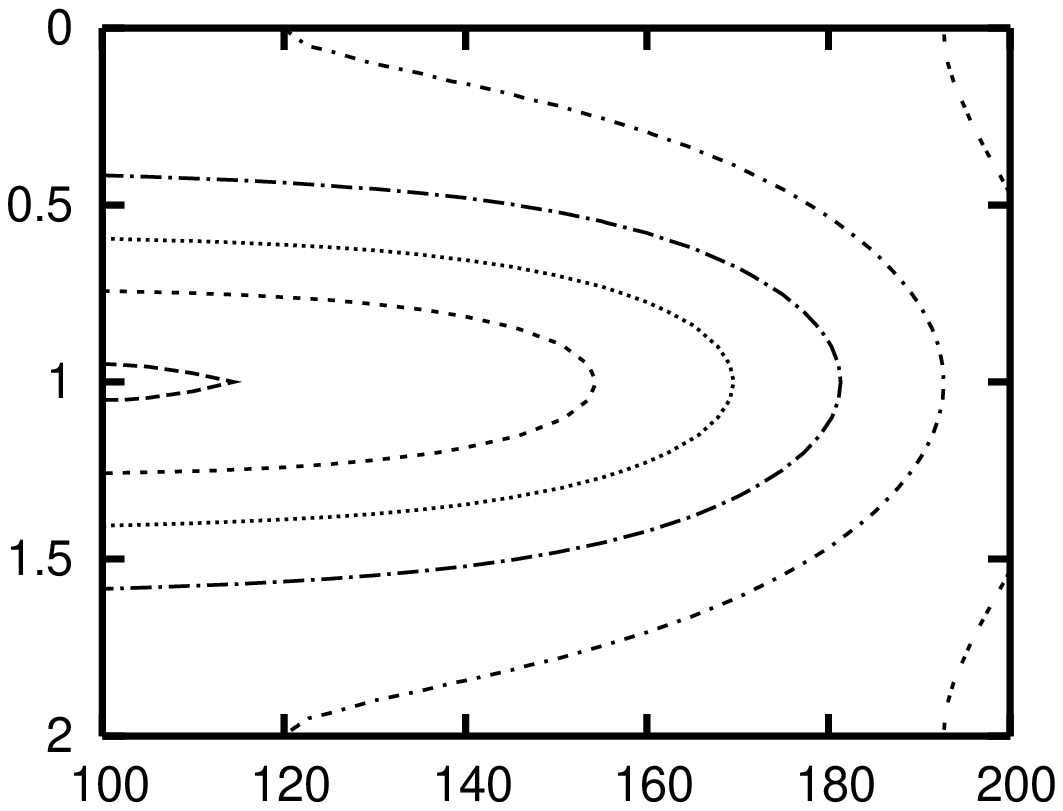,scale=0.6,angle=270}}
\put(7,9.7){\epsfig{file=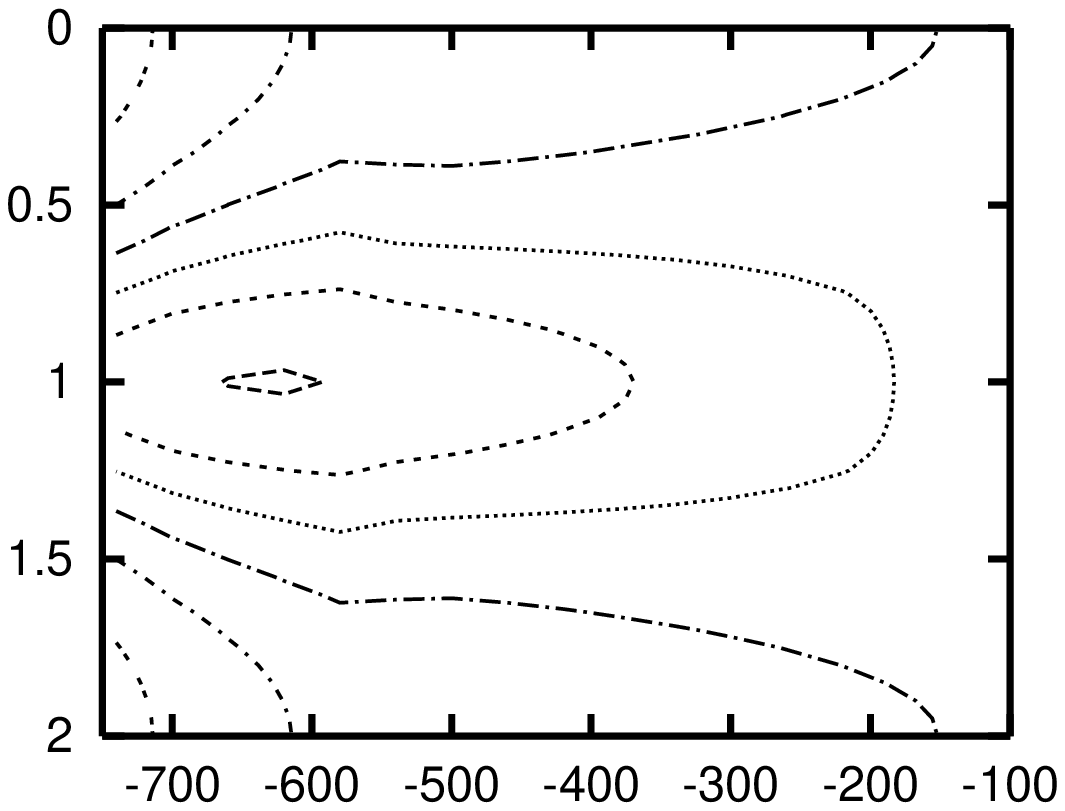,scale=0.6,angle=270}}
\end{picture}
\vspace{-3cm}
\caption{\label{fig:BRAt} Contours (in \%) of the branching ratio $BR(\tilde{t}_1 \to
  \tilde{\chi}^0_2 t)$,  in scenario A as functions of varying (a) $M_1$ and
  $\phi_{A_t}$ and (b) the common trilinear coupling $A_t=A_b=A_\tau$ and the phase of 
  the top-quark  trilinear coupling $\phi_{A_t}$.}
\end{figure}

In order to determine whether an asymmetry could be observed at the LHC, we need to
calculate the cross section for the total process. Important factors in the total
cross section are the branching ratios $BR(\tilde{t}_1 \to \tilde{\chi}^0_2 t)$ (for CP-violating case see \cite{Bartl:2003yp}) and
$BR(\tilde{\chi}^0_2 \to \tilde{\chi}^0_1 \ell^+ \ell^-)$ \cite{Bartl:2004jj}. Both of these change
considerably with $\phi_{M_1}$ and $\phi_{A_t}$, altering the statistical
significance of any measurement of $|\mathcal{A}_{T_f}|$. Analyzing first the
variation with $M_1$, seen in Figs.~\ref{fig:BRM1} and \ref{fig:BRAt}(a), 
we see that the branching ratio $BR(\tilde{t}_1 \to \tilde{\chi}^0_2
t)$ is indeed sensitive to variation of the phase, but can vary more strongly with $|M_1|$. For example, if $\phi_{M_1}=\pi$ when 
$|M_1|\approx150$~GeV then
$BR\approx11\%$, but if we keep the phase the same and change to $|M_1|\approx180$~GeV then
$BR\approx4\%$ (i.e., it drops by almost a factor of four), as seen in Fig.~\ref{fig:BRM1}(a). The general
reduction of $BR(\tilde{t}_1 \to \tilde{\chi}^0_2 t)$ as $M_1$ increases is to be expected
as the character of $\tilde{\chi}^0_2$ will be less gaugino-like. Similar large differences are
found in $BR(\tilde{\chi}^0_2 \to \tilde{\chi}^0_1\ell^+ \ell^-)$ which varies between 3\% for $M_1<135$GeV and 9\% for $M_1\approx165$GeV Fig.~\ref{fig:BRM1}(b).

The phase $\phi_{A_t}$ does not enter $BR(\tilde{\chi}^0_2 \to \tilde{\chi}^0_1 \ell^+
\ell^-)$, but can have a large effect on $BR(\tilde{t}_1 \to \tilde{\chi}^0_2 t)$. In
scenario A, we see in Fig.~\ref{fig:BRAt} that $BR\approx8\%$ at $\phi_{A_t}=0$ but increases to
$BR\approx24\%$ at $\phi_{A_t}=\pi$ (i.e.~a factor of 3 increase). The branching ratio $BR(\tilde{t}_1 \to \tilde{\chi}^0_2 t)$ also has a dependence on $|A_t|$ an this is shown in Fig.~\ref{fig:BRAt}(b). We see
that if $\phi_{A_t}=0$ then the branching vary between, $BR\approx4\%$ when $|A_t|\approx-750$GeV and
$BR\approx12\%$ when $|A_t|\approx-100$GeV.

In the range of $M_1=|M_1|e^{i\phi_{M_1}}$ and $A_t=|A_t|e^{i\phi_{A_t}}$ studied, we find
that $BR(\tilde{t}_1 \to \tilde{\chi}^0_2 t)$ varies between 4\% and 24\% and
$BR(\tilde{\chi}^0_2 \to \tilde{\chi}^0_1 \ell^+\ell^-)$ between 2.5\% and 9\% for scenario A.
Similar plots can also be produced for scenarios B and C but are not presented here. It is found
that $BR(\tilde{t}_1 \to \tilde{\chi}^0_2 t)$ varies between 4\% and 14\% for scenario B and
between 8\% and 35\% for scenario C. For $BR(\tilde{\chi}^0_2 \to \tilde{\chi}^0_1 \ell^+\ell^-)$
the variation is between 3\% and 12\% for scenario B and between 2\% and 5\% for scenario C.

\subsection{Influence of parton distribution functions (pdfs) on CP asymmetries}
\label{sec:asymmetry-at-lhc}

So far we have studied the triple-product asymmetries only when the production process is
close to threshold, and the $\st_1$ pair is produced almost at rest in its
centre-of-mass frame; triple-product
effects due to spin effects are usually greatest close to threshold. However, 
production at the LHC is not in general close to threshold, 
and we must include pdfs in our analysis to see how an
initial boost to the $\st_1$ affects the asymmetry. We focus on scenario A: similar results are obtained
in Scenarios B and C.

\begin{figure}[ht!]
\vspace{0.5cm}
\begin{picture}(16,8)
  \put(1,8){$\mathcal{A}_{\mathcal{T}_t}$, Scenario A ($\phi_{M_1}=0.9\pi$)}
  \put(9,8){$\sigma(gg\rightarrow\st_1\st_1)$, Scenario A ($\phi_{M_1}=0$)}
 \put(0,7.5){(a)}
 \put(8,7.5){(b)}
 \put(10.4,6.6){\tiny{Peak$\sim$900GeV}}
\put(0,8){\epsfig{file=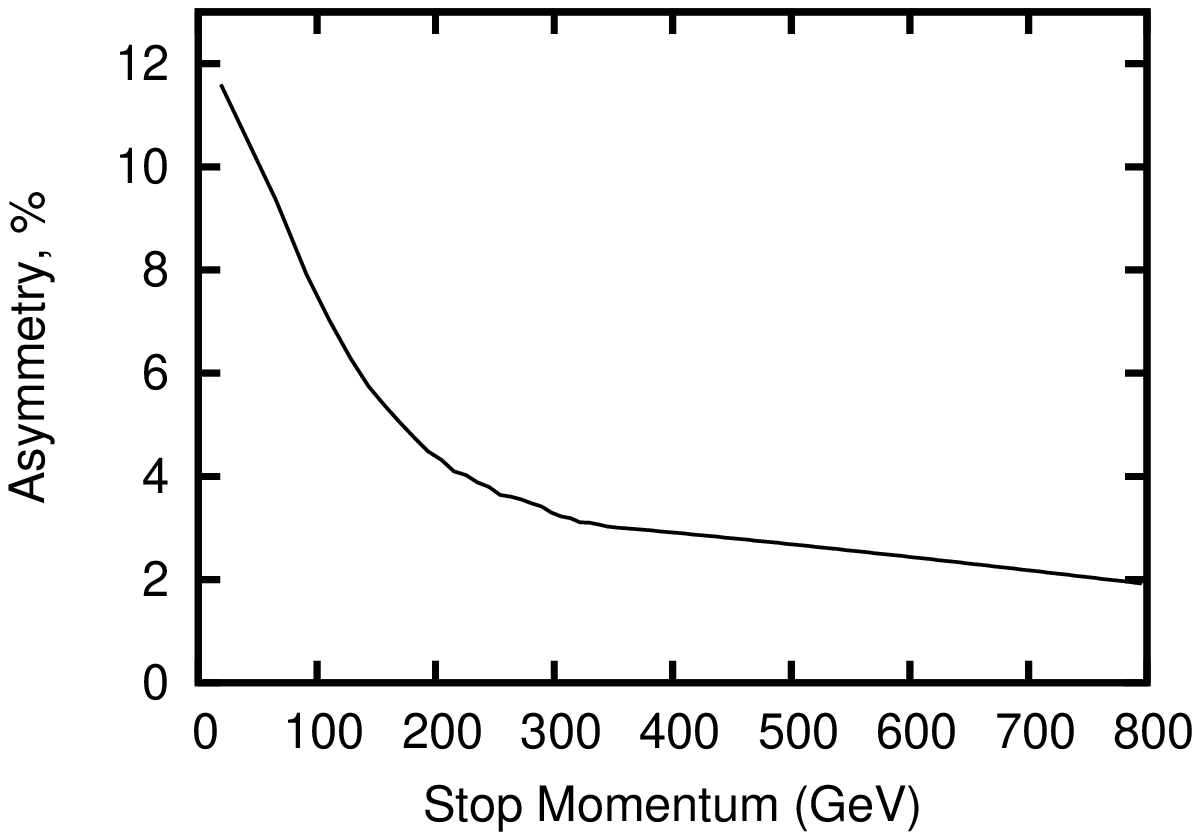,scale=0.6,angle=270}}
\put(8,8){\epsfig{file=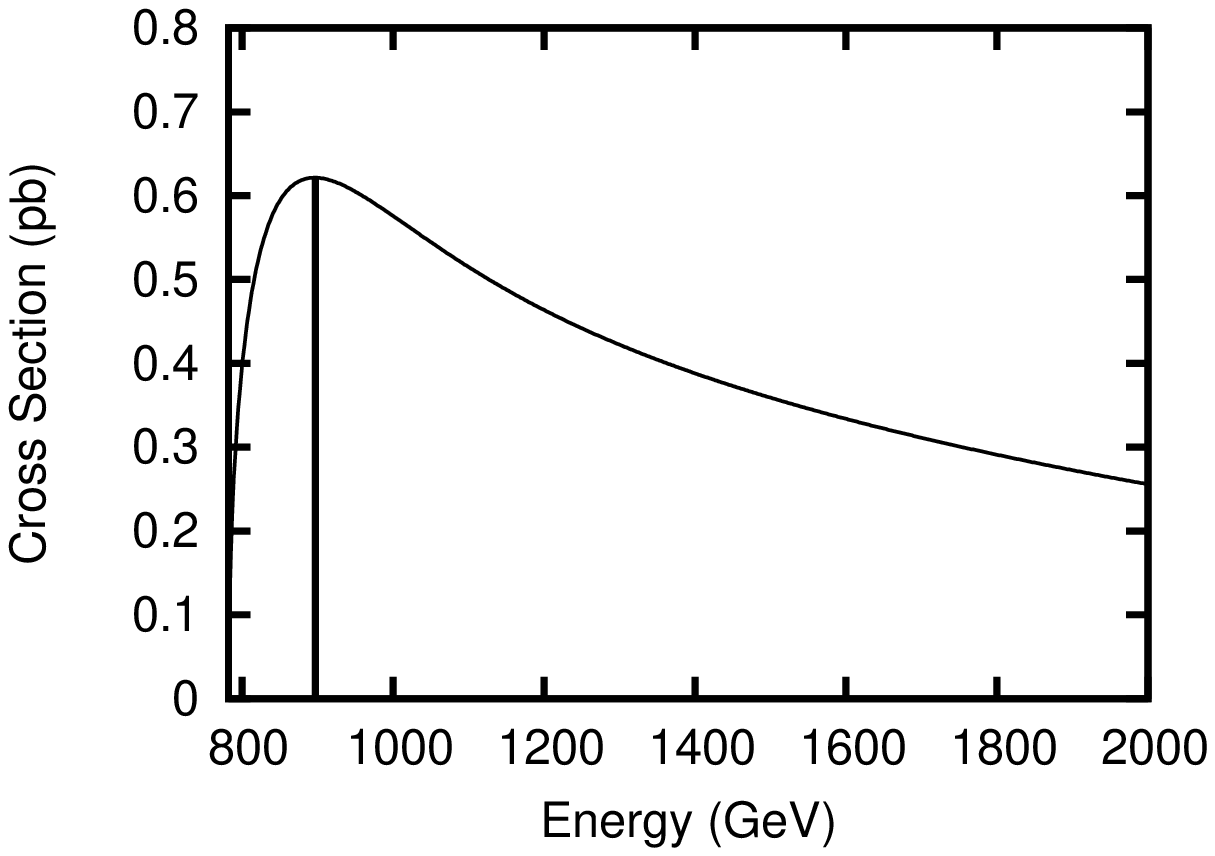,scale=0.6,angle=270}}
\end{picture}
\vspace{-3cm}
\caption{\label{fig:M1Energy} (a) Asymmetry $\mathcal{A}_{T_t}$ for scenario A as a
  function of the $\st$ momentum. (b) Total cross section for scenario A for
  $gg\longrightarrow\st\overline{\st}$ as a function of the parton-parton centre-of-mass energy.}
\end{figure}

Fig.~\ref{fig:M1Energy}(a) shows the asymmetry $|\mathcal{A}_{T_t}|$ as a function of the
$\st_1$ momentum, and shows clearly that the asymmetry is peaked at the threshold for
$\tilde t_1$ production, where the stops are produced almost at rest, and that it falls sharply as
the energy increases.  Fig.~\ref{fig:M1Energy}(b) shows the total cross section in 14~TeV
collisions at the LHC for $gg \to
\st_1 \overline{\st}_1$ as a function of the parton-parton centre-of-mass energy, 
and demonstrates that the
peak production occurs close to threshold with a long tail of production at high energy.
In addition, even when production occurs at a low parton-parton centre-of-mass energy, in the majority
of cases one gluon may be carrying significantly more momentum than the other in the
collision. Consequently the produced $\st_1$ can have a large longitudinal component to
its momenta. Both these factors mean that the asymmetry observed at the LHC will be
substantially smaller than if the all $\st_1$ were produced at threshold~\footnote{Both these effects could be overcome if one could measure the stop-stop invariant mass and tag the stop momenta,
but this is unlikely to be possible with great accuracy.}. In should be noted
that similar results were found for all asymmetries and scenarios, and this `dilution' factor is always present.

\begin{figure}[ht!]
\vspace{1cm}
\begin{picture}(16,8)
  \put(1,8.5){$\mathcal{A}_{\mathcal{T}_t}$, Scenario A ($M_1=130~\mathrm{GeV}$,}
  \put(1,8){$\phi_{A_t}=0$), $\sqrt{\hat{s}}=14~\mathrm{TeV}$}
  \put(9,8.25){$\mathcal{A}_{\mathcal{T}_{tb}}$, Scenario A ($\phi_{M_1}=0$), $\sqrt{\hat{s}}=14~\mathrm{TeV}$}
 \put(6,2.5){$\phi_{M_1}/\pi$}
 \put(0,7.5){(a)}
\put(2.5,6.8){\tiny{100fb$^{-1}$}}
\put(2.5,6.1){\tiny{500fb$^{-1}$}}
\put(2.5,5.6){\tiny{1ab$^{-1}$}}
\put(0,8){\epsfig{file=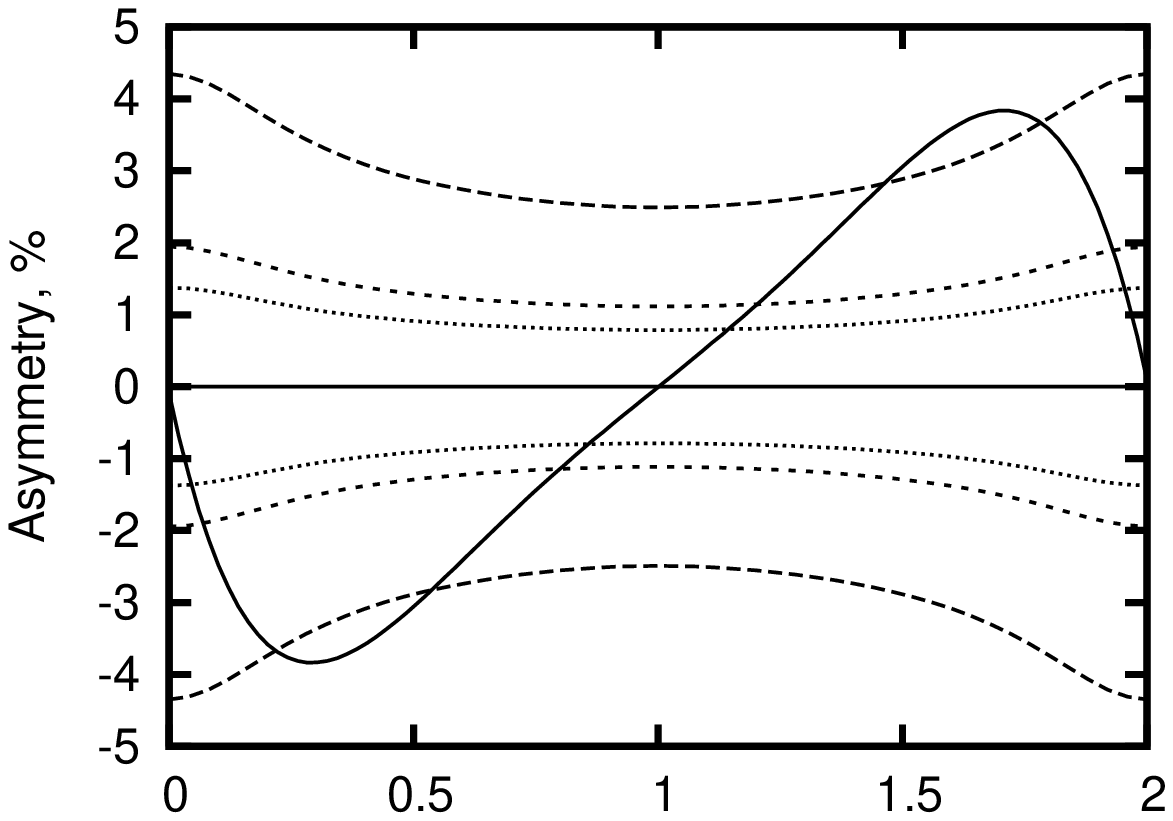,scale=0.6,angle=270}}

 \put(14,2.5){$\phi_{A_t}/\pi$}
 \put(8,7.5){(b)}
\put(12.5,6.8){\tiny{100fb$^{-1}$}}
\put(12.5,6.1){\tiny{500fb$^{-1}$}}
\put(12.5,5.6){\tiny{1ab$^{-1}$}}
\put(8,8){\epsfig{file=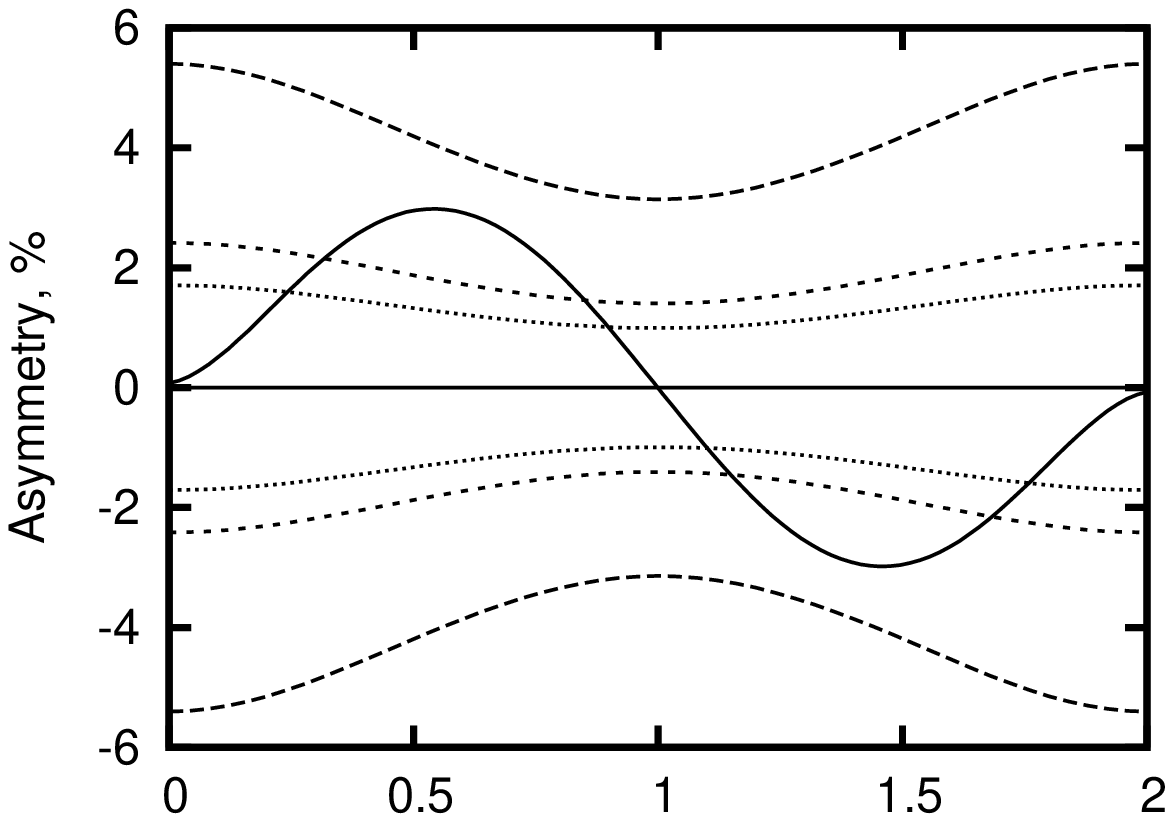,scale=0.6,angle=270}}
\end{picture}
\vspace{-3cm}
\caption{\label{fig:Sigma} Integrated asymmetries with parton density functions included in the production process.
  The dotted and dashed lines indicate the asymmetry required in order to observe a
  1$\sigma$ deviation from zero with the indicated luminosities, see the text:
  (a) $\mathcal{T}_t=\bf p_{t}\cdot(\bf
  p_{\ell^+} \times \bf p_{\ell^-})$ in scenario A as a function of $\phi_{M_1}$
  with $M_1=130$~GeV, and (b) $\mathcal{T}_{tb}=\bf p_{\ell^+}\cdot(\bf p_{t} \times \bf
  p_{b})$ in scenario A as a function of $\phi_{A_t}$ with $M_1=109$~GeV.}
\end{figure} 

\begin{figure}[ht!]
\vspace{1cm}
\begin{picture}(16,8)
  \put(5,8){$\mathcal{A}_{\mathcal{T}_{b}}$, Scenario A ($\phi_{M_1}=0$), $\sqrt{\hat{s}}=14~\mathrm{TeV}$}
 \put(10,2.5){$\phi_{A_t}/\pi$}
 \put(4,7.5){(a)}
\put(6.5,7.1){\tiny{100fb$^{-1}$}}
\put(6.5,6.2){\tiny{500fb$^{-1}$}}
\put(6.5,5.6){\tiny{1ab$^{-1}$}}
\put(4,8){\epsfig{file=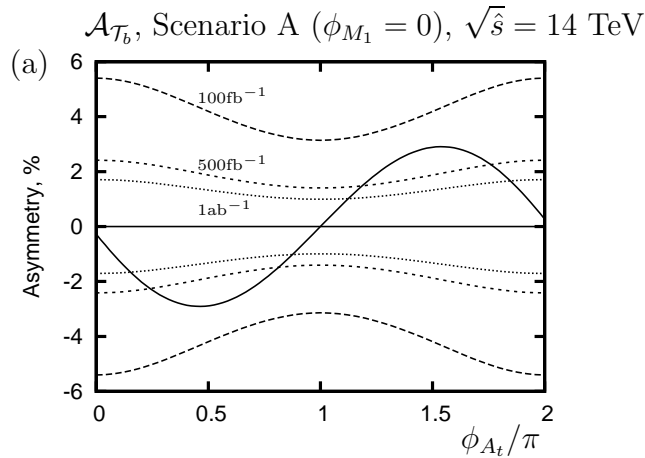,scale=0.6,angle=270}}

\end{picture}
\vspace{-3cm}
\caption{\label{fig:Sigma_b} Integrated asymmetries with parton density functions included in the production process.
  The dotted and dashed lines indicate the asymmetry required in order to observe a
  1$\sigma$ deviation from zero with the indicated luminosities, see the text:
  (a) $\mathcal{T}_{b}=\bf p_{b}\cdot(\bf p_{\ell^+} \times \bf
  p_{\ell^-})$ in scenario A as a function of $\phi_{A_t}$ with $M_1=109$~GeV.}
\end{figure} 

We use the \texttt{MRST 2004LO} pdf set~\cite{Martin:2007bv} in our analysis of the
asymmetry, and plot the integrated asymmetry
$|\mathcal{A}_{T_t}|$ as a function of $\phi_{M_1}$ and $\phi_{A_t}$ at
the LHC in Fig.~\ref{fig:Sigma}(a), as the solid line. We see that the inclusion of the pdfs
reduces the asymmetry by about a factor of four in this case. 
This reduction is unsurprising, given the reduction in asymmetry when
one moves away from threshold shown in Fig.~\ref{fig:M1Energy}(a),
though the dilution factor does depend on the scenario\footnote{These results 
have been checked independently using \texttt{Herwig++}~\cite{Bahr:2008pv,Bahr:2008tx} with 
three-body spin correlations included, a feature that is currently not available in an
  official release of the code, but will be included in a future version.}.

Using the production cross sections and branching ratios we can then estimate
the integrated luminosity required to observe an asymmetry at the LHC. We assume that
$N_{\mathcal{T}\pm}$, the numbers of events where $\mathcal{T}$ is positive and negative
as in eq.~(\ref{Asy}), are binomially distributed, giving the following statistical 
error~\cite{Desch:2006xp}:
\begin{equation}
  \label{eq:asymmerror}
  \Delta(\mathcal{A}_T)^{\rm stat}=2\sqrt{\epsilon(1-\epsilon)/N},
\end{equation}
where $\epsilon=N_{\mathcal{T}+}/(N_{\mathcal{T}+}+N_{\mathcal{T}-})=\frac 12
(1+\mathcal{A}_T)$, and $N$ is the number of selected events. This can be rearranged to
give the required number of events for a desired significance.

Figs.~\ref{fig:Sigma} (a), (b) and \ref{fig:Sigma_b}(a) show the expected levels of the
integrated asymmetries in scenario A with pdf effects included
(solid line) together with dotted and dashed lines showing the level of asymmetry one would need
with the corresponding integrated luminosity in order to obtain a statistical error
$\mathcal{A}_T>\Delta(A_T)$.  In other words, the
asymmetry could only be seen at the level of 1$\sigma$ where the solid line is above the
relevant dotted or dashed line.  For example, in scenario A after 100~fb$^{-1}$, the asymmetry
could only be seen for a small area of parameter space around $\phi_{M_1}=0.35\pi$ and
$1.7\pi$.  Figs.~\ref{fig:Sigma} (a) and (b) show that even if $\phi_{M_1}$ or
$\phi_{A_t}$ has a value that produces a maximal asymmetry, we require a substantial
integrated luminosity if we are to find a statistically significant result. In addition,
it must be noted that we have not included any detector effects into our analysis, and one
could expect that the required integrated luminosity would rise substantially
after the inclusion of backgrounds, trigger efficiencies, etc.  A measurement of an asymmetry
with an accuracy of a few \% might be possible with 100~fb$^{-1}$ of integrated
luminosity, but it would probably be insufficient to constrain significantly the model
parameter space. However, an interesting measurement could be made with an integrated luminosity
above 300~fb$^{-1}$, which is targeted by the proposed LHC luminosity upgrade.

%%%%%%%
\subsection{Determination of the CP-violating phases}

As we have shown, it will be challenging to determine the phases $\phi_{M_1}$ and
$\phi_{A_t}$ in our process using the triple-product asymmetries alone.  However, it would
be very worthwhile, as a non-zero measurement of a T-odd asymmetry would provide unique
evidence of CP violation.  In the rest of this section, we examine briefly the potential
for a measurement using other variables, again concentrating on Scenario A.

\subsubsection{Observables: masses, cross sections and CP asymmetries}

\begin{figure}[ht!]    
\vspace{0.5cm} 
\begin{picture}(16,8)
\put(2,8){$m_{\st_i}$, Scenario A ($\phi_{M_1}=0$)}
  \put(9,8){$m_{\tilde{\chi}^0_2}-m_{\tilde{\chi}^0_1}$, Scenario A ($\phi_{A_t}=0$)}
 \put(14,2.5){$M_1/\mathrm{GeV}$}
 \put(0,7.5){(a)}
  \put(8,7.5){(b)}
   \put(6,2.5){$\phi_{A_t}/\pi$}
  \put(7.8,5.4){$\phi_{M_1}/\pi$}
  \put(10.7,4.5){\tiny{60}}
  \put(11.7,5.9){\tiny{40}}
  \put(12.6,7.1){\tiny{20}}
\put(8,8){\epsfig{file=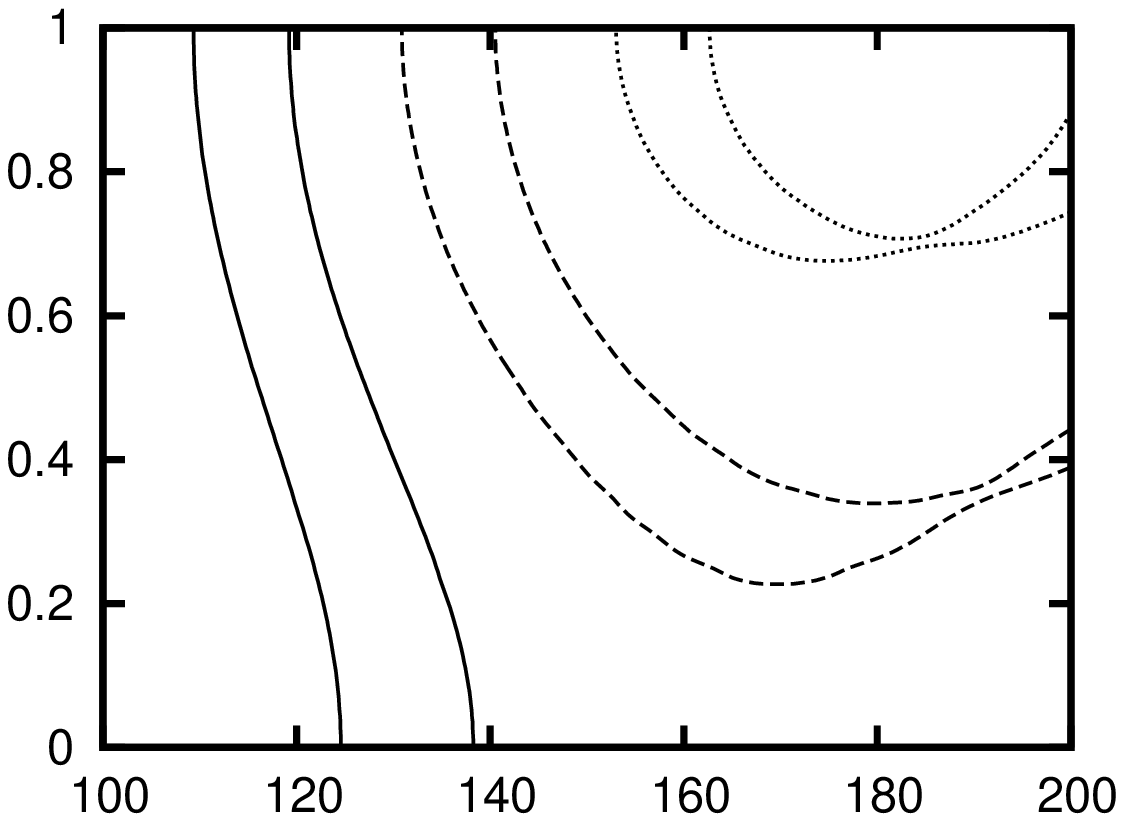,scale=0.6,angle=270}}
\put(0,8){\epsfig{file=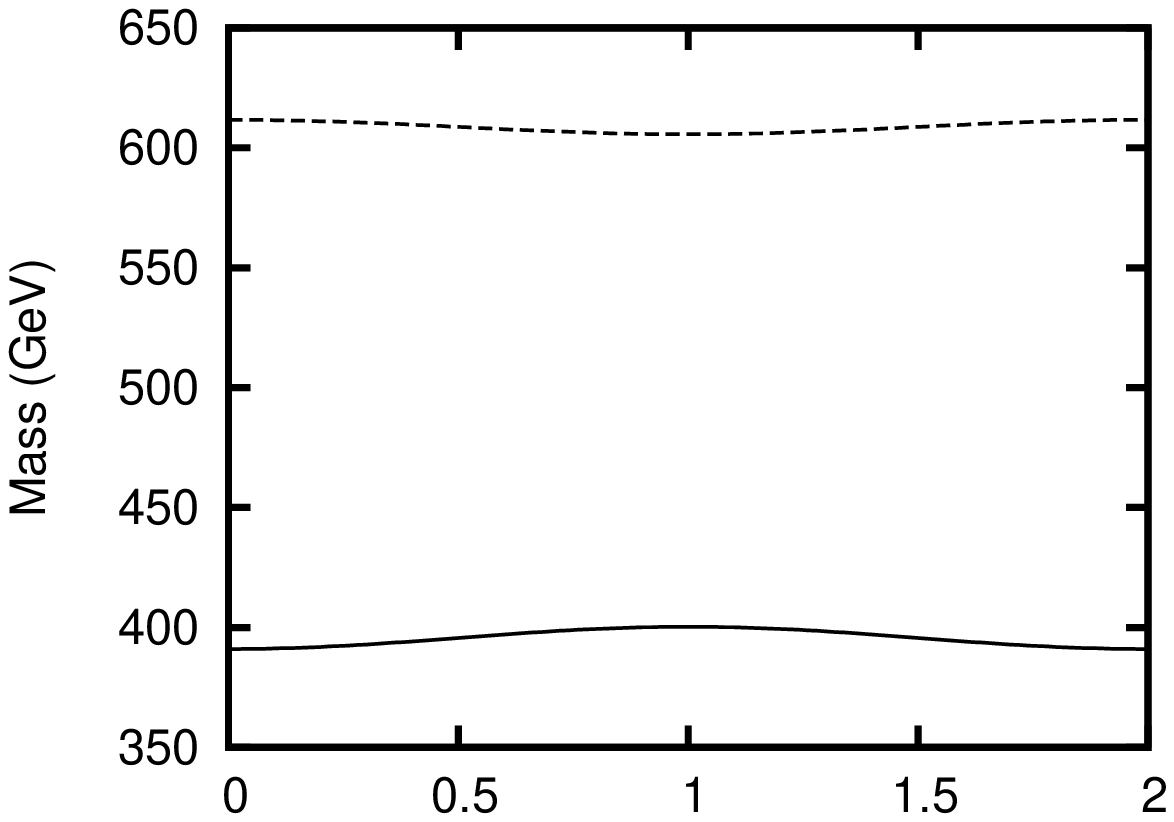,scale=0.6,angle=270}}
\end{picture}
\vspace{-3cm}
\caption{\label{fig:M1Mass} (a) The mass of the stop squarks $\tilde{t}_j$, $j$=1,2 as functions of
  $\phi_{A_t}/\pi$, and (b) contour plot showing the areas of the $(M_1,\phi_{M_1})$ parameter
  plane consistent with a mass difference between $\tilde{\chi}^0_2$ and
  $\tilde{\chi}^0_1$ of 20, 40 and 60~GeV respectively.  The bands assume a 1\% error in
  experimental measurement of the mass difference and a 5\% error in $M_2$.  }
\end{figure}

Fig.~\ref{fig:GaugeMass}(b) showed how the masses of the $\tilde{\chi}^0_i$s vary with
$\phi_{M_1}$ and Fig.~\ref{fig:M1Mass}(a) shows how the masses of the $\st_i$ vary with
$\phi_{A_t}$ in scenario A. 
Unfortunately, the variations in both of these observables are only about $1-2$
GeV, which are significantly smaller than the experimental errors expected for these
measurements. A far more accurate measurement at the LHC will be the mass difference
between $\tilde{\chi}^0_2$ and $\tilde{\chi}^0_1$, as this can be determined in our scenario with a clear
dilepton end-point. The accuracy of this measurement is expected to be $<1\%$ and, if we
assume that $M_2$ can be determined to $5\%$~\cite{Weiglein:2004hn}, 
we find the regions plotted in Fig.~\ref{fig:M1Mass}(b). At the smaller values allowed for $M_1$ 
in scenario A, we see that
this observable does not depend sensitively enough on $\phi_{M_1}$ for a measurement
to become possible.
However, as $M_1$ increases we see that the sensitivity to $\phi_{M_1}$ becomes much
clearer. Importantly, in scenario A, it is only possible to have a mass difference,
$\tilde\chi^0_2-\tilde\chi^0_1\lesssim40$ GeV if $\phi_{M_1}$ is present.

\subsubsection{Inclusion of branching ratios}

\begin{figure}[ht!]
\vspace{1cm}
\begin{picture}(16,8)
 \put(1.7,8.5){$BR(\tilde{\chi}^0_2 \to \tilde{\chi}^0_1 \ell^+ \ell^-)=0.04$,}
  \put(2,8){Scenario A ($\phi_{A_t}=0$)}
  \put(10,8.5){$BR(\tilde{t}_1 \to \tilde{\chi}^0_2 t)=0.1$,}
  \put(10,8){Scenario A ($\phi_{M_1}=0$)}
 \put(-0.2,5.4){$\phi_{M_1}/\pi$}
 \put(6,2.5){$M_1/\mathrm{GeV}$}
 \put(7.8,5.4){$\phi_{A_t}/\pi$}
 \put(14,2.5){$M_1/\mathrm{GeV}$}
 \put(6.55,3.7){\tiny{$\Delta_2$}}
 \put(6.1,4){\tiny{$\Delta_1$}}
 \put(4.2,4.45){\tiny{$\Delta_1$}}
 \put(2.6,4.4){\tiny{$\Delta_2$}}
 \put(2.25,3.85){\tiny{$\Delta_1$}}
 \put(13.8,3.6){\tiny{$\Delta_1$}}
 \put(12.3,4.1){\tiny{$\Delta_2$}}
 \put(10.9,4.3){\tiny{$\Delta_1$}}   
  \put(0,7.5){(a)}
 \put(8,7.5){(b)}
\put(0,8){\epsfig{file=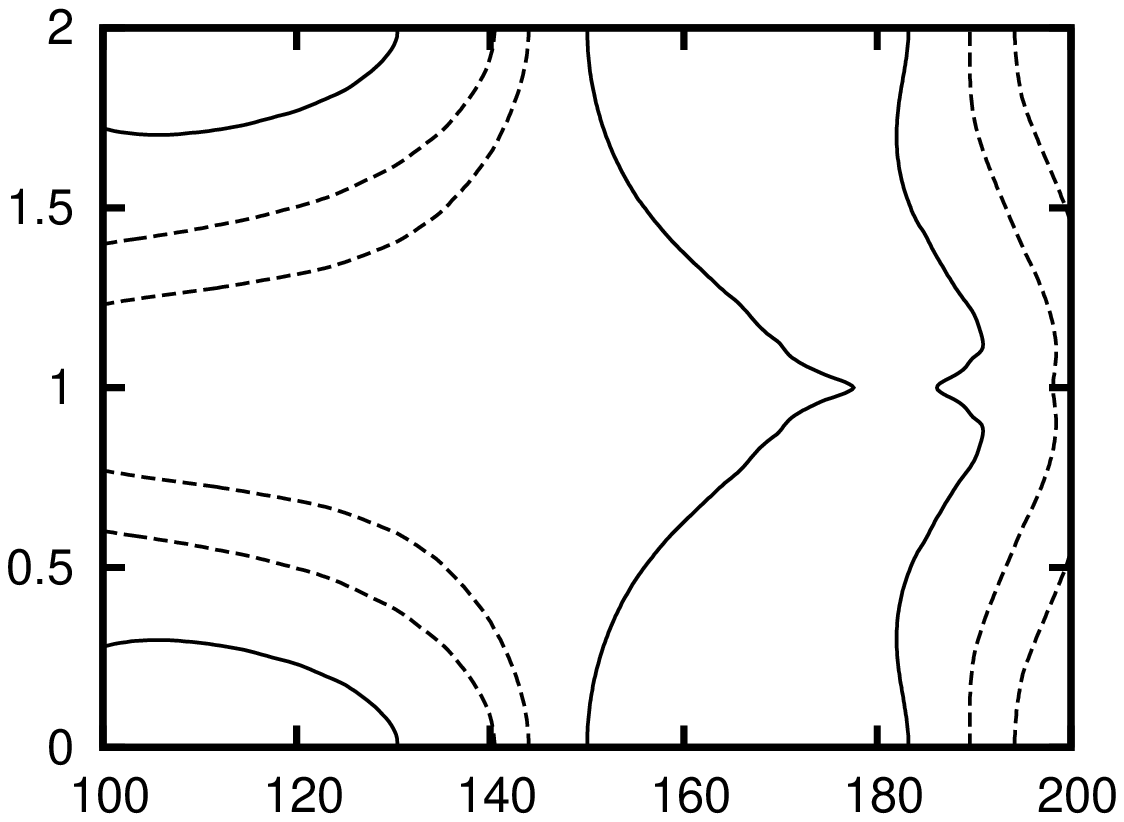,scale=0.6,angle=270}}
\put(8,8){\epsfig{file=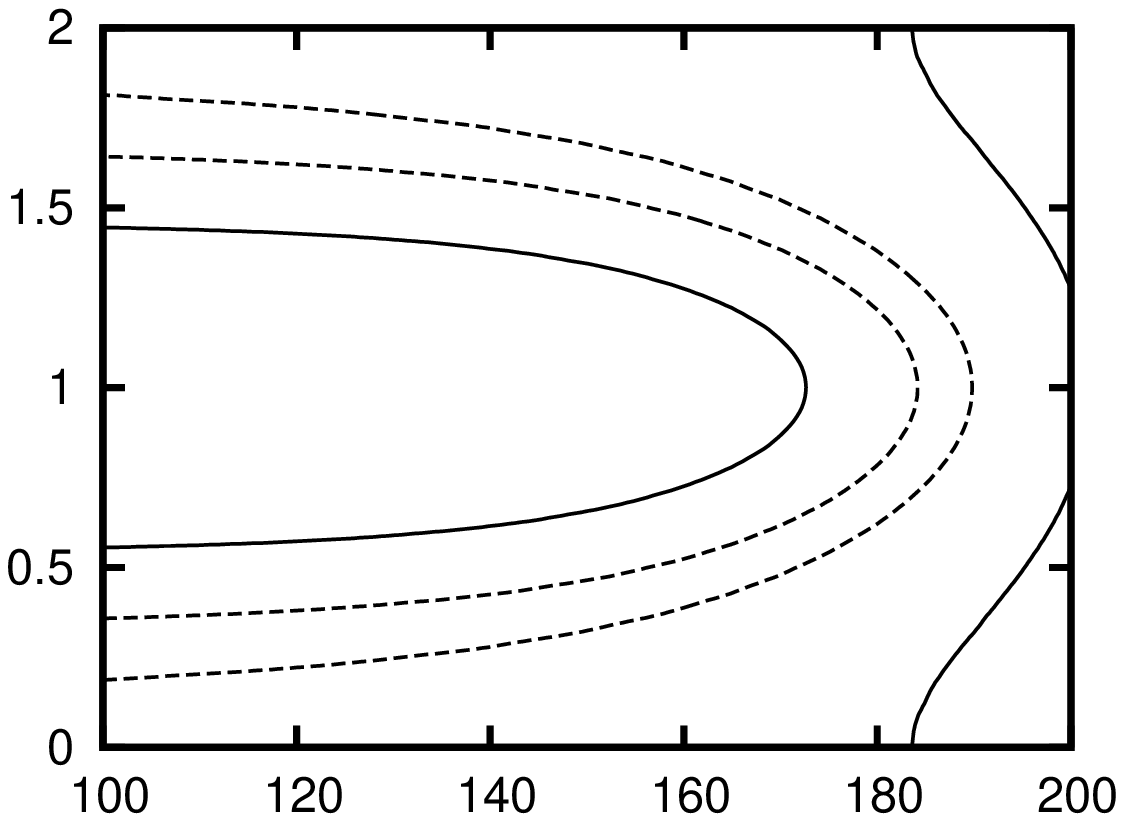,scale=0.6,angle=270}}
\end{picture}
\vspace{-3cm}
\caption{\label{fig:AreaPlot} Parameter space allowed when the experimental accuracy of
  the branching ratio measurement is 50\% ($\Delta_1$) or 10\% ($\Delta_2$) for (a)
  $BR(\tilde{\chi}^0_2 \to \tilde{\chi}^0_1 \ell^+ \ell^-)=0.04$ and (b) $BR(\tilde{t}_1
  \to \tilde{\chi}^0_2 t)=0.1$. }
\end{figure}

Other observables sensitive to the phases $\phi_{M_1}$ and $\phi_{A_t}$ are the
branching ratios $BR(\tilde{\chi}^0_2 \to \tilde{\chi}^0_1 \ell^+ \ell^-)$ and
$BR(\tilde{t}_1 \to \tilde{\chi}^0_2 t)$, as discussed in
Section~\ref{sec:depend-branch-rati}.  As is the case for the masses, though, our current
expectation of the accuracy of this measurement at the LHC looks insufficient to constrain
the phases. Fig.~\ref{fig:AreaPlot}(a) shows in the context of scenario A that,
if a measurement $BR(\tilde{\chi}^0_2
\to \tilde{\chi}^0_1 \ell^+ \ell^-)=0.4$ is made and we assume that the accuracy at the LHC is
50\% ($\Delta_1$), then the constraints on $M_1$ and $\phi_{M_1}$ are rather weak. However,
if the accuracy could be improved to 10\% ($\Delta_2$), 
a determination of $M_1$ and
$\phi_{M_1}$ looks possible if this analysis is combined with information from the
$\tilde{\chi}^0_2$, $\tilde{\chi}^0_1$ mass difference and that of the triple-product correlations. For
the branching ratio, $BR(\tilde{t}_1 \to \tilde{\chi}^0_2 t)$, the conclusion is similar, as seen in
Fig.~\ref{fig:AreaPlot}(b). With a measurement at 50\%  ($\Delta_1$), we again see that a
determination of the CP-violating parameter is not possible but,
if a measurement can be made with an accuracy of 10\%
($\Delta_2$), then a determination of $\phi_{A_t}$ would be more plausible.

Thus, we may be able to to pin the model parameters down with greater accuracy
by combining information on the CP-violating asymmetries
with this and other information.

%%%%%%%%%%%%%%%%%%%%%%%%%%%%%%%%%%%%%%%
\section{Conclusions}
%%%%%%%%%%%%%%%%%%%%%%%%%%%%%%%%%%%%%%%

We have studied direct stop production followed by the decay $\st_1 \to t\tilde\chi^0_2$, $\tilde
\chi^0_2 \to \ell^+ \ell^- \tilde \chi^0_1$, where the latter is a three-body decay and
provide compact analytical expressions for the amplitude and phase space.  We
have specifically concentrated on measuring the CP-violating phases of the parameters
$M_1$ and $A_t$.

We have provided a thorough analysis of the contributions to this process which lead to
non-zero asymmetries in the parameters $\mathcal{T}_t$, $\mathcal{T}_b$ and
$\mathcal{T}_{tb}$ formed from triple products of reconstructable final-state
particles.  These are sensitive to different combinations of the phases mentioned above.
We studied three spectra which had different neutralino characteristics at the parton
level and also studied the (large) effect of including pdfs which had previously only been roughly
estimated in the literature.

We found that with the design integrated luminosity of the LHC of 100~fb$^{-1}$, the
statistical errors would probably remain too great to distinguish a non-zero asymmetry 
measurement from
zero for most of the ranges of $\phi_{M_1}$ and $\phi_{A_t}$, and 
we recall that this initial study did
not include detector or background effects.  However, with a luminosity upgrade, the
accuracy will
improve and it could be possible either to measure a non-zero value or else to provide
limits on the possible phases.

Triple products are not the only variables sensitive to the phases of the
parameters.  We found that a good measurement of the mass difference between the $\tilde\chi^0_2$ and $\tilde \chi^0_1$ neutralinos could constrain significantly the ($M_1$,$\phi_{M_1}$)
parameter space.  It is possible that measurements of the two branching ratios
$BR(\tilde{t}_1 \to \tilde{\chi}^0_2 t)$ and  $BR(\tilde{\chi}^0_2 \to \tilde{\chi}^0_1
\ell^+ \ell^-)$ could also constrain both $\phi_{M_1}$ and $\phi_{A_t}$, although this is
heavily dependent on the experimental accuracy achieved.  However, the disadvantage of
both mass differences and branching ratios is that a non-zero value can potentially be
faked by other values of the real parameters.  This is in contrast to the asymmetries from triple
products which are uniquely due to CP violation. Therefore, even though these will be challenging
measurements at the LHC, they are worthwhile experimental objectives.

%%%%%%%%%%%%%%%%%%%%%%%%%%%%%%%%%%%%%%%%%%%%%%%%%%%%%%%%%%%%%%%%%%%%%%%%%%%%%%%%%%%%%%%%%%

%%%%%%%%%%%%%%%%%%%%%%%%%%%%%%%%%%%%%%%
\section*{Acknowledgements}
%%%%%%%%%%%%%%%%%%%%%%%%%%%%%%%%%%%%%%%
We are grateful to Martyn Gigg, David Grellscheid and Peter Richardson for their
assistance in the use of \texttt{Herwig++} for these studies.  JT thanks Martin White and
Dan Tovey for useful discussions.  JMS and JT are supported by
the UK Science and Technology Facilities Council (STFC).

%%%%%%%%%%%%%%%%%%%%%%%%%%%%%%%%%%%%%%%%

\section*{Appendices}

\begin{appendix}

\section{Lagrangian and couplings}
\label{sec:lagrangian-couplings}
The interaction Lagrangian terms for the production processes are

\begin{equation}
\mathcal{L}_{ggg} =
  g_s\partial^\nu G^a_\mu g^{\mu\rho}f_{abc}G^b_\nu G^c_\rho, 
\end{equation}
\begin{equation}
\mathcal{L}_{\tilde{q}\tilde{q}g} =
  ig_sT^a_{rs}\delta_{ij}G^a_\mu\tilde{q}^*_{jr}\stackrel{\leftrightarrow}{\partial^\mu}\tilde{q}_{is}, 
\end{equation}
\begin{equation}
\mathcal{L}_{\tilde{q}\tilde{q}gg} =
  \smaf{1}{2}g^2_s(\smaf{1}{3}\delta_{ab}+d_{abc}T^c)G^a_\mu G^{b\mu}\tilde{q}^*_{j}\tilde{q}_i.
\end{equation}
The interaction Lagrangian terms for $\tilde{\chi}^0_2$ decay are
\begin{equation}
\mathcal{L}_{Z^0\ell^+\ell^-} =
  -\frac{g}{\cos\Theta_W}Z_{\mu}\bar{\ell}\gamma^{\mu}
  [L_\ell P_L+R_\ell P_R]\ell, 
\end{equation}
\begin{equation}
\mathcal{L}_{Z^0\tilde{\chi}^0_m\tilde{\chi}^0_n} =
 \frac{1}{2}\frac{g}{\cos\Theta_W}Z_{\mu}{\bar{\tilde{\chi}}}^0_m\gamma^{\mu}
 [O''^L_{mn}P_L+O''^R_{mn}P_R]{\tilde{\chi}}^0_n,
\end{equation}
\begin{equation}
 \mathcal{L}_{\ell\tilde{\ell}\tilde{\chi}^0_k} =
 g\bar{\ell}(a^{\tilde{\ell}}_{jk} P_R + b^{\tilde{\ell}}_{jk} P_L)\tilde{\chi}^0_k\ell_j +
 \textrm{h.c.},
\end{equation}
where the couplings $a^{\tilde{\ell}}_{jk}$ and $b^{\tilde{\ell}}_{jk}$ are given by
\begin{equation}
 a^{\tilde{\ell}}_{ik}=\displaystyle\sum_{n=1}^2({\cal R}^{\tilde{\ell}}_{in})^*{\cal A}^{\ell}_{kn},\;\;\;\;
 b^{\tilde{\ell}}_{ik}=\displaystyle\sum_{n=1}^2({\cal R}^{\tilde{\ell}}_{in})^*{\cal B}^{\ell}_{kn},
\end{equation}
Here ${\cal R}^{\tilde{\ell}}_{in}$ is the mixing matrix of the squarks and 
\begin{equation}
 {\cal A}^{{\ell}}_{k}={f_{Lk}^{\ell} \choose h_{Rk}^{\ell}}, \qquad
 {\cal B}^{{\ell}}_{k}={h_{Lk}^{\ell} \choose f_{Rk}^{\ell}}, 
\end{equation}
with
 \begin{eqnarray}
  f_{Lk}^{\ell} &=& -\sqrt{2}\,e_{\ell} \sin\t_W N_{k1} 
    -\sqrt{2}\,(T_{3\ell} -e_{\ell}\sin^{2}\t_W)\smaf{N_{k2}}{\cos\t_W}, 
    \hspace{-1cm} \\
  f_{Rk}^{\ell} &=& 
    -\sqrt{2}\,e_{\ell} \sin\t_W (\tan\t_W N_{k2}^* - N_{k1}^*), \\
  h_{Lk}^{\ell} &=& -Y_{\ell} \left(N_{k3}^*\sin\b - N_{k4}^*\cos\b \right), \\
                     &=&  (h_{Rk}^{\ell})^* 
 \end{eqnarray}
\begin{equation}
O''^L_{mn}=-\smaf{1}{2}(N_{m3}N^*_{n3}-N_{m4}N^*_{n4})\cos2\beta-\smaf{1}{2}(N_{m3}N^*_{n4}+N_{m4}N^*_{n3})\sin2\beta,
\end{equation}
\begin{equation}
O''^R_{mn}=-O''^{L*}_{mn},
\end{equation}
\begin{equation}
L_\ell=T_{3\ell}-e_\ell\sin^2\Theta_W,\quad R_\ell=-e_\ell\sin^2\Theta_W ,
\end{equation}
where $P_{L,R}=\frac{1}{2}(1\mp\gamma_5)$,
$Y_t = m_t/(\sqrt{2}m_W\sin\beta)$.
Here, $g$ is the weak coupling constant
($g=e/\cos\Theta_W$, $e>0$), $e_\ell$ and 
$T_{3\ell}$ are the charge (in units of $e$) and 
the third component of the
weak isospin of the fermion $\ell$, $\Theta_W$ is the
weak mixing angle and $\tan\beta=v_2/v_1$
is the ratio of the
vacuum expectation values of the Higgs fields.
The unitary $(4\times4)$ matrix $N_{mk}$
that diagonalises the complex symmetric 
neutralino mass matrix is given in the basis
$(\tilde{\gamma}, \tilde{Z}^0, \tilde{H}^0_1, \tilde{H}^0_2)$ by
\cite{Bartl:1986hp}:

\begin{equation}
  \label{eq:Nmatrix}
  N= \left(
    \begin{array}{cccc}
      M_1 e^{i \phi_{M_1}}c^2_W+ M_2 s^2_W & (M_1 e^{i \phi_{M_1}}-M_2)s_W c_W & 0 & 0 \\
      (M_1 e^{i \phi_{M_1}}-M_2)s_W c_W & M_1 e^{i \phi_{M_1}}c^2_W+ M_2 s^2_W & M_Z & 0 \\
      0 & M_Z  & \mu e^{i\phi_\mu} s_{2\beta} & -\mu e^{i\phi_\mu} c_{2\beta} \\
      0 & 0 & -\mu e^{i\phi_\mu} c_{2\beta} & -\mu e^{i\phi_\mu} s_{2\beta} \\
    \end{array} \right).
\end{equation}
where the abbreviations $s_W=\sin\theta_W,\, c_W=\cos\theta_W,\, s_{2\beta}=\sin2\beta,\, c_{2\beta}=\cos2\beta$
have been used.

\vspace{0.5cm}

The interaction Lagrangian for $\st$ decay is
\begin{equation}
\mathcal{L}_{q\sq \tilde{\chi}^0_k} =
  g\bar{q}(a^{\sq}_{ik}P_R+b^{\sq}_{ik}P_L)\tilde{\chi}^0_k\sq_i+ \mathrm{h.c.}, 
\end{equation}
where the couplings $a^{\sq}_{ik}$ and $b^{\sq}_{ik}$ are given by
\begin{equation}
 a^{\sq}_{ik}=\displaystyle\sum_{n=1}^2({\cal R}^{\sq}_{in})^*{\cal A}^q_{kn},\;\;\;\;
 b^{\sq}_{ik}=\displaystyle\sum_{n=1}^2({\cal R}^{\sq}_{in})^*{\cal B}^q_{kn}.
\end{equation}
Here ${\cal R}^{\sq}_{in}$ is the mixing matrix of the squarks and 
\begin{equation}
 {\cal A}^{q}_{k}={f_{Lk}^q \choose h_{Rk}^q}, \qquad
 {\cal B}^{q}_{k}={h_{Lk}^q \choose f_{Rk}^q}, 
\end{equation}

with
 \begin{eqnarray}
  f_{Lk}^q &=& -\sqrt{2}\,e_q \sin\t_W N_{k1} 
    -\sqrt{2}\,(T_{3q} -e_q\sin^{2}\t_W)\smaf{N_{k2}}{\cos\t_W}, 
    \hspace{-1cm} \label{eq:fLk}\\
  f_{Rk}^q &=& 
    -\sqrt{2}\,e_q \sin\t_W (\tan\t_W N_{k2}^* - N_{k1}^*), \\
  h_{Lk}^t &=& Y_t \left(N_{k3}^*\sin\b - N_{k4}^*\cos\b \right) \\
                     &=&  (h_{Rk}^t)^* .
 \end{eqnarray}
We also use the following relations from~\cite{Bartl:2004jr}.  The left-right mixing of the stop
 squarks is described by a hermitian $2 \times 2$ mass matrix which reads as follows in the basis
 $(\tilde{t}_L,\tilde{t}_R)$:
\be{eq:stopmass}
{\mathcal{L}}_M^{\st}= -(\st_L^{\dagger},\, \st_R^{\dagger})
\left(\begin{array}{ccc}
M_{\st_{LL}}^2 & e^{-i\phi_{\st}}|M_{\st_{LR}}^2|\\[5mm]
e^{i\phi_{\st}}|M_{\st_{LR}}^2| & M_{\st_{RR}}^2
\end{array}\right)\left(
\begin{array}{ccc}
\st_L\\[5mm]
\st_R \end{array}\right),
\ee
where
\begin{eqnarray}
M_{\st_{LL}}^2 & = & M_{\tilde Q}^2+(\frac{1}{2}-\frac{2}{3}\sin^2\Theta_W)
\cos2\beta \ m_Z^2+m_t^2 ,\label{eq:mll} \\[3mm]
M_{\st_{RR}}^2 & = & M_{\tilde U}^2+\frac{2}{3}\sin^2\Theta_W\cos2\beta \
m_Z^2+m_t^2 ,\label{eq:mrr}\\[3mm]
M_{\st_{RL}}^2 & = & (M_{\st_{LR}}^2)^{\ast}=
m_t(A_t-\mu^{\ast}
\cot\beta), \label{eq:mlr}
\end{eqnarray}
\begin{equation}
\phi_{\st}  = \arg\lbrack A_{t}-\mu^{\ast}\cot\beta\rbrack .
\label{eq:phtau}
\end{equation}
Here $\tan\beta=v_2/v_1$ with $v_1 (v_2)$ being the vacuum
expectation value of the Higgs field $H_1^0 (H_2^0)$,
$m_t$ is the mass of the top quark and
$\Theta_W$ is the weak mixing angle, $\mu$ is the Higgs--higgsino mass parameter
and $M_{\ti Q}$,
$M_{\ti U}, A_t$ are the soft SUSY--breaking parameters of the stop squark
system.
The mass eigenstates $\st_i$ are $(\ti t_1, \ti t_2)=
(\st_L, \st_R) {\mathcal{R}^{\st}}^T$ with
\begin{equation}
\mathcal{R}^{\st}=\left( \begin{array}{ccc}
e^{i\phi_{\st}}\cos\theta_{\st} &
\sin\theta_{\st}\\[5mm]
-\sin\theta_{\st} &
e^{-i\phi_{\st}}\cos\theta_{\st}
\end{array}\right),
\label{eq:rtau}
\end{equation}
where
\begin{equation}
\cos\theta_{\st}=\frac{-|M_{\st_{LR}}^2|}{\sqrt{|M_{\st _{LR}}^2|^2+
(m_{\st_1}^2-M_{\st_{LL}}^2)^2}},\quad
\sin\theta_{\st}=\frac{M_{\st_{LL}}^2-m_{\st_1}^2}
{\sqrt{|M_{\st_{LR}}^2|^2+(m_{\st_1}^2-M_{\st_{LL}}^2)^2}}.
\label{eq:thtau}
\end{equation}
The mass eigenvalues are
\begin{equation}
 m_{\st_{1,2}}^2 = \frac{1}{2}\left((M_{\st_{LL}}^2+M_{\st_{RR}}^2)\mp
\sqrt{(M_{\st_{LL}}^2 - M_{\st_{RR}}^2)^2 +4|M_{\st_{LR}}^2|^2}\right).
\label{eq:m12}
\end{equation}
We note that we have $\phi_{\ti t}\approx\phi_{A_t}$ for $|A_t|\gg |\mu| \cot\beta$.

%%%%%%%
\section{Explicit expressions for the squared amplitude}
%%%%%%%
\subsection{Neutralino production $\tilde{t}_i \to \tilde{\chi}^0_j
  t$\label{sect:stopdecay}}

Here we give the analytic expression for the production density matrix:
\begin{equation}
|M(\tilde{t}_i\to \tilde{\chi}^0_j t)|^2=P(\tilde{\chi}^0_j t)+\Sigma^a_P(\tilde{\chi}^0_j)+
\Sigma^b_P(t)+\Sigma^{ab}_P(\tilde{\chi}^0_j t),
\label{eq_mq}
\end{equation}
whose spin-independent contribution reads
\begin{equation}
P(\tilde{\chi}^0_j t)=\frac{g^2}{2}\left\{(|a_{ij}|^2+|b_{ij}|^2)(p_t
p_{\tilde{\chi}^0_j})-2 m_t m_{\tilde{\chi}^0_j}
Re(a_{ij}b^*_{ij})\right\},
\end{equation}
where $p_t$ and $p_{\tilde{\chi}^0_k}$ denote the four-momenta of the $t$-quark and 
the neutralino $\tilde{\chi}^0_k$.
The coupling constants can be simplified and shown to be
\begin{eqnarray}
 (|a_{ij}|^2+|b_{ij}|^2) = |h^t_{Rj}|^2+\cos^2\t_{\st}|f^t_{Lj}|^2+\sin^2\t_{\st}|f^t_{Rj}|^2 \nonumber\\  +2\sin\t_{\st}\cos\t_{\st} Re\left[ e^{-i\phi_{\st}}h^{t*}_{Rj}(f^t_{Rj}+f^t_{Lj})\right],\\
Re(a_{ij}b^*_{ij})=\cos^2\t_{\st}Re(f^t_{Rj}h^t_{Rj})+\sin^2\t_{\st}Re(f^{t*}_{Rj}h^t_{Rj}) \nonumber\\ +\smaf{1}{2}\sin2\t_{\st}Re\left[e^{i\phi_{\st}}|h^t_{Rj}|^2+e^{-i\phi_{\st}}f^t_{Lj}f^{t*}_{Rj} \right],
\end{eqnarray}
where $\phi_{\st}$ is given in eq.~(\ref{eq:phtau}).

The spin-dependent contributions are T-even and are given by
\begin{eqnarray}
\Sigma^a_P(\tilde{\chi}^0_2) &=& \frac{g^2}{2}\left\{(|b_{ij}|^2-|a_{ij}|^2)m_{\tilde{\chi}^0_j} 
(p_t s^a(\tilde{\chi}^0_j))\right\},\label{eq_prod-ea}\\
\Sigma^b_P(t) &=& \frac{g^2}{2}\left\{(|b_{ij}|^2-|a_{ij}|^2)m_{t} 
(p_{\tilde{\chi}^0_j} s^b(t))\right\},\label{eq_prod-eb}
\end{eqnarray}
where $s^a(\tilde{\chi}^0_j)$ $(s^b(t))$ denote the spin-basis vectors of the neutralino
$\tilde{\chi}^0_j$ (t-quark). Again the coupling constant can be simplified as
\begin{eqnarray}
 (|b_{ij}|^2-|a_{ij}|^2) = \cos2\t_{\st}|h^t_{Rj}|^2-\cos^2\t_{\st}|f^t_{Lj}|^2+\sin^2\t_{\st}|f^t_{Rj}|^2 \nonumber\\  -2\sin\t_{\st}\cos\t_{\st} Re\left[ e^{-i\phi_{\st}}h^{t*}_{Rj}(f^t_{Rj}+f^t_{Lj})\right].
\end{eqnarray}

The three spin-basis four-vectors $s^1$, $s^2$ and $s^3$
form a right-handed system and provide, together with the momentum, an orthogonal basis
system.  They are chosen as:
\begin{eqnarray}
s^1(\tilde{\chi}^0_j)&=&\left(0,\frac{(\vec{p}_{\tilde{\chi}^0_j}\times \vec{p}_{\tilde{t}_i})\times \vec{p}_{\tilde{\chi}^0_j}}{|(\vec{p}_{\tilde{\chi}^0_j}\times \vec{p}_{\tilde{t}_i})\times \vec{p}_{\tilde{\chi}^0_j}|}\right), \label{eq-s1chi}\\
s^2(\tilde{\chi}^0_j)&=&\left(0,\frac{\vec{p}_{\tilde{\chi}^0_j}\times \vec{p}_{\tilde{t}_i}}{|\vec{p}_{\tilde{\chi}^0_j}\times \vec{p}_{\tilde{t}_i}|}\right),
\label{eq-s2chi}\\
s^3(\tilde{\chi}^0_j)&=&\frac{1}{m_{\tilde{\chi}^0_j}}\left( |\vec{p}_{\tilde{\chi}^0_j}|, \frac{E_{\tilde{\chi}^0_j}}{|\vec{p}_{\tilde{\chi}^0_j}|} \vec{p}_{\tilde{\chi}^0_j}\right).
\label{eq-s3chi}
\end{eqnarray}
The spin-system for the top quark has been chosen analogously:
\begin{eqnarray}
s^1(t)&=&\left(0,\frac{(\vec{p}_{t}\times \vec{p}_{\tilde{\chi}^0_j})\times \vec{p}_{t}}
{|(\vec{p}_{t}\times \vec{p}_{\tilde{\chi}^0_j})\times \vec{p}_{t}|}\right), \label{eq-s1t}\\
s^2(t)&=&\left(0,\frac{\vec{p}_{t}\times \vec{p}_{\tilde{\chi}^0_j}}{|\vec{p}_{t}\times \vec{p}_{\tilde{\chi}^0_j}|}\right),
\label{eq-s2t}\\
s^3(t)&=&\frac{1}{m_{t}}\left( |\vec{p}_{t}|, \frac{E_{t}}{|\vec{p}_{t}|} \vec{p}_{t}\right)\label{eq-s3t},
\end{eqnarray}
and $E_t$ and $E_{\tilde{\chi}^0_j}$ denote the energies of the top quark and the neutralino 
$\tilde{\chi}^0_j$, respectively.

The terms that depend simultaneously on the spin of the top quark and of the neutralino can be split into
T-even, $\Sigma^{ab,E}_P(\tilde{\chi}^0_2 t)$, and T-odd, 
$\Sigma^{ab,O}_P(\tilde{\chi}^0_2 t)$, contributions:
\begin{eqnarray}
\Sigma^{ab,E}_P(\tilde{\chi}^0_2 t)&=& \frac{g^2}{2} \Big\{2
Re(a_{ij}b^*_{ij}) [(s^a(\tilde{\chi}^0_j) p_t) (s^b(t)
p_{\tilde{\chi}^0_j}) - (p_t p_{\tilde{\chi}^0_j})
(s^a(\tilde{\chi}^0_j) s^b(t))]\nonumber\\ &&\mbox{\hspace{-2cm}} 
+ m_t m_{\tilde{\chi}^0_j} (s^a(\tilde{\chi}^0_j)s^b(t))
(|a_{ij}|^2+|b_{ij}|^2) \Big\}, \label{eq_prod-e}\\
\Sigma^{ab,O}_P(\tilde{\chi}^0_2 t)&=& - g^2
Im(a_{ij}b^*_{ij}) f_4^{ab}, \label{eq_prod-o}
\end{eqnarray}
where the T-odd kinematical factor is given by
\begin{equation}
f_4^{ab}= \epsilon_{\mu\nu\rho\sigma}s^{a,\mu}(\tilde{\chi}^0_j)p^{\nu}_{\tilde{\chi}^0_j}
s^{b,\rho}(t)p^{\sigma}_t,
\label{eq_f4}
\end{equation}
and the coupling constant by
\begin{eqnarray}
 Im(a_{ij}b^*_{ij})=\cos^2\t_{\st}Im(f^t_{Rj}h^t_{Rj})+\sin^2\t_{\st}Im(f^{t*}_{Rj}h^t_{Rj}) \nonumber\\ +\smaf{1}{2}\sin2\t_{\st}Im\left[e^{i\phi_{\st}}|h^t_{Rj}|^2+e^{-i\phi_{\st}}f^t_{Lj}f^{t*}_{Rj} \right].
\label{eq_Imab}
\end{eqnarray}

%%%%%
\subsection{Neutralino three-body decay $\tilde{\chi}^0_j \to
\tilde{\chi}^0_k \ell^+ \ell^-$ \label{sect:neutdecay}}
%%%%%
Here we give the analytical expressions for the
different contributions to the
decay density matrix
for the three-body decay, where we sum over
the spins of the final-state particles~\cite{MoortgatPick:1999di}.
The contributions independent of the polarisation of the neutralino
$\tilde{\chi}^0_j$
\begin{equation}
D(\tilde{\chi}^0_j)=D(Z Z)+ D(Z \tilde{\ell}_L)+ D(Z
\tilde{\ell}_R)+ D(\tilde{\ell}_L \tilde{\ell}_L)+
D(\tilde{\ell}_R \tilde{\ell}_R) \label{eq_sumz},
\end{equation}
are given by
\begin{eqnarray} D(Z Z)&=& 8
  \frac{g^4}{\cos^4\Theta_W} |\Delta(Z)|^2
  (L_{\ell}^2+R_{\ell}^2) \nonumber\\
 & & \Big[ |O^{''L}_{kj}|^2 (g_1+g_2) +(Re
  O^{''L}_{kj})^2 -(Im O^{''L}_{kj})^2) g_3  \Big],
  \label{eq_dzz}\\
D(Z \tilde{\ell}_L)&=&4 \frac{g^4}{\cos^2\Theta_W} L_{\ell}
  Re\Big\{\Delta(Z) \Big[f^L_{\ell j} f^{L*}_{\ell k}
  \Delta^{t*}(\tilde{\ell}_L)
  (2O^{''L}_{kj} g_1 +O^{''L*}_{kj} g_3) \nonumber\\
 & &\phantom{4 \frac{g^4}{\cos^2\Theta_W} L_{\ell}
     Re\Big\{\Delta(Z) \Big[}  
  +f^{L*}_{\ell j} f^{L}_{\ell k} \Delta^{u*}(\tilde{\ell}_L)
  (2O^{''L*}_{kj} g_2 +O^{''L}_{kj} g_3)
  \Big]\Big\},\label{eq_dzel}\\
D(\tilde{\ell}_L \tilde{\ell}_L)&=& 2 g^4 \Big[ |f^{L}_{\ell j}|^2
  |f^L_{\ell k}|^2 \big(|\Delta^{t}(\tilde{\ell}_L)|^2 g_1
  +|\Delta^{u}(\tilde{\ell}_L)|^2 g_2\big)\nonumber\\
 & &\phantom{2 g^4 \Big[} +Re\big\{(f^{L*}_{\ell j})^2 (f^L_{\ell k})^2
  \Delta^{t}(\tilde{\ell}_L) \Delta^{u*}(\tilde{\ell}_L)\big\}
  g_3 \Big],  \label{eq_delel}
\end{eqnarray}
where $\Delta(Z)$ and $\Delta^{t,u}(\tilde{\ell}_{L})$ denote the propagators of the 
virtual particles in the direct channel and in both crossed channels (labelled $t,u$, cf. 
Fig.\ref{Fig:FeynDecayA}).

The quantities $D(Z\tilde{\ell}_R), D(\tilde{\ell}_R \tilde{e}_R)$
can be derived from eqs.~(\ref{eq_dzel}), (\ref{eq_delel}) by
the substitutions
\begin{equation} \label{eq_substdecayP}
 L_{\ell}\to R_{\ell}, \quad
 \Delta^{t,u}(\tilde{\ell}_L)\to \Delta^{t,u}(\tilde{\ell}_R),\quad
 %\Delta^{u_i}(\tilde{\ell}_L)\to \Delta^{u_i}(\tilde{\ell}_R),\nonumber\\
 O^{''L}_{kj}\to O^{''R}_{kj}, \quad
 f_{\ell j,k}^L\to f_{\ell j,k}^R.
\end{equation}
The kinematical factors are
\begin{eqnarray}
g_1&=&(p_{\tilde{\chi}^0_k} p_{\ell^-})(p_{\tilde{\chi}^0_j} p_{\ell^+}),
\label{eq_dkin1}\\
g_2&=& (p_{\tilde{\chi}^0_k} p_{\ell^+})(p_{\tilde{\chi}^0_j} p_{\ell^-}),
\label{eq_dkin2}\\
g_3 &=& m_j m_k (p_{\ell^-} p_{\ell^+}).\label{eq_dkin3}
\end{eqnarray}
We can split the terms depending on the 
polarization of the neutralino into T-even and T-odd 
contributions:
%%%
\begin{equation}
\Sigma_D^{a}(\tilde{\chi}^0_j)= \Sigma_D^{a,E}(\tilde{\chi}^0_j)
+\Sigma_D^{a,O}(\tilde{\chi}^0_j).
\end{equation}
%%%
The T-even contributions depending on the polarisation of the decaying
neutralino $\tilde{\chi}^0_j$
\begin{equation}
\Sigma_D^{a,E}(\tilde{\chi}^0_j)= \Sigma_D^{a,E}(ZZ) +\Sigma_D^{a,E}(Z
\tilde{\ell}_L) +\Sigma_D^{a,E}(Z \tilde{\ell}_R)
+\Sigma_D^{a,E}(\tilde{\ell}_L \tilde{\ell}_L)
+\Sigma_D^{a,E}(\tilde{\ell}_R \tilde{\ell}_R). \label{eq_dssum-e}
\end{equation}
are
\begin{eqnarray}
\Sigma_D^{a,E}(ZZ)&=& 8 \frac{g^4}{\cos^4\Theta_W} |\Delta(Z)|^2
  (R^2_{\ell}-L_{\ell}^2) \nonumber\\
 & &\times\Big[ -[(Re O^{''L}_{kj})^2 -(Im O^{''L}_{kj})^2]g^a_3+
  |O^{''L}_{kj}|^2(g^a_1-g^a_2) \Big],\label{eq_dszz}\\
\Sigma_D^{a,E}(Z \tilde{\ell}_L)&=& \frac{4
  g^4}{\cos^2\Theta_W}L_{\ell} Re\Big\{\Delta(Z)
  \Big[ f^L_{\ell j} f^{L*}_{\ell k} \Delta^{t*}(\tilde{\ell}_L)
  \big(-2 O^{''L}_{kj} g^a_1 +O^{''L*}_{kj} g^a_3\big)\nonumber\\
 & & \phantom{\frac{4 g^4}{\cos^2\Theta_W}L_{\ell} Re\Big\{}
  + f^{L*}_{\ell j} f^{L}_{\ell k} \Delta^{u*}(\tilde{\ell}_L)
  \big(2 O^{''L*}_{kj} g^a_2 +O^{''L}_{kj} g^a_3\big)
  \Big]\Big\},\label{eq_dszel}\\
\Sigma_D^{a,E}(\tilde{\ell}_L \tilde{\ell}_L) & = & 2 g^4 \Big[
  |f^{L}_{\ell j}|^2 |f^L_{\ell k}|^2
  [|\Delta^{u}(\tilde{\ell}_L)|^2 g_2^a
  -|\Delta^{t}(\tilde{\ell}_L)|^2 g_1^a]\nonumber\\
 & & \phantom{2 g^4 \Big[}
  + Re\big\{ (f^{L*}_{\ell j})^2 (f^L_{\ell k})^2
  \Delta^{t}(\tilde{\ell}_L)
  \Delta^{u*}(\tilde{\ell}_L)g_3^a\big\}\Big],
  \label{eq_dselel}
\end{eqnarray}
where the contributions $\Sigma^{a,E}_D(Z\tilde{\ell}_R),
\Sigma^{a,E}_D(\tilde{\ell}_R \tilde{\ell}_R)$ are derived from
eqs.~(\ref{eq_dszel}), (\ref{eq_dselel}) by applying the substitutions
in eq.~(\ref{eq_substdecayP}) and in addition $g_{1,2,3}^a\to -g_{1,2,3}^a$.

The kinematical factors are
\begin{eqnarray}
g^a_1&=& m_j (p_{\tilde{\chi}^0_k} p_{\ell^-}) (p_{\ell^+} 
s^a), \label{eq422_3a}\\
g^a_2&=& m_j (p_{\tilde{\chi}^0_k} p_{\ell^+}) 
(p_{\ell^-} s^a), \label{eq_dssub4}\\
g^a_3&=& m_k [(p_{\tilde{\chi}^0_j} p_{\ell^+}) (p_{\ell^-} s^a)
-(p_{\tilde{\chi}^0_j} p_{\ell^-}) (p_{\ell^+} s^a)].
\label{eq_dssub5}
\end{eqnarray}
%%%
The T-odd contributions depending on the polarisation of the decaying
neutralino $\tilde{\chi}^0_j$
\begin{equation}
\Sigma_D^{a,O}(\tilde{\chi}^0_j)= \Sigma_D^{a,O}(ZZ) +\Sigma_D^{a,O}(Z
\tilde{\ell}_L) +\Sigma_D^{a,O}(Z \tilde{\ell}_R)
+\Sigma_D^{a,O}(\tilde{\ell}_L \tilde{\ell}_L)
+\Sigma_D^{a,O}(\tilde{\ell}_R \tilde{\ell}_R). \label{eq_dssum-o}
\end{equation}
are
\begin{eqnarray}
\Sigma_D^{a, \mathrm{O}}(ZZ) &=& 
    8 \frac{g^4}{\cos^4\Theta_W} |\Delta(Z)|^2
    (L^2_{\ell} - R_{\ell}^2) 
    \Big[ 2 Re(O^{''L}_{kj}) Im(O^{''L}_{kj}) i g_4^a \Big],
    \label{eq_dszz_to}\\
\Sigma_D^{a, \mathrm{O}}(Z \tilde{\ell}_L) & = & 
    \frac{4 g^4}{\cos^2\Theta_W}L_{\ell}
    Re\Big\{\Delta(Z)
    \Big[-f^L_{\ell j} f^{L*}_{\ell k} O^{''L*}_{kj}
    \Delta^{t*}(\tilde{\ell}_L)\nonumber\\
  & &\phantom{\frac{4 g^4}{\cos^2\Theta_W}L_{\ell}Re\Big\{\Delta(Z)\Big[}
    + f^{L*}_{\ell j} f^{L}_{\ell k} O^{''L}_{kj}
    \Delta^{u*}(\tilde{\ell}_L)
    \Big]g_4^a\Big\},\label{eq_dszel_to}\\
\Sigma_D^{a, \mathrm{O}}(\tilde{\ell}_L \tilde{\ell}_L)&=& 
    2 g^4 Re\Big\{ (f^{L*}_{\ell j})^2 (f^L_{\ell k})^2
    \Delta^{t}(\tilde{\ell}_L)
    \Delta^{u*}(\tilde{\ell}_L) g_4^a\Big\},
    \label{eq_dselel_to}
\end{eqnarray}
where the contributions $\Sigma^{a,O}_D(Z\tilde{\ell}_R),
\Sigma^{a,O}_D(\tilde{\ell}_R \tilde{\ell}_R)$ are derived from
eqs.~(\ref{eq_dszel}), (\ref{eq_dselel}) by applying the substitutions
in eq.~(\ref{eq_substdecayP}).
The kinematical factor is
\begin{eqnarray}
g^a_4 & = & i m_k \epsilon_{\mu \nu \rho \sigma}
 s^{a \mu} p_{\tilde{\chi}^0_j}^{\nu} p_{\ell^-}^{\rho} p_{\ell^+}^{\sigma}. 
\label{eq_dssub6}
\end{eqnarray}

%%%
\subsection{Top decay  $t \to W^+ b$ \label{sect:topdecay}}
%%%

We provide analytical expressions for the 2-body decay of the top quark 
into a $W$-boson and the final-state bottom quark:
\begin{equation}
D(t)=\frac{g^2}{4} \{ m_t^2 -2 m_W^2+\frac{m_t^4}{m_W^2} \}.
\end{equation}
The spin-dependent contribution is T-even and reads:
\begin{equation}
\Sigma_D^{b}(t)=-\frac{g^2}{2} m_t \{ (s^b(t) p_b) + \frac{m_t^2 -m_W^2}{m_W^2}
(s^b(t) p_W) \}.
\label{eq_stdecay}
\end{equation}

%%%%%%%%%%%%%%%%%%%%%%%%%%%%%%%%%%%%%%%%
\section{Kinematics}
%%%%%%%%%%%%%%%%%%%%%%%%%%%%%%%%%%%%%%%%
\label{sec:kinematics}
  \subsection{Phase Space}
  The complete cross section for the process can be decomposed into the production 
cross section and the branching ratios of the subsequent decays:
  \begin{eqnarray}
     d{\sigma_{Total}} &=& d{\sigma}(gg \to \st_1 \overline{\st_1})\,\frac{E_{\st_1
}}{m_{\st_1}\Gamma_{\st_1}}\,
     d\Gamma  (\widetilde t_1 \to t \tilde\chi^0_2)\,\cdot
          \nonumber \\ 
        &&~\frac{E_{\tilde{\chi}^0_2}}{m_{\tilde{\chi}^0_2} \Gamma_{\tilde{\chi}^0_2}}
        \,d\Gamma (\tilde\chi^0_2\to \tilde{\chi}^0_1 l^+ l^-)\,\frac{E_t}{m_t\Gamma_t}\
        d\Gamma (t\to W^{+}b)\, ,\label{dGamma}
      \end{eqnarray}
   where the factors $E/m \Gamma$ come from the use of the narrow-width approximation 
   for the propagators of the $\st$, $\tilde{\chi}^0_2$ and $t$. This approximation is valid for 
   $(\Gamma/m)^2 \ll 1$,
   which is satisfied for $\Gamma_t\sim1.5$ GeV \cite{Hoang:1999zc} and $\Gamma_{\st}\sim4$ GeV. It is also
   trivially satisfied in the case of $\Gamma_{\tilde{\chi}^0_2}\sim10^{-4}$ where the width is small because only the three-body 
   decay is kinematically possible. 
   
   We have:   
     \begin{eqnarray}
          d\Gamma  (\widetilde t_1 \to t  \tilde\chi^0_2)
                 &=&\frac2{E_{\ti t_1}}P(\tilde{\chi}^0_2t) \,d\Phi_{\tilde t}, \\
            d\Gamma  (\tilde{\chi}^0_2 \to \tilde{\chi}^0_1 l^+ l^-)
                 &=&\frac1{4 E_{\tilde{\chi}^0_2}}D(\tilde{\chi}^0_2)\,d\Phi_{\tilde{\chi}^0_2}, \\
           d\Gamma  (t \to W^+b)
                   &=&\frac1{4 E_{t}} D(t)\,d\Phi_{t},
     \end{eqnarray}
        where the phase-space factors in the laboratory system are given by:
      \begin{eqnarray}
          d\Phi_{\tilde t}
                 &=& \frac{1}{(2\pi)^2} \frac{|{\mathbf p}_{\tilde{\chi}^0_2}^{\pm}|^2}
        {2|E_{\st}|{\mathbf p}_{\tilde{\chi}^0_2}^{\pm}|-E_{\tilde{\chi}^0_2}^{\pm}|{\mathbf p}_{\st}
|\mathbf{cos}\theta_{\st}|}~d\Omega_{\st}, \\
            d\Phi_{\tilde{\chi}^0_2 }
                 &=&\frac1{8(2\pi)^5} \frac{E_{l^+}}{||{\mathbf p}_{\tilde{\chi}^0_2}|
\mathbf{cos}\theta_{l^+}-E_{\tilde{\chi}^0_1}-E_{l^+}-E_{l^-} \mathbf{cos}\alpha|}~E_{l^-}
dE_{l^-}d\Omega_{l^+}d\Omega_{l^-},\,\,\,\, \\
           d\Phi_{t}
                   &=&\frac1{(2\pi)^2}\frac{E_b}{2||\mathbf{p}_t| \mathbf{cos}
\theta_b-E_W-E_b|}d\Omega_{b}. 
     \end{eqnarray}     
     There is a subtlety in the phase-space calculation, namely that there can be two solutions
      for ${\mathbf p}_{\tilde{\chi}^0_2}$. If $|{\mathbf p}_{\tilde{\chi}^0_2}| < p_0$ where $p_0
=\lambda^{\frac{1}{2}}(m^2_{\st},m^2_{\tilde{\chi}^0_2},m^2_t)/2m_{\tilde{\chi}^0_2}$, then the 
decay angle, $\theta_{\st}=\measuredangle(\mathbf{p}_{\st}, \mathbf{p}_{t})$, is
 unconstrained and there is only one solution. However, if ${\mathbf p}_{\tilde{\chi}^0_
2} > p_0$, then the angle is constrained by sin$\theta^{max}_{\st}=p_0/|{\mathbf
 p}_{\tilde{\chi}^0_2}|$ and there are two physical solutions
      \begin{eqnarray}\mbox{\hspace{-1cm}}
        | {\mathbf p}_{\tilde{\chi}^0_2}|= 
        \frac{
        (m^2_{\st}+m^2_{\tilde{\chi}^0_2}-m^2_{\t})|\mathbf{p}_{\st}|\cos\theta_{\st}\pm
        E_{\st}\sqrt{\lambda(m^2_{\st},m^2_{\tilde{\chi}^0_2},m^2_{t})-
         4|\mathbf{p}_{\st}|^2~m^2_{\tilde{\chi}^0_2}~(1-\cos^2\theta_{\st})}}
        {2|\mathbf{p}_{\st}|^2 (1-\cos^2\theta_{\st})+2 m^2_{\st}}.
	\label{eq:p+-}
     \end{eqnarray}  
     For the region of phase space where two solutions exist the cross section becomes 
     a summation of the solutions for each of the subsequent decay chains.
  
     \subsection{Integration limits}
     When evaluating the phase-space integral at the parton level, kinematical limits
     need to be determined on some of the variables and these are listed below.
     
     If $|{\mathbf p}_{\tilde{\chi}^0_2}| < p_0$, where
     $p_0=\lambda^{\frac{1}{2}}(m^2_{\st},m^2_{\tilde{\chi}^0_2},m^2_t)/2m_{\tilde{\chi}^0_2}$, there are
     two solutions for ${\mathbf p}_{\tilde{\chi}^0_2}$, eq.(\ref{eq:p+-}), and the decay angle of
     the $\st$ is constrained by
     \begin{equation}
        \mathrm{sin}\theta_{\st}<\frac{\lambda^{\frac{1}{2}}(m^2_{\st},m^2_{\tilde{\chi}^0_2},m^2_t)}{2|\mathbf{p}_{\st}|m_{\tilde{\chi}^0_2}}.
     \end{equation}
 The three-body-decay phase space of the $\tilde{\chi}^0_2$ also has limits:
     \begin{eqnarray}
       E_{\ell^-}&<&\frac{m_{\tilde{\chi}^0_2}-m_{\tilde{\chi}^0_1}}{2(E_{\tilde{\chi}^0_2}-|p_{\tilde{\chi}^0_2}|)}, \\
       \mathrm{cos}\theta_{\ell^-}&<&\frac{2E_{\tilde{\chi}^0_2}E_{\ell^-}+m_{\tilde{\chi}^0_1}-m_{\tilde{\chi}^0_2}}{2E_{\ell^-}|p_{\tilde{\chi}^0_2}|}.
     \end{eqnarray}

\end{appendix}

%%%%%%%%%%%%%%%%%%%%%%%%%%%%

\bibliography{refs}

\end{document}